\documentclass[prd,twocolumn,preprintnumbers,footinbib,tightenlines,nofootinbib,superscriptaddress,aps,10pt]{revtex4-1}

\pdfoutput=1
\usepackage[english]{babel}
\usepackage{amsmath,amssymb,amsfonts, bm,bbm,slashed, subdepth}
\usepackage{graphicx}
\usepackage[sort&compress]{natbib}
\usepackage{xcolor}
\usepackage[normalem]{ulem}
\usepackage{enumerate}
\usepackage{setspace}
\usepackage{booktabs, tabularx}
\usepackage{units}
\usepackage{placeins}
\usepackage{multirow}
\usepackage{mathtools}
\usepackage[utf8]{inputenc}
\usepackage{floatrow}
\usepackage[colorlinks=true,citecolor=blue,linkcolor=blue]{hyperref}
\usepackage{cleveref}

\newfloatcommand{capbtabbox}{table}[][\FBwidth]

\newcommand{\cm}{\ensuremath{\mathrm{cm}}}

\newcommand{\eV}{\ensuremath{\mathrm{eV}}}

\newcommand{\MeV}{\ensuremath{\mathrm{MeV}}}
\newcommand{\GeV}{\ensuremath{\mathrm{GeV}}}

\begin{document}

\preprint{PI/UAN-2019-663FT}

\title{Self-interacting dark matter without prejudice}

\author{Nicol{\'a}s Bernal}
\email{nicolas.bernal@uan.edu.co}
\affiliation{Centro de Investigaciones, Universidad Antonio Nariño,
Carrera 3 Este \# 47A-15, Bogotá, Colombia}
\author{Xiaoyong Chu}
\email{xiaoyong.chu@oeaw.ac.at }
\affiliation{Institute of High Energy Physics, Austrian Academy of Sciences, Nikolsdorfer Gasse 18, 1050 Vienna, Austria
}
\author{Suchita Kulkarni}
\email{suchita.kulkarni@oeaw.ac.at}
\affiliation{Institute of High Energy Physics, Austrian Academy of Sciences, Nikolsdorfer Gasse 18, 1050 Vienna, Austria
}\author{Josef Pradler}
\email{josef.pradler@oeaw.ac.at}
\affiliation{Institute of High Energy Physics, Austrian Academy of Sciences, Nikolsdorfer Gasse 18, 1050 Vienna, Austria
}

\begin{abstract}
  The existence of dark matter particles that carry phenomenologically
  relevant self-interaction cross sections mediated by light dark
  sector states is considered to be severely constrained through a
  combination of experimental and observational data. The conclusion
  is based on the assumption of specific dark matter production
  mechanisms such as thermal freeze-out together with an extrapolation
  of a standard cosmological history beyond the epoch of primordial
  nucleosynthesis.
  In this work, we drop these assumptions and examine the
  scenario from the perspective of the current firm knowledge we have:
  results from direct and indirect dark matter searches and
  cosmological and astrophysical observations, without additional
  assumptions on dark matter genesis or the thermal state of the very early universe.
  We show that even in the minimal set-up, where dark matter particles
  self-interact via a kinetically mixed vector mediator, a significant
  amount of parameter space remains allowed. Interestingly, however,
  these parameter regions imply a meta-stable, light mediator, which
  in turn calls for modified search strategies.
\end{abstract}

\maketitle

\section{Introduction}
Cold dark matter (DM) contributes 26\% to the present day energy
budget of the universe. Whereas its existence has been firmly inferred
from its gravitational effects in a large body of astrophysical
large-scale observations,
a `mass deficit' in small-scale halos seems to persist, such as the
so-called `core vs.~cusp
problem'~\cite{Flores:1994gz,Moore:1994yx,Oh:2010mc,Walker:2011zu} and
`too-big-to-fail problem'~\cite{BoylanKolchin:2011de,
  BoylanKolchin:2011dk, Garrison-Kimmel:2014vqa,
  Papastergis:2014aba}. This deficit may be due to systematic errors
introduced in subtracting the DM distribution from visible
objects~\cite{Blok:2002tr, Rhee:2003vw, Gentile:2005de,
  Spekkens:2005ik, Valenzuela:2005dh, Dalcanton:2010bp,
  Kormendy:2014ova, 2016MNRAS.462.3628R, Maccio:2016egb,
  2017A&A...601A...1P, Brooks:2017rfe, Oman:2017vkl,
  2018MNRAS.474.1398G, Read:2018pft}, or due to the lack of
understanding of baryonic effects on halo evolution~\cite{Navarro:1996bv,
  MacLow:1998djk, Gelato:1998hb, Binney:2000zt, Gnedin:2001ec,
  ElZant:2001re, Weinberg:2001gm, Ahn:2004xt, Tonini:2006gwz,
  Governato:2009bg, Silk:2010aw, VeraCiro:2012na, Martizzi:2011aa,
  Read:2018fxs}.
A third, and not mutually excluding possibility may be that those
pertinent small-scale problems point to new dynamics in the dark
sector, in particular to the possibility that DM particles $\chi$
self-interact with sizable non-gravitational
strength~\cite{Spergel:1999mh,Wandelt:2000ad}.

Self-interactions among DM particles ---provided that they are frequent
enough--- lead to a heat transfer that initially decreases the density
contrast in the center of DM halos. As DM particles evacuate from the
central regions, cuspy density profiles turn into cored ones, overall
reducing the halo mass concentration and potentially explaining the
`mass-deficit' problem. Concretely, it is found that the required
self-scattering cross section over DM mass,
$\sigma_{\chi\chi}/m_{\chi}$, needs to be larger than $0.1-2$~cm$^2$/g
at the scale of dwarf
galaxies~\cite{Vogelsberger:2012ku,Rocha:2012jg,Peter:2012jh,Zavala:2012us,Vogelsberger:2014pda,Elbert:2014bma,Kaplinghat:2015aga,Fry:2015rta}.
On the flip side, the non-observation of an offset between the mass
distributions of DM and hot baryonic gas in the Bullet Cluster
constrains the same ratio to
$\sigma_{\chi\chi}/m_{\chi}<1.25$~cm$^2$/g at
$68\%$~CL~\cite{Clowe:2003tk,Markevitch:2003at,Randall:2007ph},
\textit{i.e.}, approximately 1~barn 
for a 1~GeV DM mass particle. This tension is further exacerbated by recent
observations of cluster collisions, implying
$\sigma_{\chi\chi}/m_{\chi}<0.2-0.5$~cm$^2$/g~\cite{Harvey:2015hha,
  Bondarenko:2017rfu, Harvey:2018uwf}.

Among particle physics models which can produce such large
self-interaction cross sections are the scenarios where
self-interactions are mediated by light
states~\cite{Spergel:1999mh,Feng:2009hw, Buckley:2009in,Tulin:2012wi,
  Tulin:2013teo}.  
Such scenarios have been tightly constrained by linking the
experimental observations with the assumption that DM particles stay
in thermal contact with the SM bath before they freeze out
(\textit{e.g.}~\cite{Kaplinghat:2013yxa, Abdullah:2014lla,
  DelNobile:2015uua, Bernal:2015ova, Bringmann:2016din,
  Duerr:2018mbd}).  However, this needs not to be the case, for
instance if the DM abundance was generated through
freeze-in~\cite{McDonald:2001vt,Hall:2009bx,Bernal:2017kxu}, if the
reheating temperature is lower than the DM
mass~\cite{deSalas:2015glj}, or if the thermal history of the universe
was non-standard~\cite{Barrow:1982ei}.

The purpose of this work is to sever this link and to be agnostic
about the production mechanism for DM in the early universe. In fact,
many alternative mechanisms to generate the observed DM abundance
beyond thermal freeze-out mechanism exist, and even modifications to
the latter are entirely possible.
In fact, DM production in the context of non-standard cosmologies has
recently gained increasing interest, see
\textit{e.g.}~\cite{Davoudiasl:2015vba, Randall:2015xza,
  Tenkanen:2016jic, Dror:2016rxc, DEramo:2017gpl, Hamdan:2017psw,
  Visinelli:2017qga, Drees:2017iod,
  Bramante:2017obj,Bernal:2018ins,DiMarco:2018bnw, DEramo:2017ecx,
  Maity:2018dgy, Garcia:2018wtq, Bernal:2018kcw, Arbey:2018uho,
  Drees:2018dsj, Betancur:2018xtj, Maldonado:2019qmp, Poulin:2019omz,
  Tenkanen:2019cik, Arias:2019uol, Chanda:2019xyl}.  Such scenarios enlarge the
parameter space relative to standard freeze-out case, allowing, for
example, for smaller annihilation cross-sections and/or higher DM
masses.

In light of this, it is the purpose of this work to explore the
self-interacting dark matter (SIDM) scenario under the following
assumptions:\footnote{We focus on symmetric DM throughout the work, and briefly comment on the asymmetric case at the end of Sec.~\ref{sec:exper-constr-kinet}.}
\begin{enumerate}
\item in lieu of firm knowledge of the thermal state of the early
  universe, \textit{i.e.}~for $T\gtrsim 1\,\MeV$, the DM production
  mechanism remains unspecified.  However,
\item we require that at $T \gtrsim 1$~eV the DM abundance has matched
  onto the CMB-observed value, while acknowledging the possibility that
  additional restrictions in the window $1\,\eV < T < 1\,\MeV$ may
  apply.
\end{enumerate}
Based on these assumptions, and requiring that the self-interaction
cross section is of suitable strength, we obtain the parameter space
that is allowed by current direct DM searches, and by astrophysical
and cosmological observations once we couple the dark sector to the
standard model (SM).  It will be shown that there is a large viable
range in the latter coupling, but generically such that it points
towards a long-lived force mediator. This in turn, calls for specific
search-strategies for indirect DM detection and provides a target for
what has been termed the `lifetime-frontier' in the search of new
physics.

The paper is organized as follows: in Sec.~\ref{sec:setup} we
introduce a simple, prototypical SIDM model featuring a fermionic DM
candidate $\chi$ and a light dark photon mediator $V$. Imposing the
required self-scattering cross section allows us in
Sec.~\ref{sec:exper-constr-kinet} to place constraints on the kinetic
mixing parameter in the plane of DM and mediator mass, the main
results of this paper.  We conclude in Sec.~\ref{sec:conclusions}.

\begin{figure}[t]
\includegraphics[height=0.94\textwidth]{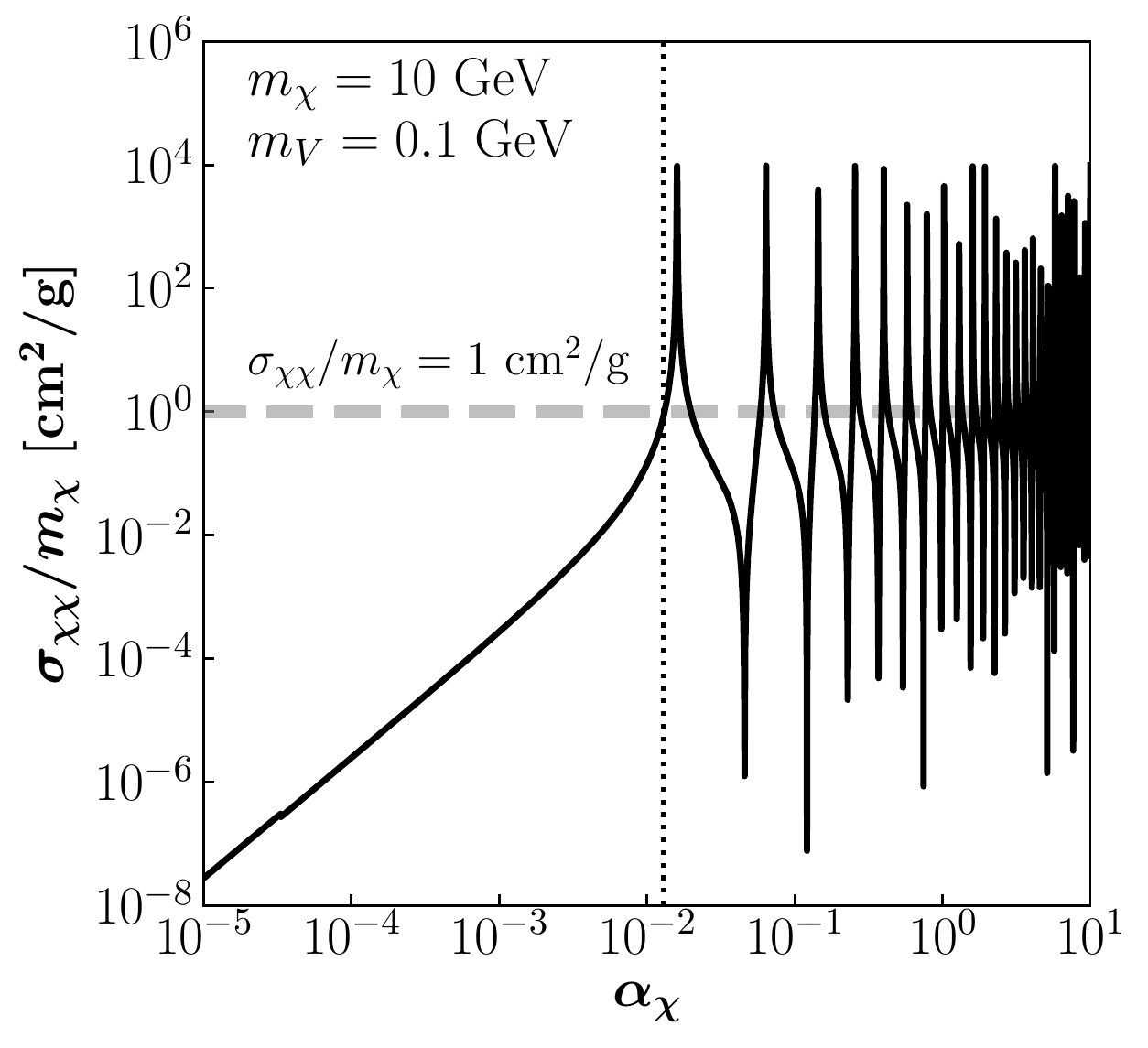}
\caption{DM
  self-scattering cross section per DM mass,
  $\sigma_{\chi\chi}/m_\chi$, as a function of $\alpha_\chi$, for
	$m_\chi=10$~GeV and $m_V=0.1$~GeV, for a relative DM
velocity $v = 10$~km/s. The horizontal dashed line
  corresponds to $\sigma_{\chi\chi}/m_{\chi} = 1$~cm$^2$/g.  The
  vertical dotted line ($\alpha_\chi\simeq1.3\times 10^{-2}$) depicts
  the first solution where the latter value of the self-scattering cross section  is met.
}
\label{fig:som_g}
\end{figure}

\section{Self-interacting Dark Matter}
\label{sec:setup}

For concreteness, in this work we consider the scenario of fermionic
DM $\chi$ that is coupled to the SM through
a dark vector portal
\begin{equation}
	{\mathcal L} \supset -\frac{\epsilon_Y}{2}V_{\mu\nu} F_Y^{\mu\nu}	 + g_\chi (\bar \chi \gamma^\mu \chi )V_\mu \, .
\end{equation}
Here, $V_{\mu\nu}$ and $F_Y^{\mu\nu}$ are the field strengths of the
dark vector $V$ and SM hypercharge, respectively. The kinetic mixing
with strength $\epsilon_Y$ induces the interaction between charged SM
fermions $f$ and $V$ via $\epsilon\, q_f e (\bar f \gamma^\mu f)V_\mu$,
where $\epsilon \equiv \epsilon_Y \cos\theta_W $ and $q_f$ is the
charge in units of the electromagnetic coupling $e$. As we will focus
on mediator masses well below the electroweak scale, we may neglect
its suppressed mixing with the SM $Z$ boson. Instead of the DM gauge
coupling $g_{\chi}$ we will characterize the strength of DM
interaction by a dark fine structure constant
$\alpha_{\chi} \equiv g_\chi^2/(4\pi)$.

SIDM may alleviate small-scale structure problems if the
self-interaction cross section per particle mass,
$\sigma_{\chi\chi}/m_{\chi}$ is in the ballpark of
\begin{equation}\label{eq:som}
	{\sigma_{\chi\chi}}/{m_\chi} = 1~\text{cm}^2/\text{g}.
\end{equation}
In this work, we use this as a requirement and fix the DM
self-interaction to the value Eq.~\eqref{eq:som} at a relative DM
velocity $v= 10$~km/s, typical for dwarf galaxies. As is well known,
the DM self interaction cross section $\sigma_{\chi\chi}$ has resonant
structure~\cite{Tulin:2013teo}, and multiple solutions to
Eq.~\eqref{eq:som} in $\alpha_{\chi}$ exist. In the following, and
throughout this work, we therefore take the minimum value of
$\alpha_{\chi}$ that satisfies the equation above for each parameter
set of $(m_\chi,\,m_V)$ in this way. This choice generates the most
conservative bounds on the parameter space, as solutions with larger
$\alpha_\chi$ lead to stronger interactions between DM and SM
particles.\footnote{We checked numerically that this is true in the
  parameter region of interest for both bounds, from direct detection
  and energy injection discussed below.}

\begin{figure}[t]
\includegraphics[height=0.94\textwidth]{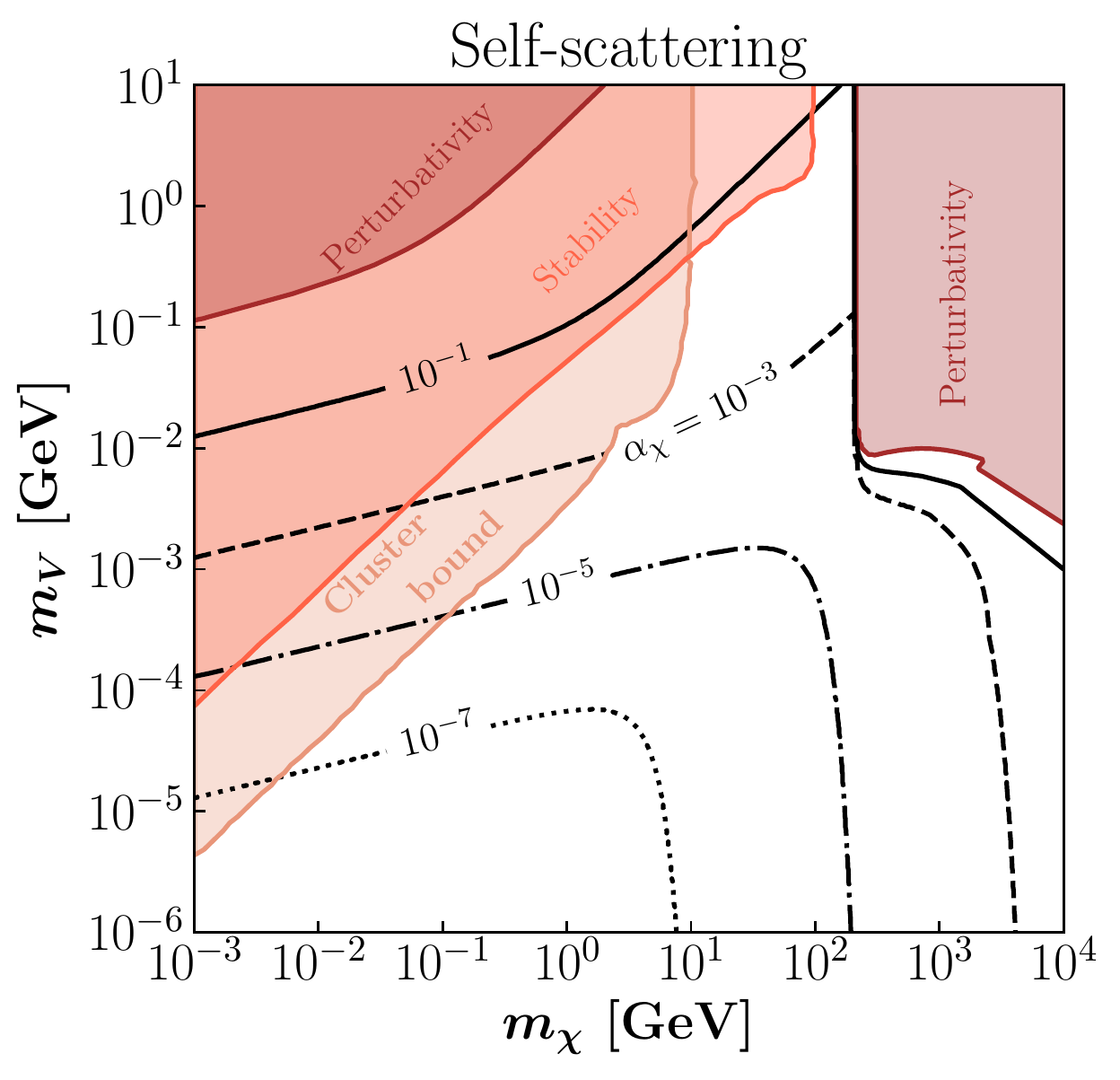}
\caption{Contours of minimal values of $\alpha_\chi$ satisfying
  Eq.~\eqref{eq:som}, \textit{i.e.}~implying a phenomenologically
  relevant DM self-scattering cross section.  Shaded regions show the
  bounds from perturbativity (dark shading), stability of DM halos
  (medium shading), and cluster bounds (light shading).}
\label{fig:contour_g}
\end{figure}

As an example, Fig.~\ref{fig:som_g} shows $\sigma_{\chi\chi}/m_\chi$
as a function of $\alpha_\chi$, for $m_\chi=10$~GeV and $m_V=0.1$~GeV, for a relative DM
velocity $v = 10$~km/s.
The peak structure is characteristic of the quantum resonant regime
where the scattering cross section has a non-trivial velocity
dependence.  For this benchmark, the corresponding solution adopted is
$\alpha_\chi\simeq 1.3\times 10^{-2}$ (dotted vertical line).
Additionally, Fig.~\ref{fig:contour_g} depicts contour levels for the
minimal values of $\alpha_\chi$ satisfying Eq.~\eqref{eq:som}.

The value of $\alpha_\chi$  needs to further satisfy several
conditions. First, the cluster bounds on SIDM constrain the cross
section to $\sigma_{\chi\chi}/m_\chi \lesssim 0.5$~cm$^2/$g at
$v=10^3$~km/s.%
\footnote{Albeit potential uncertainties in deriving the constraint
  from astrophysical data, the limit nevertheless provides a useful
  benchmark point and we use it at face value, for simplicity.}
Compatibility with Eq.~\eqref{eq:som} therefore requires the
self-scattering cross section to be velocity dependent. We are hence
in the light-mediator regime $m_\chi > m_V$ where the typical momentum
transfer in the scattering exceeds the mediator mass.  Second, we
impose that the observed DM density in the centers of dwarf-sized
halos ($\sim 10^8\,M_\odot$~kpc$^{-3}$) should not be diminished by DM
annihilation within their lifetime ($\sim 10^{10}$~yr), leading to
\begin{equation}
	\langle \sigma_\text{an} v\rangle \lesssim  4\times 10^{-19}\left(\frac{m_\chi}{\text{GeV}}\right)\,\text{cm}^3/\text{s} \, 
\end{equation}
at $v=10$~km/s. This condition also guarantees that DM
annihilation decouples at $T\gg 1$~eV, so that the CMB observations of
the DM abundance are not affected.  Finally, we require
$\alpha_\chi \lesssim 10$ as the perturbativity condition. In
practice, this excludes DM masses above the electroweak scale for
$m_V$ above tens of~MeV.

Figure~\ref{fig:contour_g}
summarizes the above constraints in the $[m_{\chi},\,m_V]$ plane. The
bounds that are violated in the shaded regions are from perturbativity
(dark red), stability of DM halos (medium red), and clusters (light
red).  On the left side of the plane, the DM mass is at or below
the GeV-scale. With decreasing $m_{\chi}$, \textit{i.e.}~with
increasing occupation number of DM particles, the constraint on
late-time annihilation (`stability') and self-scattering in clusters
(`cluster bound') become more severe, implying a maximum mediator
mass. Among the two, the cluster bound dominates, as the
self-scattering cross section must be guaranteed to remain
velocity-dependent, setting the most stringent limit on the value of
$m_V$. On the right side, the only constraint is from
perturbativity. The sharp onset of the limit at
$m_\chi\simeq 250\,\GeV$ for $m_V\gtrsim 10\,\MeV$ is because the
resonance self-scattering strength is insufficient to reach Eq.~\eqref{eq:som} with perturbative values of $\alpha_\chi$.

\begin{figure*}[t]
\begin{center}
\includegraphics[height=0.45\textwidth]{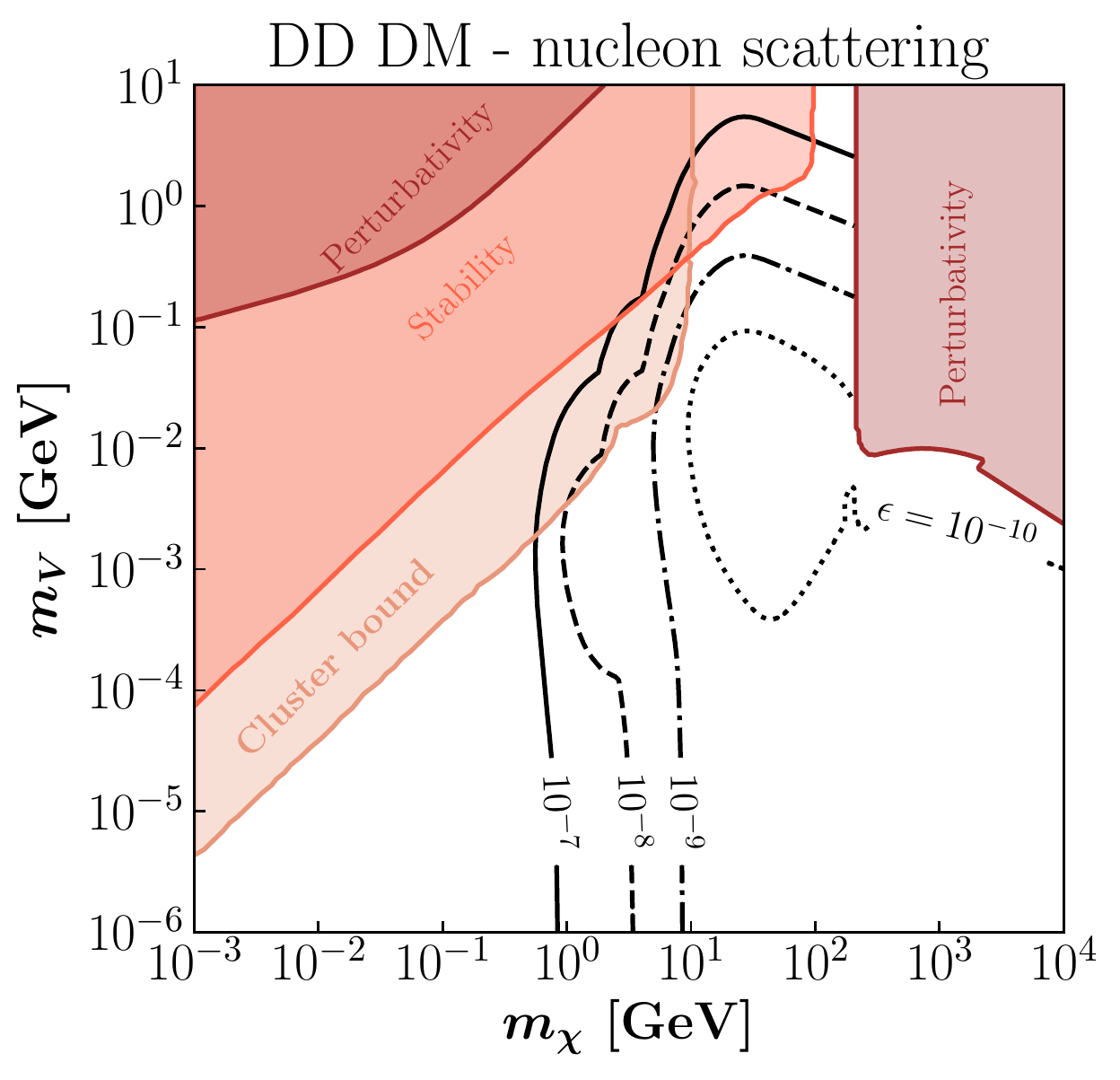}
\includegraphics[height=0.45\textwidth]{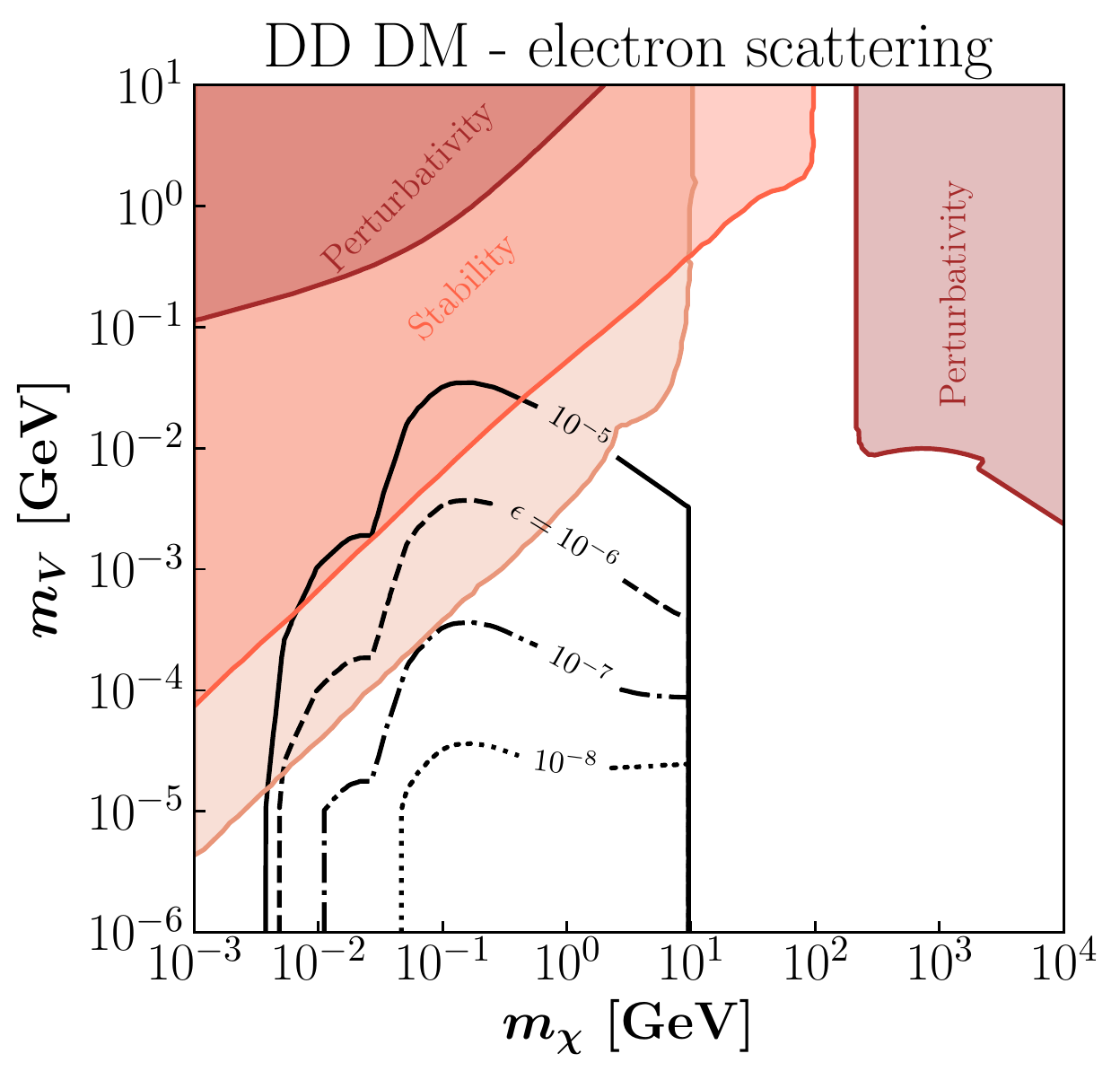}
\caption{Contours of maximally allowed values of the kinetic mixing
  parameter $\epsilon$ from a combination of current DM direct
  detection data, for DM-nucleon scattering~\cite{Angloher:2015ewa,
    Agnese:2015nto, Akerib:2016vxi, Aprile:2018dbl} in the left panel
  and DM-electron scattering~\cite{Essig:2012yx, Essig:2017kqs, Abramoff:2019dfb,
    Aprile:2019xxb} in the right panel.  }
\label{fig:eps_DD}
\end{center}
\end{figure*}

\section{Experimental constraints on kinetic mixing}
\label{sec:exper-constr-kinet}

We now discuss the interaction of DM with the visible sector, induced
by the Lagrangian term $\epsilon\,e (\bar f \gamma^\mu f)V_\mu$.
In this section we show that if thermal freeze-out for DM production
is not assumed, experimental bounds can be alleviated, and upper
bounds on the kinetic mixing parameter $\epsilon$ can instead be
extracted.

\subsection{Scattering between DM and SM particles}
For portal interactions induced by the mediator $V$, the
spin-independent differential scattering cross section between DM and
protons ($i=p$) or electrons ($i=e$) can be written in terms of
\begin{equation}
  \frac{d\sigma^{\rm{SI}}_i}{dE_R} =
  \frac{8\pi\,\alpha\,\alpha_{\chi}\,\epsilon^2\,m_i}{v^2\,(m^2_V +2m_i E_R )^2}\,,
\end{equation} 
where $v$ is the DM velocity, and $E_R$ the kinetic recoil energy of
target $i$ of mass $m_i$.
Figure~\ref{fig:eps_DD} summarizes the resulting direct detection
limits from both DM-nucleus (left panel) and DM-electron (right panel)
scatterings, which are discussed in detail below.

{\bf  Nucleon recoil direct detection.}
We constrain the size of the DM-proton scattering cross section
$\sigma_p$ by taking into account results from four different
experiments.  Among these are the low threshold experiments CDMSlite
with an exposure of 70 kg days, which constrains DM masses from 1.5~GeV
onwards~\cite{Agnese:2015nto} and CRESST-II experiment with 52 kg live
days, which constrains DM mass above 0.5~GeV~\cite{Angloher:2015ewa}. For
detectors with higher threshold, the LUX experiment with exposure of
$3.35\times10^4$~kg~day sets limits for masses above
7~GeV~\cite{Akerib:2016vxi}, while the latest XENON1T results from
1~ton~year exposure limits DM mass above 6~GeV~\cite{Aprile:2018dbl}.

It should also be noted that the direct detection limits depend on
the mediator mass. If $m_V\lesssim 10^{-3} m_{\chi}$, the momentum
exchange in the dark vector propagator is resolved. In the case of
DM-proton collisions, the effect of the light mediator mass has been
implemented by using the public tool DDCalc~\cite{Workgroup:2017lvb,
  Athron:2018hpc}. Also note that in the translation of general
constraints on the DM-nucleon cross section $\sigma_n$, the relation
to $\sigma_p$ is $\sigma_n = (Z/A)^2\,\sigma_p$ where $A$ ($Z$) are
the atomic mass (charge) number of the target nucleus. 

{\bf Electron recoil direct detection.}  For DM-electron scattering,
we use the latest results from the SENSEI
experiment~\cite{Abramoff:2019dfb} together with data from the XENON
experiment.  We further use the limits derived from reanalyzing
XENON10 data~\cite{Essig:2012yx, Essig:2017kqs}, and the latest
official XENON1T limits~\cite{Aprile:2019xxb}. The  limits are
obtained using the so-called `S2-only' analysis, where
electroluminescence produces a secondary scintillation. The three
limits cover complementary DM mass ranges. At every point in parameter
space, we use the most stringent upper limit on scattering cross
section given the mass of the DM.

The DM-electron scattering cross section is conventionally expressed
as the cross section on a free electron at a reference momentum
transfer $q_\text{ref}\equiv \alpha\,m_e$,
\begin{equation}
  \bar \sigma^{\rm{SI}}_e = \frac{16\pi\,\alpha\,\alpha_{\chi}\,
    \epsilon^2\,\mu^2_{\chi e}}{(m^2_V+q^2_\text{ref})^2}\,,
\end{equation}
where $\mu_{\chi e}$ is the DM-electron reduced
mass~\cite{Emken:2019tni}. The heavy mediator limit
$m_V \gg q_\text{ref}$ applies to most of the considered parameter
region. For $m_V$ below 10~keV, a DM  form factor
\begin{equation}
	F_\text{DM}(E_R) = \frac{m_V^2 + q^2_\text{ref}}{m_V^2 + 2m_e E_R }
\end{equation}
should be taken into account in the standard calculation of recoil
events~\cite{Essig:2011nj}. Here, we instead conservatively assume the
bound  on $\epsilon$ at $m_V= 10$~keV applies to smaller $m_V$ as well.

{\bf  DM-SM scattering in astrophysics.}
Beside the direct detection bounds, DM-SM interactions are especially
constrained from astrophysical considerations. These include
cosmic-ray attenuation by scattering with DM~\cite{Cappiello:2018hsu},
CMB and large-scale structures modified by momentum-transfer with
DM~(\textit{e.g.}~\cite{Boehm:2000gq, Boehm:2004th, Dvorkin:2013cea,Xu:2018efh,
  Boddy:2018wzy, Nadler:2019zrb}), among other probes.%
\footnote{We do not consider the limit $m_V\to 0$, for which the resulting
  long-range force can be constrained differently; see
  \textit{e.g.}~\cite{McDermott:2010pa, Kadota:2016tqq, Stebbins:2019xjr}.}

Most of them are typically weaker than the bounds from nucleon (electron)
direct detection experiments for GeV (MeV) scale DM. One example is
the limit derived with direct detection and neutrino experiments,
utilizing solar- or cosmic ray-upscattered DM~\cite{An:2017ojc, Emken:2017hnp,
  Bringmann:2018cvk,Ema:2018bih, Cappiello:2019qsw}. For
velocity-independent scatterings, the upper bound on $\sigma_{\chi N}$
can be as stringent as $10^{-31}$\,cm$^2$ for MeV
DM~\cite{Bringmann:2018cvk, Cappiello:2019qsw}, but becomes much
weaker in presence of a light
mediator~\cite{Bondarenko:2019vrb}. This is because the upscattering process with cosmic rays is dominated
by energy-exchange much larger than that in direct detection
experiments, and thus does not get enhanced even if the mediator
particle is much lighter than the reduced mass of the system. Taking
$m_\chi=10$~MeV and $m_V = 10$~keV, we have calculated that the upscattered DM flux spectrum is approximately
\begin{equation}
	\frac{d\Phi_{\chi}}{dE_{\chi}} \simeq 10^2\left(\frac{\epsilon^2\,\alpha_\chi}{\alpha}\right)\left(\frac{\text{keV}}{E_\chi - m_\chi}\right)^2 \,\text{cm}^{-2}\text{sec}^{-1}\text{keV}^{-1}\,
\end{equation}
taking into account electron, proton, and helium cosmic rays, which
results in $\epsilon \lesssim 10^{-2}$ using the electron-recoil data
from XENON1T; nuclear recoil data does not lead to a useful
constraint.

\subsection{Energy injection from dark particles}

\begin{figure}[t]
\begin{center}
\includegraphics[height=0.94\textwidth]{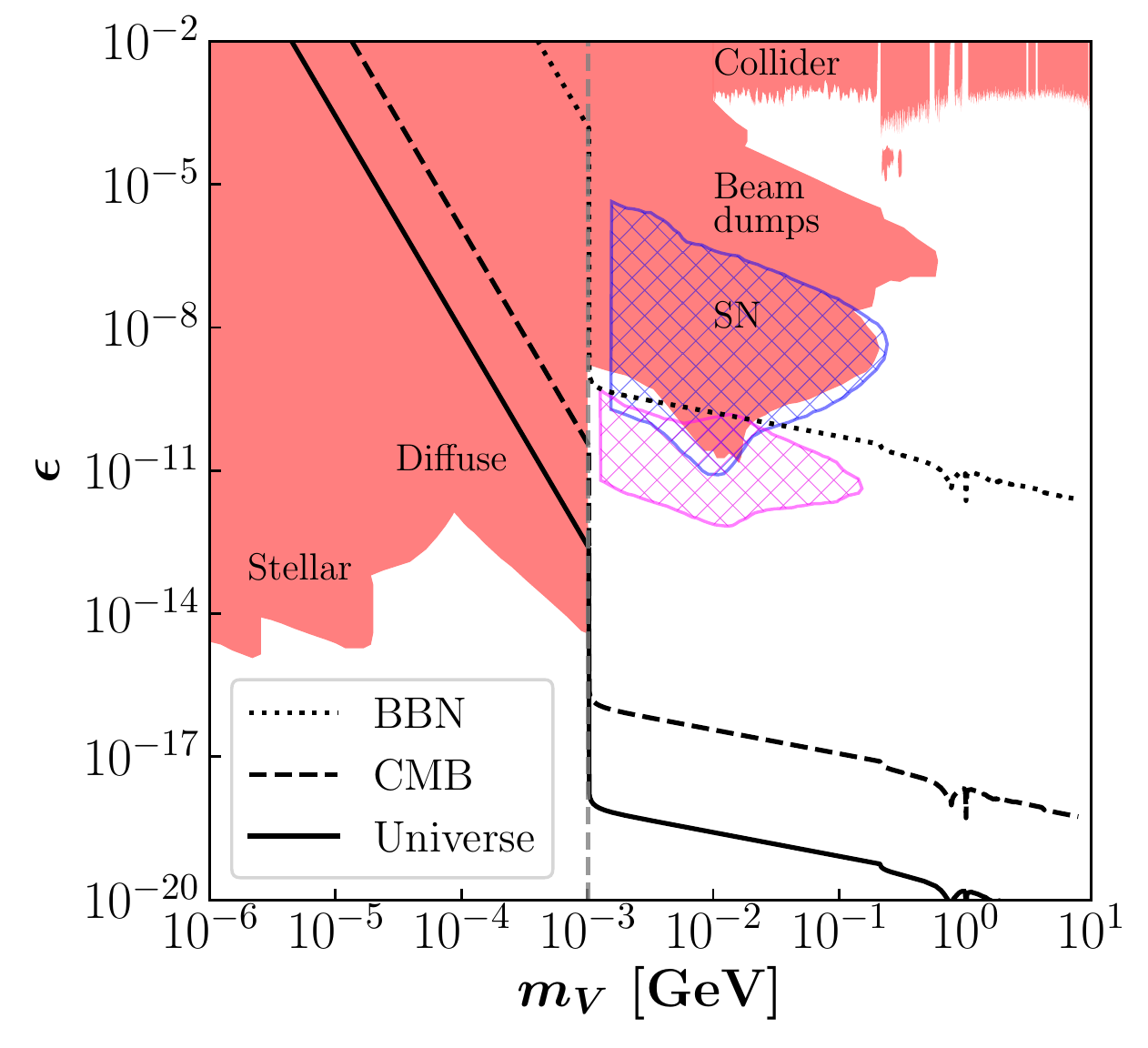}
\caption{Summary of existing constraints of the dark photon
  $[\epsilon,\,m_V]$ parameter space. The various labels summarize
  principal detection strategies, astrophysical constraints from
  stellar energy loss (`stellar')~\cite{An:2013yfc}, and diffuse
  $\gamma$-ray background (`diffuse')~\cite{Redondo:2008ec}, from
  cooling of the proto-neutron star of SN1987A
  (`SN')~\cite{Chang:2016ntp}, from beam-dump and collider
  experiments, see \textit{e.g.}~\cite{Ilten:2018crw,Aaij:2019bvg} and
  references therein. Hatched regions show recent exclusions derived
  from gamma-ray signatures from SN~\cite{DeRocco:2019njg} (blue) and
  energy-transfer argument inside SN~\cite{Sung:2019xie} (pink). In
  addition, contours of mediator lifetime $\tau_V$ equal to 1~sec
  (BBN), $10^{13}$~sec (CMB), as well as $t_U$ (universe) are shown.
  The vertical dashed gray line corresponds to
  $m_V=2\,m_e$. Additional cosmological constraints from BBN and CMB
  for $m_V > 1\,\MeV$ are not shown, as they rely on a $V$-abundance
  created at $T> 1\,\MeV$~\cite{Fradette:2014sza}.  }
\label{fig:eps_DP}
\end{center}
\end{figure}

In the chosen scenario, the effects associated with energy injection
are induced by the produced abundance of $V$ particles and their
subsequent decay with lifetime $\tau_V$. The dark vectors may
\textit{e.g.}~be produced via freeze-in from the SM
sector~\cite{Redondo:2008ec,An:2013yfc,Fradette:2014sza} or via DM
annihilation~\cite{Pospelov:2007mp}. In accordance with our
assumptions made in the introduction, we neglect any population of $V$
that may have emerged prior to $T=1\,\MeV$; we however do take into
account an abundance of $V$ that arises from (residual) $\bar\chi\chi$
annihilation and from direct production of $V$ particles from the SM
bath below a photon temperature of $1\,\MeV$.

Potential cosmological constraints on late $V$ decay are governed by
the $V$ lifetime. They are located in the $[m_V,\,\epsilon]$ parameter
plane along the contours of constant lifetime, shown in
Fig.~\ref{fig:eps_DP}.  Solid, dashed and dotted lines correspond to
$\tau_V=t_U\simeq 4\times 10^{17}$~sec (age of the universe),
$\tau_V=t_\text{CMB}=1.2\times 10^{13}$~sec, and
$\tau_V=t_\text{BBN}=1$~sec, respectively.  For masses lighter than
$m_V=2\,m_e\sim 1$~MeV (vertical gray line), the mediator can only
decay via (loop-induced) processes into three photons or two
neutrinos.%
\footnote{See~\cite{McDermott:2017qcg} for a more precise calculation of the width for a dark photon decaying to three photons  for $m_V \sim 2\,m_e$.}
The relevant bounds on dark photon production from the SM sector
reproduced in Fig.~\ref{fig:eps_DP} are taken
from~\cite{DeRocco:2019njg,Sung:2019xie,Ilten:2018crw,Aaij:2019bvg,An:2013yfc,Chang:2016ntp}. Bounds from a
freeze-in production of $V$ at $T>1\,\MeV$ derived
in~\cite{Fradette:2014sza} are not applied, in accordance with our
assumptions.

The mediator abundance produced from DM annihilation for
$T\leq 1\, \MeV$ is governed by the non-relativistic annihilation
cross section of $\chi \bar\chi\to VV$,
\begin{equation}\label{eq:anni}
	\langle\sigma_\text{an} v\rangle = S(v) \times \frac{\pi\,\alpha_\chi^2}{m_\chi^2}\sqrt{1-\frac{m_V^2}{m_\chi^2}}\,,
\end{equation}
where $S(v)$ is the Sommerfeld enhancement factor~\cite{Cassel:2009wt, Tulin:2013teo}, expressed as  
\begin{equation}
	S(v)=\frac{2\pi\,\alpha_\chi\,\sinh\left(\frac{\pi\,m_\chi\,v}{m_V\,\kappa}\right)}{v\cosh\left(\frac{\pi  m_\chi v}{m_V \kappa } \right)  - v \cos\left(2 \pi \sqrt{\frac{\alpha_\chi m_\chi}{m_V \kappa} -\frac{m_\chi^2 v^2}{4 m_V^2 \kappa^2 }}\right)}\,,
\end{equation}
where $\kappa = 1.6$, consistent with our scattering cross section
formula adopted from \cite{Tulin:2013teo}.  This enhancement only
plays a role for $m_\chi \gtrsim 10$~GeV, and we have checked
that a more careful treatment of the enhancement~\cite{Blum:2016nrz}
does not affect our result qualitatively.
Moreover, although radiative bound state formation can happen for
$m_\chi \gtrsim 1$~TeV and
$m_V\le\frac14\alpha_\chi^2\,m_\chi$~\cite{Pospelov:2008jd}, this
process only improves the bounds mildly~\cite{Petraki:2016cnz,
  Cirelli:2016rnw}, and will not be further considered here.
Equation~\eqref{eq:anni} then allows to study the effects of energy
injection from DM annihilation at different epochs, and to estimate
the corresponding constraints.

{\bf Energy injection during the BBN epoch.} Energy injection from
dark particles after $T\lesssim 1\,\MeV$ may affect the primordial
abundances of light elements, such as $^4$He and deuterium, see
\textit{e.g.}~\cite{Pospelov:2010hj} and references
therein.
If a population of $\chi$-particles is already present at $T=1\,\MeV$,
their annihilation $\chi\bar\chi\to VV$ leads to an accumulation of
$V$-particles,
$Y_V|_{> 1~{\rm sec}} = 2 \int_{1\,{\rm sec}} dt\, \langle
\sigma_\text{an} v \rangle Y_{\chi}^2 s $, which ---when they
decay--- inject energy into the primordial plasma; $Y_i\equiv n_i/s$ where
$s$ is the entropy density. However, as is well known, BBN sensitivity
falls short to probe a thermal annihilation cross section of 1~pb.
In addition, %
the corresponding bound on $s$-wave (or velocity-enhanced) DM
annihilation is much weaker than that from CMB observations; see
\textit{e.g.}~\cite{Kawasaki:2015yya, Depta:2019lbe}.
In a similar vein, the freeze-in of $V$-particles post $T=1\,\MeV$ will lead to
a very weak constraint on $\epsilon$ that is already excluded
otherwise. As a result, we will not further consider the effect of
energy injection during BBN in our final bounds.

\begin{figure*}[t]
\begin{center}
\includegraphics[height=0.45\textwidth]{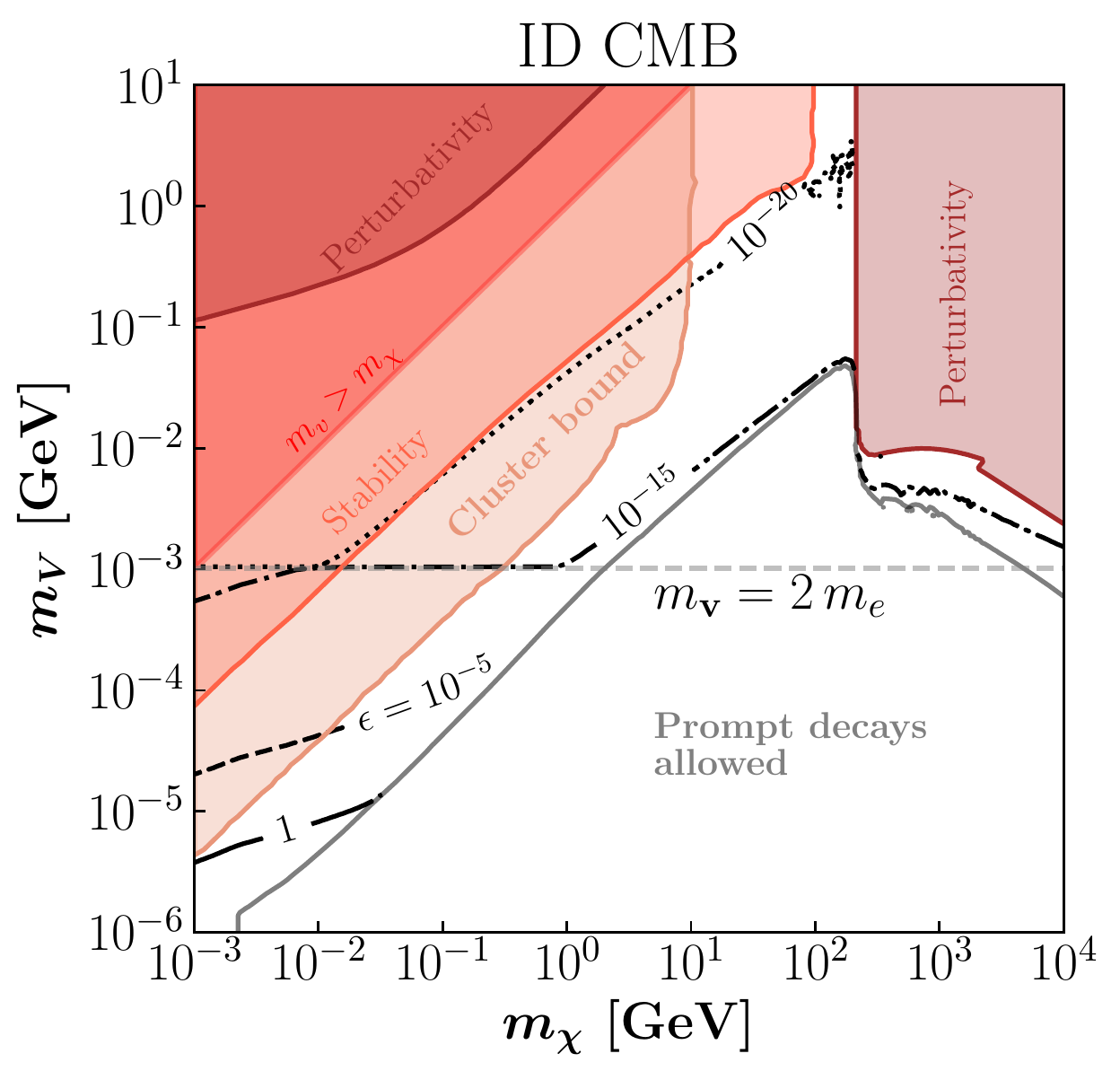}~~~
\includegraphics[height=0.45\textwidth]{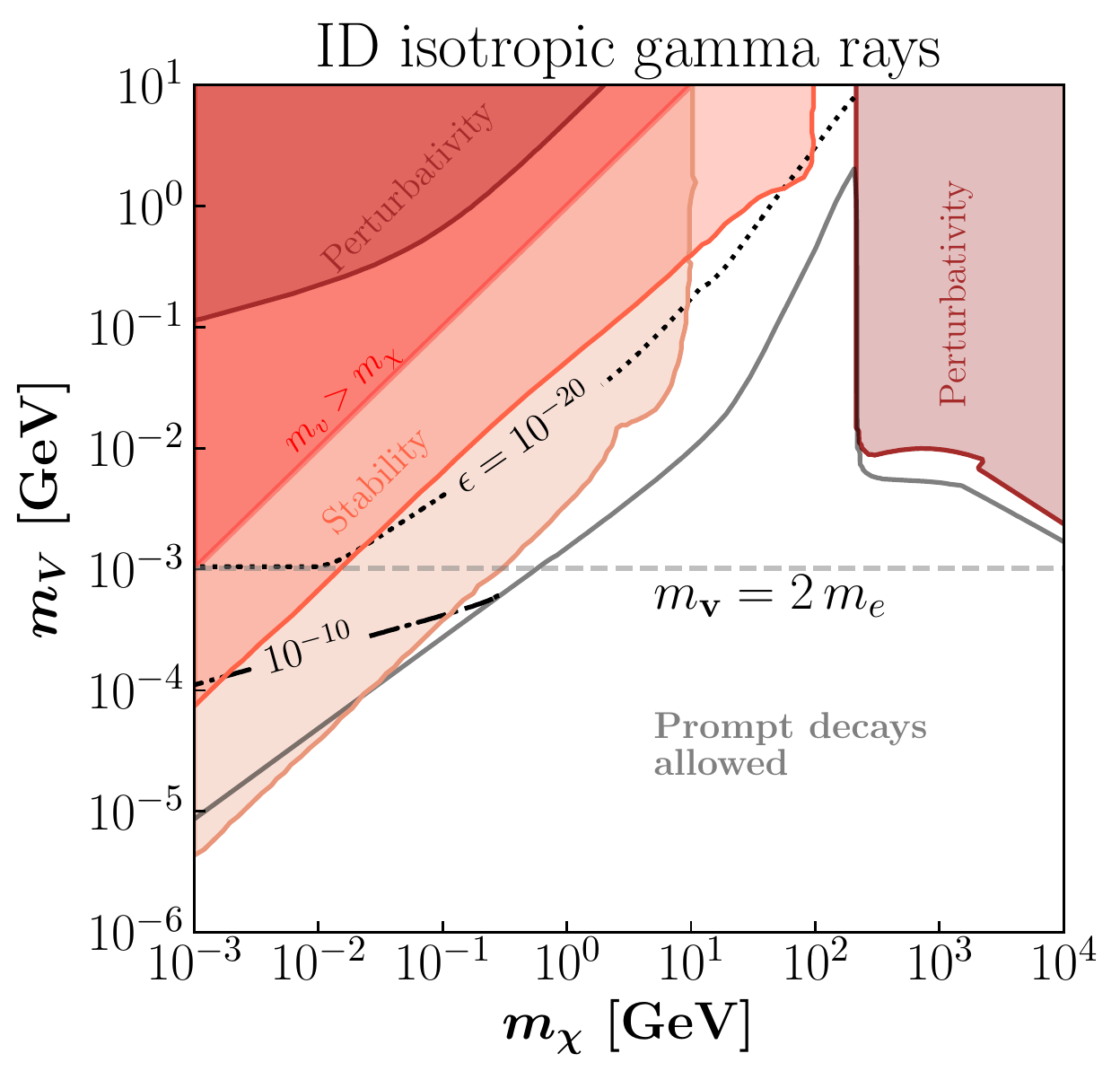}
\caption{Maximum values of $\epsilon$ allowed from CMB
  observations~\cite{Slatyer:2015jla, Ade:2015xua, Elor:2015bho} (left panel) and isotropic
  gamma-rays~\cite{Ackermann:2015tah, Liu:2016ngs} (right
  panel). 
  Below the solid gray line no bounds can be extracted as
  Eq.~(\ref{eq:sv}) is always satisfied.  The horizontal gray dashed
  line marks $m_V=2\,m_e$.  }
\label{fig:eps_ID}
\end{center}
\end{figure*}

{\bf Energy injection during the CMB epoch.} Here we interpret the
Planck constraints on DM annihilation from~\cite{Slatyer:2015jla,
  Ade:2015xua, Elor:2015bho} as upper bounds on the total
electromagnetic (EM) energy injection at the time of
recombination.\footnote{While energy injection from mediator decay
  well after decoupling may also modify the CMB predictions, its
  contribution diminishes with time~\cite{Slatyer:2016qyl}, and will
  not affect our bounds
  qualitatively. This can also be seen in Figs. 7 and 9 of \cite{Slatyer:2012yq}. } %
Similarly to the BBN case above, one may start the investigation from
the redshift-dependent production rate of the $V$ abundance from DM
annihilation, $d Y_V (z) / dz $, following it from an arbitrary
earlier redshift, $z_\text{I}$, down to the CMB time.
  In the calculation of the accumulated $V$ abundance, we may then
  safely neglect mediator decays prior to the CMB epoch, since we are
  interested in parameter regions where the $V$ lifetime in-flight is
  considerably larger than $t_\text{CMB}$.

  In this way, the EM energy density injected from $V$-decays at
  $z_\text{CMB}$ is found from the redshift integral,
  \begin{eqnarray}
    \label{eq:Etot}
	  E_\text{tot}  & \ge  &{2\rho_{\chi,0}^2 \over m_\chi^2} \, (1+z_\text{CMB})^3 \,\text{Br}_\text{EM}  \int^{z_\text{I}}_{z_\text{CMB}} dz \, E_{z,\,\text{dep}} \notag\\
	  &&\times { (1+z)^2 \langle \sigma_\text{an} v \rangle  \over H(z)} \,\left[1- e^{-{t_\text{CMB}\over t(E_{z,\text{dep}}) }}\right]\,,
\end{eqnarray}
where $\rho_{\chi,0}$ is the DM energy density at present,
$E_{z,\text{dep}}\simeq \max[m_V, { m_\chi (1+z_\text{CMB})/ (1+z)}]$
is the (relativistic) energy of $V$ produced at redshift $z$,
$t(E_V) = (E_V/m_V)\tau_V$ is the decay time in-flight, and $H(z)$ is
the Hubble rate at redshift $z$. Here $\text{Br}_\text{EM}$ denotes
the effective branching ratio of $V$ into electromagnetic energy.  For
$m_V\gtrsim
4$~keV, %
$\text{Br}_\text{EM} \simeq 0.4-1$, and we adopt its inferred value as
a function of $m_V$ from~\cite{Fradette:2014sza}.  In turn, for
$m_V\lesssim 4$~keV, the mediator dominantly decays into neutrinos
($\text{Br}_\text{EM}\simeq 0$), and therefore the bounds on energy
injection vanish. The inequality sign in (\ref{eq:Etot}) is owed to
the fact, that we have taken the smallest, \textit{i.e.}  the relic
value of the DM abundance for sourcing $V$ through annihilation.

In the limit $m_\chi \gg m_V$ and
$t(E_{z,\text{dep}}) \gg t_\text{CMB} $, one finds that the injected
EM energy density, $E_\text{tot}$, is well approximated by the $V$
abundance produced at the CMB time. This holds true even without any
Sommerfeld enhancement of the annihilation cross section. Hence, in
practice, the bounds on the case of mediator-decay with a finite
lifetime can be obtained by rescaling from the existing constraints on
WIMP DM annihilation. Therefore, we  impose as requirement,
\begin{equation}\label{eq:sv}
\langle \sigma_\text{an} v \rangle \times \text{Br}_\text{EM}  \left(1- \exp\left[-\frac{t_*}{t(m_\chi) }\right]\right)  <	\langle \sigma_\text{an} v \rangle_*\,,
\end{equation}
where $t_*=t_\text{CMB}$ corresponds to the characteristic time of the
cosmological epoch, and $t(m_{\chi}) = (m_{\chi}/m_V)\tau_V$. For
concreteness, we set the DM velocity to $10^{-6}c$ and use for
$\langle\sigma_\text{an} v\rangle_*$ the limiting value derived from
Planck data~\cite{Ade:2015xua, Elor:2015bho}. As long as
$\langle \sigma_\text{an} v \rangle \, \text{Br}_\text{EM}
\ge\langle\sigma_\text{an} v \rangle_*$, Eq.~\eqref{eq:sv} implies an
upper bound on $\epsilon$, which is shown in the left panel of
Fig.~\ref{fig:eps_ID}.

{\bf Diffuse gamma-ray background.} Similar to the CMB bound, we use
Eq.~\eqref{eq:sv} with $t_*=t_{U}$ to constrain late-time DM
annihilation through the isotropic extra-galactic gamma-ray bounds
obtained by Fermi-LAT for $m_\chi\ge $10\,GeV~\cite{Ackermann:2015tah, Liu:2016ngs}.  %
Furthermore, it is conservatively assumed that the bound on DM
annihilation cross section $\langle \sigma_\text{an}v\rangle$ is
independent of $m_\chi$ below 10~GeV, while it is expected to become
even stronger with the deceasing $m_\chi$ (\textit{e.g.} see Fig.~14 of
\cite{Essig:2013goa}).  We also set the SE factor to unity, to obtain
the upper bounds on $\epsilon$ shown in the right panel of
Fig.~\ref{fig:eps_ID}.  We note in passing that a cosmologically
long-lived mediator would strengthen the bound because of a
correspondingly smaller optical depth of the gamma ray
signal. Neglecting this effect yields a conservative limit.  Finally,
the diffuse gamma-ray bound on $\epsilon$ from the freeze-in of $V$
below $T=1\,\MeV$ is taken from~\cite{Redondo:2008ec}, and 
incorporated in Fig.~\ref{fig:eps_DP} (`Diffuse').

{\bf Indirect searches from the galactic center.} At last, for the
heavy DM mass regime, where the isotropic extra-galactic gamma-ray
bound weakens, measurements from of the galactic center by
H.E.S.S.~\cite{Abdallah:2016ygi} become relevant. In the parameter
region where
$\langle \sigma_\text{an} v \rangle \, \text{Br}_\text{EM}$ is larger
than the limiting value of the H.E.S.S. bound, we require
$c\,t(m_\chi) \ge$ 1~kpc to put an upper bound on $\epsilon$.
For larger in-flight decay lengths, the morphology of the spectrum is
distorted~\cite{Chu:2017vao}, and thus the above bound cannot be
applied directly.  Overall, the ensuing limits are only competitive in
the region of high mass $m_\chi\gtrsim 100$~GeV, and do not improve the diffuse gamma-ray bound above noticeably.

The derived upper limits on $\epsilon$ are shown in
Fig.~\ref{fig:eps_ID}, where the left and right panels correspond to
CMB and isotropic gamma-rays observations, respectively.  Below the
gray line the inequality in Eq.~\eqref{eq:sv} is always satisfied and
therefore no limits can be extracted.  For $m_V$ below the
horizontal dashed gray line ($m_V=2\,m_e$), the mediator can only
decay via loop-induced processes into three photons or two neutrinos.
Both decay channels are rather suppressed, and therefore bounds
coming from electromagnetic energy injection weaken significantly.

\begin{figure*}[t!]
\begin{center}
\includegraphics[height=0.4\textwidth]{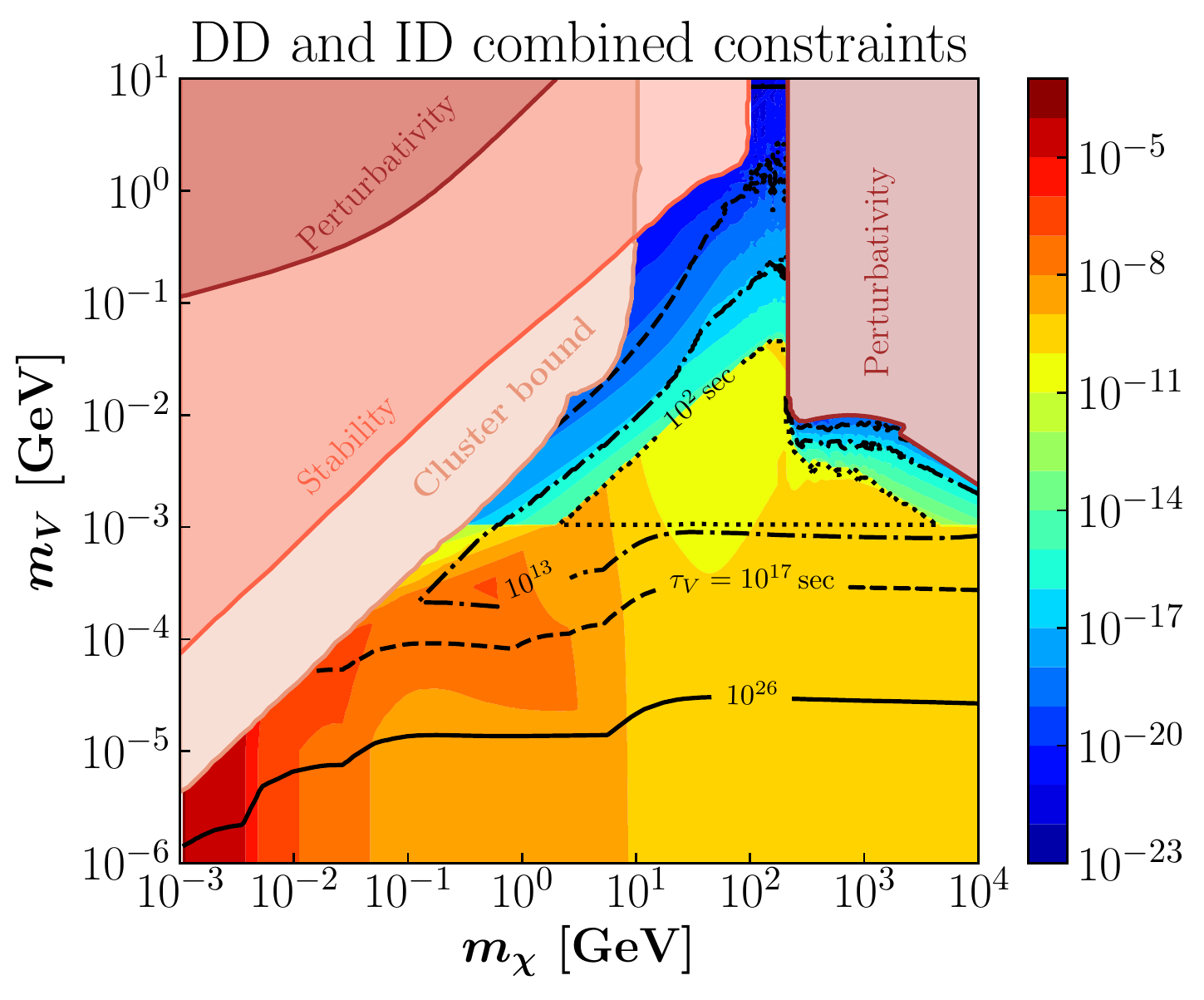}
\includegraphics[height=0.4\textwidth]{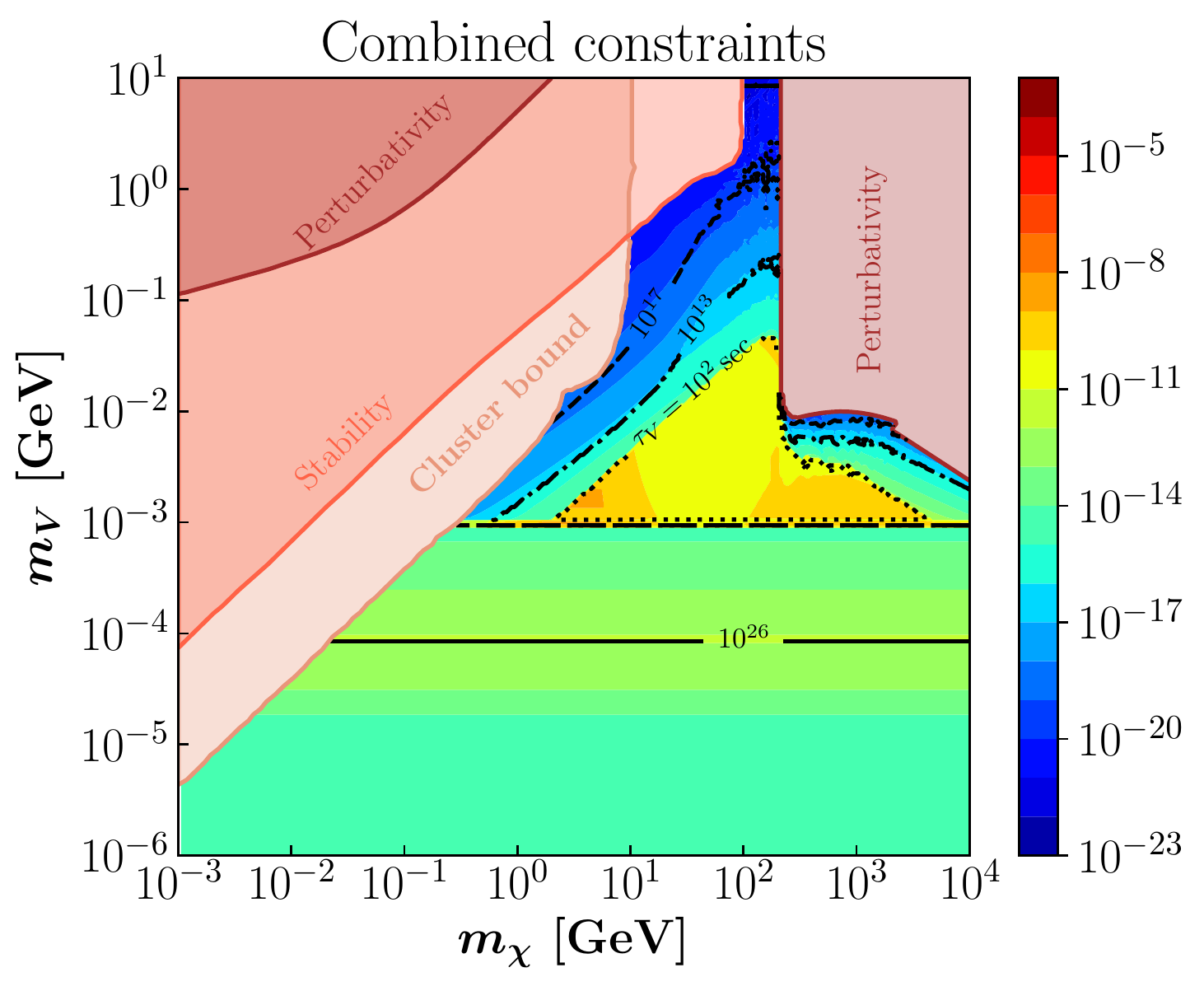}
\caption{Combination of upper limits on $\epsilon$ presented in
  Figs.~\ref{fig:eps_DD} and \ref{fig:eps_ID} (left panel) and
  combination of constraints in
  Figs.~\ref{fig:eps_DD},~\ref{fig:eps_DP} and \ref{fig:eps_ID} (right
panel). At each point $(m_{\chi},\,m_V)$ the maximum permissible
  value of $\epsilon $ has been chosen,
  $\epsilon=\min \{\epsilon_i \}$. The enormous range in $\epsilon$ as
  seen from the color legend is owed to the fact that the direct
  (Fig.~\ref{fig:eps_DD}) and cosmological constraints
  (Fig.~\ref{fig:eps_ID}) apply to largely complementary regions in
  parameter space with significantly different sensitivity to the
  value of $\epsilon$ (left panel); this sensitivity is 
  altered for $m_V < 1~\rm{MeV}$ once constraints from
  Fig.~\ref{fig:eps_DP} are taken into account (right panel).}
\label{fig:eps_all}
\end{center}
\end{figure*}

\vspace{.2cm}
\subsection{Potentially complementary observables}

There exist other observables that can be used to constrain the
parameter region of this scenario. While they currently do not lead to
stronger bounds than the ones discussed above, they may become
relevant in the future, or provide competitive bounds in other SIDM
scenarios.

First, DM particles heavier than several GeV can be captured by the
Sun and annihilate into light mediators, which escape and finally
decay. Depending on the assumptions, the bounds from Fermi-LAT and
HAWC vary between $10^{-42}$~cm$^2$ and $10^{-46}$~cm$^2$ on
spin-dependent DM-nucleon scattering cross section for
$m_\chi \ge 4$~GeV~\cite{Ajello:2011dq, Feng:2016ijc, Leane:2017vag,
  Robertson:2017hdw, Albert:2018jwh}.\footnote{If SIDM captured by the
  Sun via self-scattering is taken into account, an enhancement of
  $10-10^3$ may be achieved, potentially leading to stronger
  bounds~\cite{Zentner:2009is, Kouvaris:2016ltf,Robertson:2017hdw,
    Gaidau:2018yws}.} However, they only apply to a very narrow
parameter region in $m_V$, about a few MeV. This is
due to the fact that for the relevant values of DM-nucleon scattering
cross section, the mediator either decays inside the Sun
($m_V \gg 1$~MeV) or becomes very long-lived ($m_V \le 1$~MeV).
  
It is also possible to directly search for a boosted (meta-)stable
dark mediator.
Similarly to the boosted DM scenario from DM
annihilation~\cite{Agashe:2014yua} or decay~\cite{Cui:2017ytb}, a
long-lived mediator produced relativistically by DM annihilation could
induce a signal in a DM or neutrino detector. This possibility is
mostly interesting for sub-MeV mediators with ensuing cosmological
long lifetime. For thermal DM annihilation
cross sections ($\sim$~pb), \cite{Agashe:2014yua} estimates the
bound as
\begin{equation}
	\sigma_{\text{SM}-\text{V}} \lesssim 10^{-33} - 10^{-34}\,\text{cm}^2,
\end{equation}
which, in our case, is predominantly driven by the inelastic channel
$e+V \to e +\gamma $.  Simple dimensional analysis of the relevant
cross sections then reveals that this bound is evaded for
$\epsilon \le 10^{-4}-10^{-5}$.  In addition, if any of the dark
particles is light enough, they can be produced in supernovae.  Since
this thermally produced population is much hotter than the halo DM, it
could be observed with direct detection
experiments~\cite{DeRocco:2019jti}.
 
Upper bounds on dark couplings can also be derived from dissipation in
dark halos. Cooling of SIDM via inelastic scattering (and radiative
bound state formation in a limited parameter subspace) may
happen. After BBN, such processes are typically strongly suppressed
due to the low DM relative velocities.  At low redshift, it may affect
the halo dynamics, but the bound turns out to be
weak~\cite{Essig:2018pzq, Chang:2018bgx}. This can be understood from the SIDM
requirement that each DM particle only scatters elastically
${\mathcal O}(1)$ times during the whole lifetime of dwarf halos, and
the frequency of inelastic scattering is naturally much smaller.

\subsection{Discussions on combined constraints}

Figure~\ref{fig:eps_all} (left panel) presents upper limits on
$\epsilon$ coming from the combination of the direct
(Fig.~\ref{fig:eps_DD}) and indirect/cosmological constraints
(Fig.~\ref{fig:eps_ID}) derived in this
work. %
At each point $(m_{\chi},\,m_V)$ the maximum permissible value of
$\epsilon $ is chosen, $\epsilon=\min \{\epsilon_i \}$. The enormous
range in $\epsilon$ as seen from the color legend is owed to the fact
that the direct detection and indirect/cosmological searches apply to
largely complementary regions in parameter space with significantly
different sensitivity to the value of $\epsilon$.  The limits are
further improved once the bounds independent of the DM abundance
(Fig.~\ref{fig:eps_DP}) are taken into account. This is especially
true in the region of $m_V < 1~\rm{MeV}$ and is summarized in right
panel of
Fig.~\ref{fig:eps_all}. 

Our summary figures show that the minimal mediator lifetime that is
allowed  is greater than  $1~\rm{sec}$,  found
in the region $m_V \sim 10~\rm{MeV}$ and $\epsilon \sim 10^{-10}$.
Furthermore, taking into account the newly-derived bound from
\cite{Sung:2019xie} would require the mediator to have a lifetime
longer than $\mathcal{O}(10^5)~\rm{sec}$. Together with the BBN
observations, it suggests that a thermalization between the dark and
visible sectors at $T\sim$\,MeV is very unlikely.

At last, it is worth pointing out that if DM is asymmetric, the above
bounds from energy injection can be relaxed. However, a larger DM
annihilation cross section may be needed to erase the symmetric DM
component in early universe. This adds a layer of complication when
the premise is to study SIDM with the least set of assumption about
the early universe DM history, as spelled out in the introduction.  We
leave an exploration of this possibility for future work.

\section{Conclusions}
\label{sec:conclusions}

In this work we %
analyze the parameter space
for self-interacting dark matter (DM) without assuming any specific
mechanism for relic density generation and the thermalization between
the dark and visible sectors. The motivation for it is that,
currently, no firm conclusions can be drawn about the thermal state
and the particle content of the early universe above a photon
temperature of several MeV ---despite a number of circumstantial hints
of what the sequence of events may have been.

For concreteness, we analyze a symmetric fermionic DM candidate
$\chi$, where the self-interaction is mediated by a dark photon $V$
that kinetically mixes with the SM counterpart with strength
$\epsilon$. Throughout the work we require a self-scattering cross
section over DM mass of $1\,\cm^2/{\rm g}$ as the fiducial value that
is able to address the astrophysical small-scale problems of
$\Lambda$CDM. For chosen DM and mediator masses, this fixes the dark
gauge coupling and completely determines the hidden sector parameters.
The link to SM is controlled by the kinetic mixing parameter
$\epsilon$ and the velocity dependence that is introduced in this
model through Sommerfeld enhancement in DM annihilation makes it also
interesting from the astrophysical point of view. 

Despite relaxing the assumptions on the early universe history, a
number of strong constraints remain applicable to this model. Instead
of ruling out the model parameter space, these constraints imply upper
limits on the value of $\epsilon$ and hence lower limits on the
lifetime of $V$.
The limits on $\epsilon$ are derived from direct and indirect
detection as well as from cosmology.  The direct detection of DM
through nuclear or electron recoil signatures offer complementary
limits on $\epsilon$. While DM-electron scattering strongly constrains
the low mass region ($\sim m_{\chi}< 1$~GeV), the nuclear recoil
signature is sensitive to heavier DM
($\sim m_{\chi} > 1$~GeV).  The constraints on the kinetic mixing from
DM-nucleus scattering are comparable to those from beam dump experiments at the laboratory.
The indirect  searches and the CMB limits on the other hand
are sensitive to a wide range of symmetric DM and mediator mass
combinations and set some of the most stringent constraints for
heavier mediators. While literature results are derived under the assumption
of prompt energy injection, here we find that these limits need to be
rescaled in order to take into account the mediator's finite (boosted)
lifetime.
For heavier mediators, the limits on $\epsilon$ arising from indirect
searches usually supersede any laboratory limits and imply mediators
to be (meta-)stable at a cosmological time scale.

Our analysis suggests a long-lived mediator, and thus the need for
analyzing displaced vertices in the sky and small-scale anisotropies
in cosmic rays from DM annihilation~\cite{Rothstein:2009pm,
  Chu:2017vao}, instead of exclusively focusing on astrophysical
objects that are DM-dense. For instance, a MeV-mass mediator with
a lifetime of $10^{10}$\,sec ($\epsilon \sim 10^{-14}$), produced in
GeV-mass DM annihilation in the Galactic center, preferentially decays
outside our Galaxy. Given that we have been agnostic about the
DM-generation mechanism, the `model-independent' bounds on $\epsilon$
that we derive call for being further interpreted and tested in SIDM
scenarios with spelled-out relic density mechanisms beyond the
standard case. Such scenarios include feebly-interacting DM or DM in
the context of a modified early universe thermal history.

\acknowledgments
This project has received funding from the European Union's Horizon 2020 research and innovation programme under the Marie Sklodowska-Curie grant agreements 674896 and 690575.
NB is partially supported by Spanish MINECO under Grant FPA2017-84543-P, by Universidad Antonio Nariño grants 2018204, 2019101 and 2019248, and by the `Joint Excellence in Science and Humanities' (JESH) program of the Austrian Academy of Sciences. 
NB thanks the Institut Pascal at Université Paris-Saclay with the support of the P2I and SPU research departments and the P2IO Laboratory of Excellence (program `Investissements d’avenir' ANR-11-IDEX-0003-01 Paris-Saclay and ANR-10-LABX-0038), as well as the IPhT.
XC and JP are supported by the `New Frontiers' program of the Austrian Academy of Sciences.  
SK is supported by Elise-Richter grant project number V592-N27.
The authors thank the Erwin Schr\"odinger International Institute for  hospitality during this work.  
This research made use of IPython~\cite{Perez:2007emg}, Matplotlib~\cite{Hunter:2007ouj} and SciPy~\cite{SciPy}.

\bibliography{biblio}

\begin{thebibliography}{163}%
\makeatletter
\providecommand \@ifxundefined [1]{%
 \@ifx{#1\undefined}
}%
\providecommand \@ifnum [1]{%
 \ifnum #1\expandafter \@firstoftwo
 \else \expandafter \@secondoftwo
 \fi
}%
\providecommand \@ifx [1]{%
 \ifx #1\expandafter \@firstoftwo
 \else \expandafter \@secondoftwo
 \fi
}%
\providecommand \natexlab [1]{#1}%
\providecommand \enquote  [1]{``#1''}%
\providecommand \bibnamefont  [1]{#1}%
\providecommand \bibfnamefont [1]{#1}%
\providecommand \citenamefont [1]{#1}%
\providecommand \href@noop [0]{\@secondoftwo}%
\providecommand \href [0]{\begingroup \@sanitize@url \@href}%
\providecommand \@href[1]{\@@startlink{#1}\@@href}%
\providecommand \@@href[1]{\endgroup#1\@@endlink}%
\providecommand \@sanitize@url [0]{\catcode `\\12\catcode `\$12\catcode
  `\&12\catcode `\#12\catcode `\^12\catcode `\_12\catcode `\%12\relax}%
\providecommand \@@startlink[1]{}%
\providecommand \@@endlink[0]{}%
\providecommand \url  [0]{\begingroup\@sanitize@url \@url }%
\providecommand \@url [1]{\endgroup\@href {#1}{\urlprefix }}%
\providecommand \urlprefix  [0]{URL }%
\providecommand \Eprint [0]{\href }%
\providecommand \doibase [0]{http://dx.doi.org/}%
\providecommand \selectlanguage [0]{\@gobble}%
\providecommand \bibinfo  [0]{\@secondoftwo}%
\providecommand \bibfield  [0]{\@secondoftwo}%
\providecommand \translation [1]{[#1]}%
\providecommand \BibitemOpen [0]{}%
\providecommand \bibitemStop [0]{}%
\providecommand \bibitemNoStop [0]{.\EOS\space}%
\providecommand \EOS [0]{\spacefactor3000\relax}%
\providecommand \BibitemShut  [1]{\csname bibitem#1\endcsname}%
\let\auto@bib@innerbib\@empty
%</preamble>
\bibitem [{\citenamefont {Flores}\ and\ \citenamefont
  {Primack}(1994)}]{Flores:1994gz}%
  \BibitemOpen
  \bibfield  {author} {\bibinfo {author} {\bibfnamefont {R.~A.}\ \bibnamefont
  {Flores}}\ and\ \bibinfo {author} {\bibfnamefont {J.~R.}\ \bibnamefont
  {Primack}},\ }\href {\doibase 10.1086/187350} {\bibfield  {journal} {\bibinfo
   {journal} {Astrophys. J.}\ }\textbf {\bibinfo {volume} {427}},\ \bibinfo
  {pages} {L1} (\bibinfo {year} {1994})},\ \Eprint
  {http://arxiv.org/abs/astro-ph/9402004} {arXiv:astro-ph/9402004 [astro-ph]}
  \BibitemShut {NoStop}%
%%CITATION = ASTRO-PH/9402004;%%
\bibitem [{\citenamefont {Moore}(1994)}]{Moore:1994yx}%
  \BibitemOpen
  \bibfield  {author} {\bibinfo {author} {\bibfnamefont {B.}~\bibnamefont
  {Moore}},\ }\href {\doibase 10.1038/370629a0} {\bibfield  {journal} {\bibinfo
   {journal} {Nature}\ }\textbf {\bibinfo {volume} {370}},\ \bibinfo {pages}
  {629} (\bibinfo {year} {1994})}\BibitemShut {NoStop}%
%%CITATION = NATUA,370,629;%%
\bibitem [{\citenamefont {Oh}\ \emph {et~al.}(2011)\citenamefont {Oh},
  \citenamefont {Brook}, \citenamefont {Governato}, \citenamefont {Brinks},
  \citenamefont {Mayer}, \citenamefont {de~Blok}, \citenamefont {Brooks},\ and\
  \citenamefont {Walter}}]{Oh:2010mc}%
  \BibitemOpen
  \bibfield  {author} {\bibinfo {author} {\bibfnamefont {S.-H.}\ \bibnamefont
  {Oh}}, \bibinfo {author} {\bibfnamefont {C.}~\bibnamefont {Brook}}, \bibinfo
  {author} {\bibfnamefont {F.}~\bibnamefont {Governato}}, \bibinfo {author}
  {\bibfnamefont {E.}~\bibnamefont {Brinks}}, \bibinfo {author} {\bibfnamefont
  {L.}~\bibnamefont {Mayer}}, \bibinfo {author} {\bibfnamefont {W.~J.~G.}\
  \bibnamefont {de~Blok}}, \bibinfo {author} {\bibfnamefont {A.}~\bibnamefont
  {Brooks}}, \ and\ \bibinfo {author} {\bibfnamefont {F.}~\bibnamefont
  {Walter}},\ }\href {\doibase 10.1088/0004-6256/142/1/24} {\bibfield
  {journal} {\bibinfo  {journal} {Astron. J.}\ }\textbf {\bibinfo {volume}
  {142}},\ \bibinfo {pages} {24} (\bibinfo {year} {2011})},\ \Eprint
  {http://arxiv.org/abs/1011.2777} {arXiv:1011.2777 [astro-ph.CO]} \BibitemShut
  {NoStop}%
%%CITATION = ARXIV:1011.2777;%%
\bibitem [{\citenamefont {Walker}\ and\ \citenamefont
  {Peñarrubia}(2011)}]{Walker:2011zu}%
  \BibitemOpen
  \bibfield  {author} {\bibinfo {author} {\bibfnamefont {M.~G.}\ \bibnamefont
  {Walker}}\ and\ \bibinfo {author} {\bibfnamefont {J.}~\bibnamefont
  {Peñarrubia}},\ }\href {\doibase 10.1088/0004-637X/742/1/20} {\bibfield
  {journal} {\bibinfo  {journal} {Astrophys. J.}\ }\textbf {\bibinfo {volume}
  {742}},\ \bibinfo {pages} {20} (\bibinfo {year} {2011})},\ \Eprint
  {http://arxiv.org/abs/1108.2404} {arXiv:1108.2404 [astro-ph.CO]} \BibitemShut
  {NoStop}%
%%CITATION = ARXIV:1108.2404;%%
\bibitem [{\citenamefont {Boylan-Kolchin}\ \emph {et~al.}(2011)\citenamefont
  {Boylan-Kolchin}, \citenamefont {Bullock},\ and\ \citenamefont
  {Kaplinghat}}]{BoylanKolchin:2011de}%
  \BibitemOpen
  \bibfield  {author} {\bibinfo {author} {\bibfnamefont {M.}~\bibnamefont
  {Boylan-Kolchin}}, \bibinfo {author} {\bibfnamefont {J.~S.}\ \bibnamefont
  {Bullock}}, \ and\ \bibinfo {author} {\bibfnamefont {M.}~\bibnamefont
  {Kaplinghat}},\ }\href {\doibase 10.1111/j.1745-3933.2011.01074.x} {\bibfield
   {journal} {\bibinfo  {journal} {Mon. Not. Roy. Astron. Soc.}\ }\textbf
  {\bibinfo {volume} {415}},\ \bibinfo {pages} {L40} (\bibinfo {year}
  {2011})},\ \Eprint {http://arxiv.org/abs/1103.0007} {arXiv:1103.0007
  [astro-ph.CO]} \BibitemShut {NoStop}%
%%CITATION = ARXIV:1103.0007;%%
\bibitem [{\citenamefont {Boylan-Kolchin}\ \emph {et~al.}(2012)\citenamefont
  {Boylan-Kolchin}, \citenamefont {Bullock},\ and\ \citenamefont
  {Kaplinghat}}]{BoylanKolchin:2011dk}%
  \BibitemOpen
  \bibfield  {author} {\bibinfo {author} {\bibfnamefont {M.}~\bibnamefont
  {Boylan-Kolchin}}, \bibinfo {author} {\bibfnamefont {J.~S.}\ \bibnamefont
  {Bullock}}, \ and\ \bibinfo {author} {\bibfnamefont {M.}~\bibnamefont
  {Kaplinghat}},\ }\href {\doibase 10.1111/j.1365-2966.2012.20695.x} {\bibfield
   {journal} {\bibinfo  {journal} {Mon. Not. Roy. Astron. Soc.}\ }\textbf
  {\bibinfo {volume} {422}},\ \bibinfo {pages} {1203} (\bibinfo {year}
  {2012})},\ \Eprint {http://arxiv.org/abs/1111.2048} {arXiv:1111.2048
  [astro-ph.CO]} \BibitemShut {NoStop}%
%%CITATION = ARXIV:1111.2048;%%
\bibitem [{\citenamefont {Garrison-Kimmel}\ \emph {et~al.}(2014)\citenamefont
  {Garrison-Kimmel}, \citenamefont {Boylan-Kolchin}, \citenamefont {Bullock},\
  and\ \citenamefont {Kirby}}]{Garrison-Kimmel:2014vqa}%
  \BibitemOpen
  \bibfield  {author} {\bibinfo {author} {\bibfnamefont {S.}~\bibnamefont
  {Garrison-Kimmel}}, \bibinfo {author} {\bibfnamefont {M.}~\bibnamefont
  {Boylan-Kolchin}}, \bibinfo {author} {\bibfnamefont {J.~S.}\ \bibnamefont
  {Bullock}}, \ and\ \bibinfo {author} {\bibfnamefont {E.~N.}\ \bibnamefont
  {Kirby}},\ }\href {\doibase 10.1093/mnras/stu1477} {\bibfield  {journal}
  {\bibinfo  {journal} {Mon. Not. Roy. Astron. Soc.}\ }\textbf {\bibinfo
  {volume} {444}},\ \bibinfo {pages} {222} (\bibinfo {year} {2014})},\ \Eprint
  {http://arxiv.org/abs/1404.5313} {arXiv:1404.5313 [astro-ph.GA]} \BibitemShut
  {NoStop}%
%%CITATION = ARXIV:1404.5313;%%
\bibitem [{\citenamefont {Papastergis}\ \emph {et~al.}(2015)\citenamefont
  {Papastergis}, \citenamefont {Giovanelli}, \citenamefont {Haynes},\ and\
  \citenamefont {Shankar}}]{Papastergis:2014aba}%
  \BibitemOpen
  \bibfield  {author} {\bibinfo {author} {\bibfnamefont {E.}~\bibnamefont
  {Papastergis}}, \bibinfo {author} {\bibfnamefont {R.}~\bibnamefont
  {Giovanelli}}, \bibinfo {author} {\bibfnamefont {M.~P.}\ \bibnamefont
  {Haynes}}, \ and\ \bibinfo {author} {\bibfnamefont {F.}~\bibnamefont
  {Shankar}},\ }\href {\doibase 10.1051/0004-6361/201424909} {\bibfield
  {journal} {\bibinfo  {journal} {Astron. Astrophys.}\ }\textbf {\bibinfo
  {volume} {574}},\ \bibinfo {pages} {A113} (\bibinfo {year} {2015})},\ \Eprint
  {http://arxiv.org/abs/1407.4665} {arXiv:1407.4665 [astro-ph.GA]} \BibitemShut
  {NoStop}%
%%CITATION = ARXIV:1407.4665;%%
\bibitem [{\citenamefont {Blok}\ \emph {et~al.}(2003)\citenamefont {Blok},
  \citenamefont {Bosma},\ and\ \citenamefont {McGaugh}}]{Blok:2002tr}%
  \BibitemOpen
  \bibfield  {author} {\bibinfo {author} {\bibfnamefont {W.~J. G.~d.}\
  \bibnamefont {Blok}}, \bibinfo {author} {\bibfnamefont {A.}~\bibnamefont
  {Bosma}}, \ and\ \bibinfo {author} {\bibfnamefont {S.~S.}\ \bibnamefont
  {McGaugh}},\ }\href {\doibase 10.1046/j.1365-8711.2003.06330.x} {\bibfield
  {journal} {\bibinfo  {journal} {Mon. Not. Roy. Astron. Soc.}\ }\textbf
  {\bibinfo {volume} {340}},\ \bibinfo {pages} {657} (\bibinfo {year}
  {2003})},\ \Eprint {http://arxiv.org/abs/astro-ph/0212102}
  {arXiv:astro-ph/0212102 [astro-ph]} \BibitemShut {NoStop}%
%%CITATION = ASTRO-PH/0212102;%%
\bibitem [{\citenamefont {Rhee}\ \emph {et~al.}(2004)\citenamefont {Rhee},
  \citenamefont {Klypin},\ and\ \citenamefont {Valenzuela}}]{Rhee:2003vw}%
  \BibitemOpen
  \bibfield  {author} {\bibinfo {author} {\bibfnamefont {G.}~\bibnamefont
  {Rhee}}, \bibinfo {author} {\bibfnamefont {A.}~\bibnamefont {Klypin}}, \ and\
  \bibinfo {author} {\bibfnamefont {O.}~\bibnamefont {Valenzuela}},\ }\href
  {\doibase 10.1086/425565} {\bibfield  {journal} {\bibinfo  {journal}
  {Astrophys. J.}\ }\textbf {\bibinfo {volume} {617}},\ \bibinfo {pages} {1059}
  (\bibinfo {year} {2004})},\ \Eprint {http://arxiv.org/abs/astro-ph/0311020}
  {arXiv:astro-ph/0311020 [astro-ph]} \BibitemShut {NoStop}%
%%CITATION = ASTRO-PH/0311020;%%
\bibitem [{\citenamefont {Gentile}\ \emph {et~al.}(2005)\citenamefont
  {Gentile}, \citenamefont {Burkert}, \citenamefont {Salucci}, \citenamefont
  {Klein},\ and\ \citenamefont {Walter}}]{Gentile:2005de}%
  \BibitemOpen
  \bibfield  {author} {\bibinfo {author} {\bibfnamefont {G.}~\bibnamefont
  {Gentile}}, \bibinfo {author} {\bibfnamefont {A.}~\bibnamefont {Burkert}},
  \bibinfo {author} {\bibfnamefont {P.}~\bibnamefont {Salucci}}, \bibinfo
  {author} {\bibfnamefont {U.}~\bibnamefont {Klein}}, \ and\ \bibinfo {author}
  {\bibfnamefont {F.}~\bibnamefont {Walter}},\ }\href {\doibase 10.1086/498939}
  {\bibfield  {journal} {\bibinfo  {journal} {Astrophys. J. Lett.}\ }\textbf
  {\bibinfo {volume} {634}},\ \bibinfo {pages} {L145} (\bibinfo {year}
  {2005})},\ \Eprint {http://arxiv.org/abs/astro-ph/0506538}
  {arXiv:astro-ph/0506538 [astro-ph]} \BibitemShut {NoStop}%
%%CITATION = ASTRO-PH/0506538;%%
\bibitem [{\citenamefont {Spekkens}\ and\ \citenamefont
  {Giovanelli}(2005)}]{Spekkens:2005ik}%
  \BibitemOpen
  \bibfield  {author} {\bibinfo {author} {\bibfnamefont {K.}~\bibnamefont
  {Spekkens}}\ and\ \bibinfo {author} {\bibfnamefont {R.}~\bibnamefont
  {Giovanelli}},\ }\href {\doibase 10.1086/429592} {\bibfield  {journal}
  {\bibinfo  {journal} {Astron. J.}\ }\textbf {\bibinfo {volume} {129}},\
  \bibinfo {pages} {2119} (\bibinfo {year} {2005})},\ \Eprint
  {http://arxiv.org/abs/astro-ph/0502166} {arXiv:astro-ph/0502166 [astro-ph]}
  \BibitemShut {NoStop}%
%%CITATION = ASTRO-PH/0502166;%%
\bibitem [{\citenamefont {Valenzuela}\ \emph {et~al.}(2007)\citenamefont
  {Valenzuela}, \citenamefont {Rhee}, \citenamefont {Klypin}, \citenamefont
  {Governato}, \citenamefont {Stinson}, \citenamefont {Quinn},\ and\
  \citenamefont {Wadsley}}]{Valenzuela:2005dh}%
  \BibitemOpen
  \bibfield  {author} {\bibinfo {author} {\bibfnamefont {O.}~\bibnamefont
  {Valenzuela}}, \bibinfo {author} {\bibfnamefont {G.}~\bibnamefont {Rhee}},
  \bibinfo {author} {\bibfnamefont {A.}~\bibnamefont {Klypin}}, \bibinfo
  {author} {\bibfnamefont {F.}~\bibnamefont {Governato}}, \bibinfo {author}
  {\bibfnamefont {G.}~\bibnamefont {Stinson}}, \bibinfo {author} {\bibfnamefont
  {T.~R.}\ \bibnamefont {Quinn}}, \ and\ \bibinfo {author} {\bibfnamefont
  {J.}~\bibnamefont {Wadsley}},\ }\href {\doibase 10.1086/508674} {\bibfield
  {journal} {\bibinfo  {journal} {Astrophys. J.}\ }\textbf {\bibinfo {volume}
  {657}},\ \bibinfo {pages} {773} (\bibinfo {year} {2007})},\ \Eprint
  {http://arxiv.org/abs/astro-ph/0509644} {arXiv:astro-ph/0509644 [astro-ph]}
  \BibitemShut {NoStop}%
%%CITATION = ASTRO-PH/0509644;%%
\bibitem [{\citenamefont {Dalcanton}\ and\ \citenamefont
  {Stilp}(2010)}]{Dalcanton:2010bp}%
  \BibitemOpen
  \bibfield  {author} {\bibinfo {author} {\bibfnamefont {J.~J.}\ \bibnamefont
  {Dalcanton}}\ and\ \bibinfo {author} {\bibfnamefont {A.}~\bibnamefont
  {Stilp}},\ }\href {\doibase 10.1088/0004-637X/721/1/547} {\bibfield
  {journal} {\bibinfo  {journal} {Astrophys. J.}\ }\textbf {\bibinfo {volume}
  {721}},\ \bibinfo {pages} {547} (\bibinfo {year} {2010})},\ \Eprint
  {http://arxiv.org/abs/1007.2535} {arXiv:1007.2535 [astro-ph.CO]} \BibitemShut
  {NoStop}%
%%CITATION = ARXIV:1007.2535;%%
\bibitem [{\citenamefont {Kormendy}\ and\ \citenamefont
  {Freeman}(2016)}]{Kormendy:2014ova}%
  \BibitemOpen
  \bibfield  {author} {\bibinfo {author} {\bibfnamefont {J.}~\bibnamefont
  {Kormendy}}\ and\ \bibinfo {author} {\bibfnamefont {K.~C.}\ \bibnamefont
  {Freeman}},\ }\href {\doibase 10.3847/0004-637X/817/2/84} {\bibfield
  {journal} {\bibinfo  {journal} {Astrophys. J.}\ }\textbf {\bibinfo {volume}
  {817}},\ \bibinfo {pages} {84} (\bibinfo {year} {2016})},\ \Eprint
  {http://arxiv.org/abs/1411.2170} {arXiv:1411.2170 [astro-ph.GA]} \BibitemShut
  {NoStop}%
%%CITATION = ARXIV:1411.2170;%%
\bibitem [{\citenamefont {{Read}}\ \emph {et~al.}(2016)\citenamefont {{Read}},
  \citenamefont {{Iorio}}, \citenamefont {{Agertz}},\ and\ \citenamefont
  {{Fraternali}}}]{2016MNRAS.462.3628R}%
  \BibitemOpen
  \bibfield  {author} {\bibinfo {author} {\bibfnamefont {J.~I.}\ \bibnamefont
  {{Read}}}, \bibinfo {author} {\bibfnamefont {G.}~\bibnamefont {{Iorio}}},
  \bibinfo {author} {\bibfnamefont {O.}~\bibnamefont {{Agertz}}}, \ and\
  \bibinfo {author} {\bibfnamefont {F.}~\bibnamefont {{Fraternali}}},\ }\href
  {\doibase 10.1093/mnras/stw1876} {\bibfield  {journal} {\bibinfo  {journal}
  {Mon. Not. Roy. Astron. Soc.}\ }\textbf {\bibinfo {volume} {462}},\ \bibinfo
  {pages} {3628} (\bibinfo {year} {2016})},\ \Eprint
  {http://arxiv.org/abs/1601.05821} {arXiv:1601.05821} \BibitemShut {NoStop}%
\bibitem [{\citenamefont {Macci\`{o}}\ \emph {et~al.}(2016)\citenamefont
  {Macci\`{o}}, \citenamefont {Udrescu}, \citenamefont {Dutton}, \citenamefont
  {Obreja}, \citenamefont {Wang}, \citenamefont {Stinson},\ and\ \citenamefont
  {Kang}}]{Maccio:2016egb}%
  \BibitemOpen
  \bibfield  {author} {\bibinfo {author} {\bibfnamefont {A.~V.}\ \bibnamefont
  {Macci\`{o}}}, \bibinfo {author} {\bibfnamefont {S.~M.}\ \bibnamefont
  {Udrescu}}, \bibinfo {author} {\bibfnamefont {A.~A.}\ \bibnamefont {Dutton}},
  \bibinfo {author} {\bibfnamefont {A.}~\bibnamefont {Obreja}}, \bibinfo
  {author} {\bibfnamefont {L.}~\bibnamefont {Wang}}, \bibinfo {author}
  {\bibfnamefont {G.~R.}\ \bibnamefont {Stinson}}, \ and\ \bibinfo {author}
  {\bibfnamefont {X.}~\bibnamefont {Kang}},\ }\href {\doibase
  10.1093/mnrasl/slw147} {\bibfield  {journal} {\bibinfo  {journal} {Mon. Not.
  Roy. Astron. Soc.}\ }\textbf {\bibinfo {volume} {463}},\ \bibinfo {pages}
  {L69} (\bibinfo {year} {2016})},\ \Eprint {http://arxiv.org/abs/1607.01028}
  {arXiv:1607.01028 [astro-ph.GA]} \BibitemShut {NoStop}%
%%CITATION = ARXIV:1607.01028;%%
\bibitem [{\citenamefont {{Papastergis}}\ and\ \citenamefont
  {{Ponomareva}}(2017)}]{2017A&A...601A...1P}%
  \BibitemOpen
  \bibfield  {author} {\bibinfo {author} {\bibfnamefont {E.}~\bibnamefont
  {{Papastergis}}}\ and\ \bibinfo {author} {\bibfnamefont {A.~A.}\ \bibnamefont
  {{Ponomareva}}},\ }\href {\doibase 10.1051/0004-6361/201629546} {\bibfield
  {journal} {\bibinfo  {journal} {Astronomy \& Astrophysics}\ }\textbf
  {\bibinfo {volume} {601}},\ \bibinfo {eid} {A1} (\bibinfo {year} {2017})},\
  \Eprint {http://arxiv.org/abs/1608.05214} {arXiv:1608.05214} \BibitemShut
  {NoStop}%
\bibitem [{\citenamefont {Brooks}\ \emph {et~al.}(2017)\citenamefont {Brooks},
  \citenamefont {Papastergis}, \citenamefont {Christensen}, \citenamefont
  {Governato}, \citenamefont {Stilp}, \citenamefont {Quinn},\ and\
  \citenamefont {Wadsley}}]{Brooks:2017rfe}%
  \BibitemOpen
  \bibfield  {author} {\bibinfo {author} {\bibfnamefont {A.~M.}\ \bibnamefont
  {Brooks}}, \bibinfo {author} {\bibfnamefont {E.}~\bibnamefont {Papastergis}},
  \bibinfo {author} {\bibfnamefont {C.~R.}\ \bibnamefont {Christensen}},
  \bibinfo {author} {\bibfnamefont {F.}~\bibnamefont {Governato}}, \bibinfo
  {author} {\bibfnamefont {A.}~\bibnamefont {Stilp}}, \bibinfo {author}
  {\bibfnamefont {T.~R.}\ \bibnamefont {Quinn}}, \ and\ \bibinfo {author}
  {\bibfnamefont {J.}~\bibnamefont {Wadsley}},\ }\href {\doibase
  10.3847/1538-4357/aa9576} {\bibfield  {journal} {\bibinfo  {journal}
  {Astrophys. J.}\ }\textbf {\bibinfo {volume} {850}},\ \bibinfo {pages} {97}
  (\bibinfo {year} {2017})},\ \Eprint {http://arxiv.org/abs/1701.07835}
  {arXiv:1701.07835 [astro-ph.GA]} \BibitemShut {NoStop}%
%%CITATION = ARXIV:1701.07835;%%
\bibitem [{\citenamefont {Oman}\ \emph {et~al.}(2019)\citenamefont {Oman},
  \citenamefont {Marasco}, \citenamefont {Navarro}, \citenamefont {Frenk},
  \citenamefont {Schaye},\ and\ \citenamefont
  {Benítez-Llambay}}]{Oman:2017vkl}%
  \BibitemOpen
  \bibfield  {author} {\bibinfo {author} {\bibfnamefont {K.~A.}\ \bibnamefont
  {Oman}}, \bibinfo {author} {\bibfnamefont {A.}~\bibnamefont {Marasco}},
  \bibinfo {author} {\bibfnamefont {J.~F.}\ \bibnamefont {Navarro}}, \bibinfo
  {author} {\bibfnamefont {C.~S.}\ \bibnamefont {Frenk}}, \bibinfo {author}
  {\bibfnamefont {J.}~\bibnamefont {Schaye}}, \ and\ \bibinfo {author}
  {\bibfnamefont {A.}~\bibnamefont {Benítez-Llambay}},\ }\href {\doibase
  10.1093/mnras/sty2687} {\bibfield  {journal} {\bibinfo  {journal} {Mon. Not.
  Roy. Astron. Soc.}\ }\textbf {\bibinfo {volume} {482}},\ \bibinfo {pages}
  {821} (\bibinfo {year} {2019})},\ \Eprint {http://arxiv.org/abs/1706.07478}
  {arXiv:1706.07478 [astro-ph.GA]} \BibitemShut {NoStop}%
%%CITATION = ARXIV:1706.07478;%%
\bibitem [{\citenamefont {{Genina}}\ \emph {et~al.}(2018)\citenamefont
  {{Genina}}, \citenamefont {{Ben{\'{\i}}tez-Llambay}}, \citenamefont
  {{Frenk}}, \citenamefont {{Cole}}, \citenamefont {{Fattahi}}, \citenamefont
  {{Navarro}}, \citenamefont {{Oman}}, \citenamefont {{Sawala}},\ and\
  \citenamefont {{Theuns}}}]{2018MNRAS.474.1398G}%
  \BibitemOpen
  \bibfield  {author} {\bibinfo {author} {\bibfnamefont {A.}~\bibnamefont
  {{Genina}}}, \bibinfo {author} {\bibfnamefont {A.}~\bibnamefont
  {{Ben{\'{\i}}tez-Llambay}}}, \bibinfo {author} {\bibfnamefont {C.~S.}\
  \bibnamefont {{Frenk}}}, \bibinfo {author} {\bibfnamefont {S.}~\bibnamefont
  {{Cole}}}, \bibinfo {author} {\bibfnamefont {A.}~\bibnamefont {{Fattahi}}},
  \bibinfo {author} {\bibfnamefont {J.~F.}\ \bibnamefont {{Navarro}}}, \bibinfo
  {author} {\bibfnamefont {K.~A.}\ \bibnamefont {{Oman}}}, \bibinfo {author}
  {\bibfnamefont {T.}~\bibnamefont {{Sawala}}}, \ and\ \bibinfo {author}
  {\bibfnamefont {T.}~\bibnamefont {{Theuns}}},\ }\href {\doibase
  10.1093/mnras/stx2855} {\bibfield  {journal} {\bibinfo  {journal} {Mon. Not.
  Roy. Astron. Soc.}\ }\textbf {\bibinfo {volume} {474}},\ \bibinfo {pages}
  {1398} (\bibinfo {year} {2018})}\BibitemShut {NoStop}%
\bibitem [{\citenamefont {Read}\ \emph {et~al.}(2018)\citenamefont {Read},
  \citenamefont {Walker},\ and\ \citenamefont {Steger}}]{Read:2018pft}%
  \BibitemOpen
  \bibfield  {author} {\bibinfo {author} {\bibfnamefont {J.~I.}\ \bibnamefont
  {Read}}, \bibinfo {author} {\bibfnamefont {M.~G.}\ \bibnamefont {Walker}}, \
  and\ \bibinfo {author} {\bibfnamefont {P.}~\bibnamefont {Steger}},\ }\href
  {\doibase 10.1093/mnras/sty2286} {\bibfield  {journal} {\bibinfo  {journal}
  {Mon. Not. Roy. Astron. Soc.}\ }\textbf {\bibinfo {volume} {481}},\ \bibinfo
  {pages} {860} (\bibinfo {year} {2018})},\ \Eprint
  {http://arxiv.org/abs/1805.06934} {arXiv:1805.06934 [astro-ph.GA]}
  \BibitemShut {NoStop}%
%%CITATION = ARXIV:1805.06934;%%
\bibitem [{\citenamefont {Navarro}\ \emph {et~al.}(1996)\citenamefont
  {Navarro}, \citenamefont {Eke},\ and\ \citenamefont
  {Frenk}}]{Navarro:1996bv}%
  \BibitemOpen
  \bibfield  {author} {\bibinfo {author} {\bibfnamefont {J.~F.}\ \bibnamefont
  {Navarro}}, \bibinfo {author} {\bibfnamefont {V.~R.}\ \bibnamefont {Eke}}, \
  and\ \bibinfo {author} {\bibfnamefont {C.~S.}\ \bibnamefont {Frenk}},\ }\href
  {\doibase 10.1093/mnras/283.3.72L, 10.1093/mnras/283.3.L72} {\bibfield
  {journal} {\bibinfo  {journal} {Mon. Not. Roy. Astron. Soc.}\ }\textbf
  {\bibinfo {volume} {283}},\ \bibinfo {pages} {L72} (\bibinfo {year}
  {1996})},\ \Eprint {http://arxiv.org/abs/astro-ph/9610187}
  {arXiv:astro-ph/9610187 [astro-ph]} \BibitemShut {NoStop}%
%%CITATION = ASTRO-PH/9610187;%%
\bibitem [{\citenamefont {Mac~Low}\ and\ \citenamefont
  {Ferrara}(1999)}]{MacLow:1998djk}%
  \BibitemOpen
  \bibfield  {author} {\bibinfo {author} {\bibfnamefont {M.-M.}\ \bibnamefont
  {Mac~Low}}\ and\ \bibinfo {author} {\bibfnamefont {A.}~\bibnamefont
  {Ferrara}},\ }\href {\doibase 10.1086/306832} {\bibfield  {journal} {\bibinfo
   {journal} {Astrophys. J.}\ }\textbf {\bibinfo {volume} {513}},\ \bibinfo
  {pages} {142} (\bibinfo {year} {1999})},\ \Eprint
  {http://arxiv.org/abs/astro-ph/9801237} {arXiv:astro-ph/9801237 [astro-ph]}
  \BibitemShut {NoStop}%
%%CITATION = ASTRO-PH/9801237;%%
\bibitem [{\citenamefont {Gelato}\ and\ \citenamefont
  {Sommer-Larsen}(1999)}]{Gelato:1998hb}%
  \BibitemOpen
  \bibfield  {author} {\bibinfo {author} {\bibfnamefont {S.}~\bibnamefont
  {Gelato}}\ and\ \bibinfo {author} {\bibfnamefont {J.}~\bibnamefont
  {Sommer-Larsen}},\ }\href {\doibase 10.1046/j.1365-8711.1999.02223.x}
  {\bibfield  {journal} {\bibinfo  {journal} {Mon. Not. Roy. Astron. Soc.}\
  }\textbf {\bibinfo {volume} {303}},\ \bibinfo {pages} {321} (\bibinfo {year}
  {1999})},\ \Eprint {http://arxiv.org/abs/astro-ph/9806289}
  {arXiv:astro-ph/9806289 [astro-ph]} \BibitemShut {NoStop}%
%%CITATION = ASTRO-PH/9806289;%%
\bibitem [{\citenamefont {Binney}\ \emph {et~al.}(2001)\citenamefont {Binney},
  \citenamefont {Gerhard},\ and\ \citenamefont {Silk}}]{Binney:2000zt}%
  \BibitemOpen
  \bibfield  {author} {\bibinfo {author} {\bibfnamefont {J.}~\bibnamefont
  {Binney}}, \bibinfo {author} {\bibfnamefont {O.}~\bibnamefont {Gerhard}}, \
  and\ \bibinfo {author} {\bibfnamefont {J.}~\bibnamefont {Silk}},\ }\href
  {\doibase 10.1046/j.1365-8711.2001.04024.x} {\bibfield  {journal} {\bibinfo
  {journal} {Mon. Not. Roy. Astron. Soc.}\ }\textbf {\bibinfo {volume} {321}},\
  \bibinfo {pages} {471} (\bibinfo {year} {2001})},\ \Eprint
  {http://arxiv.org/abs/astro-ph/0003199} {arXiv:astro-ph/0003199 [astro-ph]}
  \BibitemShut {NoStop}%
%%CITATION = ASTRO-PH/0003199;%%
\bibitem [{\citenamefont {Gnedin}\ and\ \citenamefont
  {Zhao}(2002)}]{Gnedin:2001ec}%
  \BibitemOpen
  \bibfield  {author} {\bibinfo {author} {\bibfnamefont {O.~Y.}\ \bibnamefont
  {Gnedin}}\ and\ \bibinfo {author} {\bibfnamefont {H.}~\bibnamefont {Zhao}},\
  }\href {\doibase 10.1046/j.1365-8711.2002.05361.x} {\bibfield  {journal}
  {\bibinfo  {journal} {Mon. Not. Roy. Astron. Soc.}\ }\textbf {\bibinfo
  {volume} {333}},\ \bibinfo {pages} {299} (\bibinfo {year} {2002})},\ \Eprint
  {http://arxiv.org/abs/astro-ph/0108108} {arXiv:astro-ph/0108108 [astro-ph]}
  \BibitemShut {NoStop}%
%%CITATION = ASTRO-PH/0108108;%%
\bibitem [{\citenamefont {El-Zant}\ \emph {et~al.}(2001)\citenamefont
  {El-Zant}, \citenamefont {Shlosman},\ and\ \citenamefont
  {Hoffman}}]{ElZant:2001re}%
  \BibitemOpen
  \bibfield  {author} {\bibinfo {author} {\bibfnamefont {A.}~\bibnamefont
  {El-Zant}}, \bibinfo {author} {\bibfnamefont {I.}~\bibnamefont {Shlosman}}, \
  and\ \bibinfo {author} {\bibfnamefont {Y.}~\bibnamefont {Hoffman}},\ }\href
  {\doibase 10.1086/322516} {\bibfield  {journal} {\bibinfo  {journal}
  {Astrophys. J.}\ }\textbf {\bibinfo {volume} {560}},\ \bibinfo {pages} {636}
  (\bibinfo {year} {2001})},\ \Eprint {http://arxiv.org/abs/astro-ph/0103386}
  {arXiv:astro-ph/0103386 [astro-ph]} \BibitemShut {NoStop}%
%%CITATION = ASTRO-PH/0103386;%%
\bibitem [{\citenamefont {Weinberg}\ and\ \citenamefont
  {Katz}(2002)}]{Weinberg:2001gm}%
  \BibitemOpen
  \bibfield  {author} {\bibinfo {author} {\bibfnamefont {M.~D.}\ \bibnamefont
  {Weinberg}}\ and\ \bibinfo {author} {\bibfnamefont {N.}~\bibnamefont
  {Katz}},\ }\href {\doibase 10.1086/343847} {\bibfield  {journal} {\bibinfo
  {journal} {Astrophys. J.}\ }\textbf {\bibinfo {volume} {580}},\ \bibinfo
  {pages} {627} (\bibinfo {year} {2002})},\ \Eprint
  {http://arxiv.org/abs/astro-ph/0110632} {arXiv:astro-ph/0110632 [astro-ph]}
  \BibitemShut {NoStop}%
%%CITATION = ASTRO-PH/0110632;%%
\bibitem [{\citenamefont {Ahn}\ and\ \citenamefont
  {Shapiro}(2005)}]{Ahn:2004xt}%
  \BibitemOpen
  \bibfield  {author} {\bibinfo {author} {\bibfnamefont {K.-J.}\ \bibnamefont
  {Ahn}}\ and\ \bibinfo {author} {\bibfnamefont {P.~R.}\ \bibnamefont
  {Shapiro}},\ }\href {\doibase 10.1111/j.1365-2966.2005.09492.x} {\bibfield
  {journal} {\bibinfo  {journal} {Mon. Not. Roy. Astron. Soc.}\ }\textbf
  {\bibinfo {volume} {363}},\ \bibinfo {pages} {1092} (\bibinfo {year}
  {2005})},\ \Eprint {http://arxiv.org/abs/astro-ph/0412169}
  {arXiv:astro-ph/0412169 [astro-ph]} \BibitemShut {NoStop}%
%%CITATION = ASTRO-PH/0412169;%%
\bibitem [{\citenamefont {Tonini}\ and\ \citenamefont
  {Lapi}(2006)}]{Tonini:2006gwz}%
  \BibitemOpen
  \bibfield  {author} {\bibinfo {author} {\bibfnamefont {C.}~\bibnamefont
  {Tonini}}\ and\ \bibinfo {author} {\bibfnamefont {A.}~\bibnamefont {Lapi}},\
  }\href {\doibase 10.1086/506431} {\bibfield  {journal} {\bibinfo  {journal}
  {Astrophys. J.}\ }\textbf {\bibinfo {volume} {649}},\ \bibinfo {pages} {591}
  (\bibinfo {year} {2006})},\ \Eprint {http://arxiv.org/abs/astro-ph/0603051}
  {arXiv:astro-ph/0603051 [astro-ph]} \BibitemShut {NoStop}%
%%CITATION = ASTRO-PH/0603051;%%
\bibitem [{\citenamefont {Governato}\ \emph {et~al.}(2010)\citenamefont
  {Governato} \emph {et~al.}}]{Governato:2009bg}%
  \BibitemOpen
  \bibfield  {author} {\bibinfo {author} {\bibfnamefont {F.}~\bibnamefont
  {Governato}} \emph {et~al.},\ }\href {\doibase 10.1038/nature08640}
  {\bibfield  {journal} {\bibinfo  {journal} {Nature}\ }\textbf {\bibinfo
  {volume} {463}},\ \bibinfo {pages} {203} (\bibinfo {year} {2010})},\ \Eprint
  {http://arxiv.org/abs/0911.2237} {arXiv:0911.2237 [astro-ph.CO]} \BibitemShut
  {NoStop}%
%%CITATION = ARXIV:0911.2237;%%
\bibitem [{\citenamefont {Silk}\ and\ \citenamefont
  {Nusser}(2010)}]{Silk:2010aw}%
  \BibitemOpen
  \bibfield  {author} {\bibinfo {author} {\bibfnamefont {J.}~\bibnamefont
  {Silk}}\ and\ \bibinfo {author} {\bibfnamefont {A.}~\bibnamefont {Nusser}},\
  }\href {\doibase 10.1088/0004-637X/725/1/556} {\bibfield  {journal} {\bibinfo
   {journal} {Astrophys. J.}\ }\textbf {\bibinfo {volume} {725}},\ \bibinfo
  {pages} {556} (\bibinfo {year} {2010})},\ \Eprint
  {http://arxiv.org/abs/1004.0857} {arXiv:1004.0857 [astro-ph.CO]} \BibitemShut
  {NoStop}%
%%CITATION = ARXIV:1004.0857;%%
\bibitem [{\citenamefont {Vera-Ciro}\ \emph {et~al.}(2013)\citenamefont
  {Vera-Ciro}, \citenamefont {Helmi}, \citenamefont {Starkenburg},\ and\
  \citenamefont {Breddels}}]{VeraCiro:2012na}%
  \BibitemOpen
  \bibfield  {author} {\bibinfo {author} {\bibfnamefont {C.~A.}\ \bibnamefont
  {Vera-Ciro}}, \bibinfo {author} {\bibfnamefont {A.}~\bibnamefont {Helmi}},
  \bibinfo {author} {\bibfnamefont {E.}~\bibnamefont {Starkenburg}}, \ and\
  \bibinfo {author} {\bibfnamefont {M.~A.}\ \bibnamefont {Breddels}},\ }\href
  {\doibase 10.1093/mnras/sts148} {\bibfield  {journal} {\bibinfo  {journal}
  {Mon. Not. Roy. Astron. Soc.}\ }\textbf {\bibinfo {volume} {428}},\ \bibinfo
  {pages} {1696} (\bibinfo {year} {2013})},\ \Eprint
  {http://arxiv.org/abs/1202.6061} {arXiv:1202.6061 [astro-ph.CO]} \BibitemShut
  {NoStop}%
%%CITATION = ARXIV:1202.6061;%%
\bibitem [{\citenamefont {Martizzi}\ \emph {et~al.}(2012)\citenamefont
  {Martizzi}, \citenamefont {Teyssier}, \citenamefont {Moore},\ and\
  \citenamefont {Wentz}}]{Martizzi:2011aa}%
  \BibitemOpen
  \bibfield  {author} {\bibinfo {author} {\bibfnamefont {D.}~\bibnamefont
  {Martizzi}}, \bibinfo {author} {\bibfnamefont {R.}~\bibnamefont {Teyssier}},
  \bibinfo {author} {\bibfnamefont {B.}~\bibnamefont {Moore}}, \ and\ \bibinfo
  {author} {\bibfnamefont {T.}~\bibnamefont {Wentz}},\ }\href {\doibase
  10.1111/j.1365-2966.2012.20879.x} {\bibfield  {journal} {\bibinfo  {journal}
  {Mon. Not. Roy. Astron. Soc.}\ }\textbf {\bibinfo {volume} {422}},\ \bibinfo
  {pages} {3081} (\bibinfo {year} {2012})},\ \Eprint
  {http://arxiv.org/abs/1112.2752} {arXiv:1112.2752 [astro-ph.CO]} \BibitemShut
  {NoStop}%
%%CITATION = ARXIV:1112.2752;%%
\bibitem [{\citenamefont {Read}\ \emph {et~al.}(2019)\citenamefont {Read},
  \citenamefont {Walker},\ and\ \citenamefont {Steger}}]{Read:2018fxs}%
  \BibitemOpen
  \bibfield  {author} {\bibinfo {author} {\bibfnamefont {J.~I.}\ \bibnamefont
  {Read}}, \bibinfo {author} {\bibfnamefont {M.~G.}\ \bibnamefont {Walker}}, \
  and\ \bibinfo {author} {\bibfnamefont {P.}~\bibnamefont {Steger}},\ }\href
  {\doibase 10.1093/mnras/sty3404} {\bibfield  {journal} {\bibinfo  {journal}
  {Mon. Not. Roy. Astron. Soc.}\ }\textbf {\bibinfo {volume} {484}},\ \bibinfo
  {pages} {1401} (\bibinfo {year} {2019})},\ \Eprint
  {http://arxiv.org/abs/1808.06634} {arXiv:1808.06634 [astro-ph.GA]}
  \BibitemShut {NoStop}%
%%CITATION = ARXIV:1808.06634;%%
\bibitem [{\citenamefont {Spergel}\ and\ \citenamefont
  {Steinhardt}(2000)}]{Spergel:1999mh}%
  \BibitemOpen
  \bibfield  {author} {\bibinfo {author} {\bibfnamefont {D.~N.}\ \bibnamefont
  {Spergel}}\ and\ \bibinfo {author} {\bibfnamefont {P.~J.}\ \bibnamefont
  {Steinhardt}},\ }\href {\doibase 10.1103/PhysRevLett.84.3760} {\bibfield
  {journal} {\bibinfo  {journal} {Phys. Rev. Lett.}\ }\textbf {\bibinfo
  {volume} {84}},\ \bibinfo {pages} {3760} (\bibinfo {year} {2000})},\ \Eprint
  {http://arxiv.org/abs/astro-ph/9909386} {arXiv:astro-ph/9909386 [astro-ph]}
  \BibitemShut {NoStop}%
%%CITATION = ASTRO-PH/9909386;%%
\bibitem [{\citenamefont {Wandelt}\ \emph {et~al.}(2000)\citenamefont
  {Wandelt}, \citenamefont {Dave}, \citenamefont {Farrar}, \citenamefont
  {McGuire}, \citenamefont {Spergel},\ and\ \citenamefont
  {Steinhardt}}]{Wandelt:2000ad}%
  \BibitemOpen
  \bibfield  {author} {\bibinfo {author} {\bibfnamefont {B.~D.}\ \bibnamefont
  {Wandelt}}, \bibinfo {author} {\bibfnamefont {R.}~\bibnamefont {Dave}},
  \bibinfo {author} {\bibfnamefont {G.~R.}\ \bibnamefont {Farrar}}, \bibinfo
  {author} {\bibfnamefont {P.~C.}\ \bibnamefont {McGuire}}, \bibinfo {author}
  {\bibfnamefont {D.~N.}\ \bibnamefont {Spergel}}, \ and\ \bibinfo {author}
  {\bibfnamefont {P.~J.}\ \bibnamefont {Steinhardt}},\ }in\ \href
  {http://www.slac.stanford.edu/spires/find/books/www?cl=QB461:I57:2000} {\emph
  {\bibinfo {booktitle} {{Sources and detection of dark matter and dark energy
  in the universe. Proceedings, 4th International Symposium, DM 2000, Marina
  del Rey, USA, February 23-25, 2000}}}}\ (\bibinfo {year} {2000})\ pp.\
  \bibinfo {pages} {263--274},\ \Eprint {http://arxiv.org/abs/astro-ph/0006344}
  {arXiv:astro-ph/0006344 [astro-ph]} \BibitemShut {NoStop}%
%%CITATION = ASTRO-PH/0006344;%%
\bibitem [{\citenamefont {Vogelsberger}\ \emph {et~al.}(2012)\citenamefont
  {Vogelsberger}, \citenamefont {Zavala},\ and\ \citenamefont
  {Loeb}}]{Vogelsberger:2012ku}%
  \BibitemOpen
  \bibfield  {author} {\bibinfo {author} {\bibfnamefont {M.}~\bibnamefont
  {Vogelsberger}}, \bibinfo {author} {\bibfnamefont {J.}~\bibnamefont
  {Zavala}}, \ and\ \bibinfo {author} {\bibfnamefont {A.}~\bibnamefont
  {Loeb}},\ }\href {\doibase 10.1111/j.1365-2966.2012.21182.x} {\bibfield
  {journal} {\bibinfo  {journal} {Mon. Not. Roy. Astron. Soc.}\ }\textbf
  {\bibinfo {volume} {423}},\ \bibinfo {pages} {3740} (\bibinfo {year}
  {2012})},\ \Eprint {http://arxiv.org/abs/1201.5892} {arXiv:1201.5892
  [astro-ph.CO]} \BibitemShut {NoStop}%
%%CITATION = ARXIV:1201.5892;%%
\bibitem [{\citenamefont {Rocha}\ \emph {et~al.}(2013)\citenamefont {Rocha},
  \citenamefont {Peter}, \citenamefont {Bullock}, \citenamefont {Kaplinghat},
  \citenamefont {Garrison-Kimmel}, \citenamefont {Onorbe},\ and\ \citenamefont
  {Moustakas}}]{Rocha:2012jg}%
  \BibitemOpen
  \bibfield  {author} {\bibinfo {author} {\bibfnamefont {M.}~\bibnamefont
  {Rocha}}, \bibinfo {author} {\bibfnamefont {A.~H.~G.}\ \bibnamefont {Peter}},
  \bibinfo {author} {\bibfnamefont {J.~S.}\ \bibnamefont {Bullock}}, \bibinfo
  {author} {\bibfnamefont {M.}~\bibnamefont {Kaplinghat}}, \bibinfo {author}
  {\bibfnamefont {S.}~\bibnamefont {Garrison-Kimmel}}, \bibinfo {author}
  {\bibfnamefont {J.}~\bibnamefont {Onorbe}}, \ and\ \bibinfo {author}
  {\bibfnamefont {L.~A.}\ \bibnamefont {Moustakas}},\ }\href {\doibase
  10.1093/mnras/sts514} {\bibfield  {journal} {\bibinfo  {journal} {Mon. Not.
  Roy. Astron. Soc.}\ }\textbf {\bibinfo {volume} {430}},\ \bibinfo {pages}
  {81} (\bibinfo {year} {2013})},\ \Eprint {http://arxiv.org/abs/1208.3025}
  {arXiv:1208.3025 [astro-ph.CO]} \BibitemShut {NoStop}%
%%CITATION = ARXIV:1208.3025;%%
\bibitem [{\citenamefont {Peter}\ \emph {et~al.}(2013)\citenamefont {Peter},
  \citenamefont {Rocha}, \citenamefont {Bullock},\ and\ \citenamefont
  {Kaplinghat}}]{Peter:2012jh}%
  \BibitemOpen
  \bibfield  {author} {\bibinfo {author} {\bibfnamefont {A.~H.~G.}\
  \bibnamefont {Peter}}, \bibinfo {author} {\bibfnamefont {M.}~\bibnamefont
  {Rocha}}, \bibinfo {author} {\bibfnamefont {J.~S.}\ \bibnamefont {Bullock}},
  \ and\ \bibinfo {author} {\bibfnamefont {M.}~\bibnamefont {Kaplinghat}},\
  }\href {\doibase 10.1093/mnras/sts535} {\bibfield  {journal} {\bibinfo
  {journal} {Mon. Not. Roy. Astron. Soc.}\ }\textbf {\bibinfo {volume} {430}},\
  \bibinfo {pages} {105} (\bibinfo {year} {2013})},\ \Eprint
  {http://arxiv.org/abs/1208.3026} {arXiv:1208.3026 [astro-ph.CO]} \BibitemShut
  {NoStop}%
%%CITATION = ARXIV:1208.3026;%%
\bibitem [{\citenamefont {Zavala}\ \emph {et~al.}(2013)\citenamefont {Zavala},
  \citenamefont {Vogelsberger},\ and\ \citenamefont {Walker}}]{Zavala:2012us}%
  \BibitemOpen
  \bibfield  {author} {\bibinfo {author} {\bibfnamefont {J.}~\bibnamefont
  {Zavala}}, \bibinfo {author} {\bibfnamefont {M.}~\bibnamefont
  {Vogelsberger}}, \ and\ \bibinfo {author} {\bibfnamefont {M.~G.}\
  \bibnamefont {Walker}},\ }\href {\doibase 10.1093/mnrasl/sls053} {\bibfield
  {journal} {\bibinfo  {journal} {Mon. Not. Roy. Astron. Soc.}\ }\textbf
  {\bibinfo {volume} {431}},\ \bibinfo {pages} {L20} (\bibinfo {year}
  {2013})},\ \Eprint {http://arxiv.org/abs/1211.6426} {arXiv:1211.6426
  [astro-ph.CO]} \BibitemShut {NoStop}%
%%CITATION = ARXIV:1211.6426;%%
\bibitem [{\citenamefont {Vogelsberger}\ \emph {et~al.}(2014)\citenamefont
  {Vogelsberger}, \citenamefont {Zavala}, \citenamefont {Simpson},\ and\
  \citenamefont {Jenkins}}]{Vogelsberger:2014pda}%
  \BibitemOpen
  \bibfield  {author} {\bibinfo {author} {\bibfnamefont {M.}~\bibnamefont
  {Vogelsberger}}, \bibinfo {author} {\bibfnamefont {J.}~\bibnamefont
  {Zavala}}, \bibinfo {author} {\bibfnamefont {C.}~\bibnamefont {Simpson}}, \
  and\ \bibinfo {author} {\bibfnamefont {A.}~\bibnamefont {Jenkins}},\ }\href
  {\doibase 10.1093/mnras/stu1713} {\bibfield  {journal} {\bibinfo  {journal}
  {Mon. Not. Roy. Astron. Soc.}\ }\textbf {\bibinfo {volume} {444}},\ \bibinfo
  {pages} {3684} (\bibinfo {year} {2014})},\ \Eprint
  {http://arxiv.org/abs/1405.5216} {arXiv:1405.5216 [astro-ph.CO]} \BibitemShut
  {NoStop}%
%%CITATION = ARXIV:1405.5216;%%
\bibitem [{\citenamefont {Elbert}\ \emph {et~al.}(2015)\citenamefont {Elbert},
  \citenamefont {Bullock}, \citenamefont {Garrison-Kimmel}, \citenamefont
  {Rocha}, \citenamefont {Oñorbe},\ and\ \citenamefont
  {Peter}}]{Elbert:2014bma}%
  \BibitemOpen
  \bibfield  {author} {\bibinfo {author} {\bibfnamefont {O.~D.}\ \bibnamefont
  {Elbert}}, \bibinfo {author} {\bibfnamefont {J.~S.}\ \bibnamefont {Bullock}},
  \bibinfo {author} {\bibfnamefont {S.}~\bibnamefont {Garrison-Kimmel}},
  \bibinfo {author} {\bibfnamefont {M.}~\bibnamefont {Rocha}}, \bibinfo
  {author} {\bibfnamefont {J.}~\bibnamefont {Oñorbe}}, \ and\ \bibinfo
  {author} {\bibfnamefont {A.~H.~G.}\ \bibnamefont {Peter}},\ }\href {\doibase
  10.1093/mnras/stv1470} {\bibfield  {journal} {\bibinfo  {journal} {Mon. Not.
  Roy. Astron. Soc.}\ }\textbf {\bibinfo {volume} {453}},\ \bibinfo {pages}
  {29} (\bibinfo {year} {2015})},\ \Eprint {http://arxiv.org/abs/1412.1477}
  {arXiv:1412.1477 [astro-ph.GA]} \BibitemShut {NoStop}%
%%CITATION = ARXIV:1412.1477;%%
\bibitem [{\citenamefont {Kaplinghat}\ \emph {et~al.}(2016)\citenamefont
  {Kaplinghat}, \citenamefont {Tulin},\ and\ \citenamefont
  {Yu}}]{Kaplinghat:2015aga}%
  \BibitemOpen
  \bibfield  {author} {\bibinfo {author} {\bibfnamefont {M.}~\bibnamefont
  {Kaplinghat}}, \bibinfo {author} {\bibfnamefont {S.}~\bibnamefont {Tulin}}, \
  and\ \bibinfo {author} {\bibfnamefont {H.-B.}\ \bibnamefont {Yu}},\ }\href
  {\doibase 10.1103/PhysRevLett.116.041302} {\bibfield  {journal} {\bibinfo
  {journal} {Phys. Rev. Lett.}\ }\textbf {\bibinfo {volume} {116}},\ \bibinfo
  {pages} {041302} (\bibinfo {year} {2016})},\ \Eprint
  {http://arxiv.org/abs/1508.03339} {arXiv:1508.03339 [astro-ph.CO]}
  \BibitemShut {NoStop}%
%%CITATION = ARXIV:1508.03339;%%
\bibitem [{\citenamefont {Fry}\ \emph {et~al.}(2015)\citenamefont {Fry},
  \citenamefont {Governato}, \citenamefont {Pontzen}, \citenamefont {Quinn},
  \citenamefont {Tremmel}, \citenamefont {Anderson}, \citenamefont {Menon},
  \citenamefont {Brooks},\ and\ \citenamefont {Wadsley}}]{Fry:2015rta}%
  \BibitemOpen
  \bibfield  {author} {\bibinfo {author} {\bibfnamefont {A.~B.}\ \bibnamefont
  {Fry}}, \bibinfo {author} {\bibfnamefont {F.}~\bibnamefont {Governato}},
  \bibinfo {author} {\bibfnamefont {A.}~\bibnamefont {Pontzen}}, \bibinfo
  {author} {\bibfnamefont {T.}~\bibnamefont {Quinn}}, \bibinfo {author}
  {\bibfnamefont {M.}~\bibnamefont {Tremmel}}, \bibinfo {author} {\bibfnamefont
  {L.}~\bibnamefont {Anderson}}, \bibinfo {author} {\bibfnamefont
  {H.}~\bibnamefont {Menon}}, \bibinfo {author} {\bibfnamefont {A.~M.}\
  \bibnamefont {Brooks}}, \ and\ \bibinfo {author} {\bibfnamefont
  {J.}~\bibnamefont {Wadsley}},\ }\href {\doibase 10.1093/mnras/stv1330}
  {\bibfield  {journal} {\bibinfo  {journal} {Mon. Not. Roy. Astron. Soc.}\
  }\textbf {\bibinfo {volume} {452}},\ \bibinfo {pages} {1468} (\bibinfo {year}
  {2015})},\ \Eprint {http://arxiv.org/abs/1501.00497} {arXiv:1501.00497
  [astro-ph.CO]} \BibitemShut {NoStop}%
%%CITATION = ARXIV:1501.00497;%%
\bibitem [{\citenamefont {Clowe}\ \emph {et~al.}(2004)\citenamefont {Clowe},
  \citenamefont {Gonzalez},\ and\ \citenamefont {Markevitch}}]{Clowe:2003tk}%
  \BibitemOpen
  \bibfield  {author} {\bibinfo {author} {\bibfnamefont {D.}~\bibnamefont
  {Clowe}}, \bibinfo {author} {\bibfnamefont {A.}~\bibnamefont {Gonzalez}}, \
  and\ \bibinfo {author} {\bibfnamefont {M.}~\bibnamefont {Markevitch}},\
  }\href {\doibase 10.1086/381970} {\bibfield  {journal} {\bibinfo  {journal}
  {Astrophys. J.}\ }\textbf {\bibinfo {volume} {604}},\ \bibinfo {pages} {596}
  (\bibinfo {year} {2004})},\ \Eprint {http://arxiv.org/abs/astro-ph/0312273}
  {arXiv:astro-ph/0312273 [astro-ph]} \BibitemShut {NoStop}%
%%CITATION = ASTRO-PH/0312273;%%
\bibitem [{\citenamefont {Markevitch}\ \emph {et~al.}(2004)\citenamefont
  {Markevitch}, \citenamefont {Gonzalez}, \citenamefont {Clowe}, \citenamefont
  {Vikhlinin}, \citenamefont {David}, \citenamefont {Forman}, \citenamefont
  {Jones}, \citenamefont {Murray},\ and\ \citenamefont
  {Tucker}}]{Markevitch:2003at}%
  \BibitemOpen
  \bibfield  {author} {\bibinfo {author} {\bibfnamefont {M.}~\bibnamefont
  {Markevitch}}, \bibinfo {author} {\bibfnamefont {A.~H.}\ \bibnamefont
  {Gonzalez}}, \bibinfo {author} {\bibfnamefont {D.}~\bibnamefont {Clowe}},
  \bibinfo {author} {\bibfnamefont {A.}~\bibnamefont {Vikhlinin}}, \bibinfo
  {author} {\bibfnamefont {L.}~\bibnamefont {David}}, \bibinfo {author}
  {\bibfnamefont {W.}~\bibnamefont {Forman}}, \bibinfo {author} {\bibfnamefont
  {C.}~\bibnamefont {Jones}}, \bibinfo {author} {\bibfnamefont
  {S.}~\bibnamefont {Murray}}, \ and\ \bibinfo {author} {\bibfnamefont
  {W.}~\bibnamefont {Tucker}},\ }\href {\doibase 10.1086/383178} {\bibfield
  {journal} {\bibinfo  {journal} {Astrophys. J.}\ }\textbf {\bibinfo {volume}
  {606}},\ \bibinfo {pages} {819} (\bibinfo {year} {2004})},\ \Eprint
  {http://arxiv.org/abs/astro-ph/0309303} {arXiv:astro-ph/0309303 [astro-ph]}
  \BibitemShut {NoStop}%
%%CITATION = ASTRO-PH/0309303;%%
\bibitem [{\citenamefont {Randall}\ \emph {et~al.}(2008)\citenamefont
  {Randall}, \citenamefont {Markevitch}, \citenamefont {Clowe}, \citenamefont
  {Gonzalez},\ and\ \citenamefont {Bradac}}]{Randall:2007ph}%
  \BibitemOpen
  \bibfield  {author} {\bibinfo {author} {\bibfnamefont {S.~W.}\ \bibnamefont
  {Randall}}, \bibinfo {author} {\bibfnamefont {M.}~\bibnamefont {Markevitch}},
  \bibinfo {author} {\bibfnamefont {D.}~\bibnamefont {Clowe}}, \bibinfo
  {author} {\bibfnamefont {A.~H.}\ \bibnamefont {Gonzalez}}, \ and\ \bibinfo
  {author} {\bibfnamefont {M.}~\bibnamefont {Bradac}},\ }\href {\doibase
  10.1086/587859} {\bibfield  {journal} {\bibinfo  {journal} {Astrophys. J.}\
  }\textbf {\bibinfo {volume} {679}},\ \bibinfo {pages} {1173} (\bibinfo {year}
  {2008})},\ \Eprint {http://arxiv.org/abs/0704.0261} {arXiv:0704.0261
  [astro-ph]} \BibitemShut {NoStop}%
%%CITATION = ARXIV:0704.0261;%%
\bibitem [{\citenamefont {Harvey}\ \emph {et~al.}(2015)\citenamefont {Harvey},
  \citenamefont {Massey}, \citenamefont {Kitching}, \citenamefont {Taylor},\
  and\ \citenamefont {Tittley}}]{Harvey:2015hha}%
  \BibitemOpen
  \bibfield  {author} {\bibinfo {author} {\bibfnamefont {D.}~\bibnamefont
  {Harvey}}, \bibinfo {author} {\bibfnamefont {R.}~\bibnamefont {Massey}},
  \bibinfo {author} {\bibfnamefont {T.}~\bibnamefont {Kitching}}, \bibinfo
  {author} {\bibfnamefont {A.}~\bibnamefont {Taylor}}, \ and\ \bibinfo {author}
  {\bibfnamefont {E.}~\bibnamefont {Tittley}},\ }\href {\doibase
  10.1126/science.1261381} {\bibfield  {journal} {\bibinfo  {journal}
  {Science}\ }\textbf {\bibinfo {volume} {347}},\ \bibinfo {pages} {1462}
  (\bibinfo {year} {2015})},\ \Eprint {http://arxiv.org/abs/1503.07675}
  {arXiv:1503.07675 [astro-ph.CO]} \BibitemShut {NoStop}%
%%CITATION = ARXIV:1503.07675;%%
\bibitem [{\citenamefont {Bondarenko}\ \emph {et~al.}(2018)\citenamefont
  {Bondarenko}, \citenamefont {Boyarsky}, \citenamefont {Bringmann},\ and\
  \citenamefont {Sokolenko}}]{Bondarenko:2017rfu}%
  \BibitemOpen
  \bibfield  {author} {\bibinfo {author} {\bibfnamefont {K.}~\bibnamefont
  {Bondarenko}}, \bibinfo {author} {\bibfnamefont {A.}~\bibnamefont
  {Boyarsky}}, \bibinfo {author} {\bibfnamefont {T.}~\bibnamefont {Bringmann}},
  \ and\ \bibinfo {author} {\bibfnamefont {A.}~\bibnamefont {Sokolenko}},\
  }\href {\doibase 10.1088/1475-7516/2018/04/049} {\bibfield  {journal}
  {\bibinfo  {journal} {JCAP}\ }\textbf {\bibinfo {volume} {1804}},\ \bibinfo
  {pages} {049} (\bibinfo {year} {2018})},\ \Eprint
  {http://arxiv.org/abs/1712.06602} {arXiv:1712.06602 [astro-ph.CO]}
  \BibitemShut {NoStop}%
%%CITATION = ARXIV:1712.06602;%%
\bibitem [{\citenamefont {Harvey}\ \emph {et~al.}(2019)\citenamefont {Harvey},
  \citenamefont {Robertson}, \citenamefont {Massey},\ and\ \citenamefont
  {McCarthy}}]{Harvey:2018uwf}%
  \BibitemOpen
  \bibfield  {author} {\bibinfo {author} {\bibfnamefont {D.}~\bibnamefont
  {Harvey}}, \bibinfo {author} {\bibfnamefont {A.}~\bibnamefont {Robertson}},
  \bibinfo {author} {\bibfnamefont {R.}~\bibnamefont {Massey}}, \ and\ \bibinfo
  {author} {\bibfnamefont {I.~G.}\ \bibnamefont {McCarthy}},\ }\href {\doibase
  10.1093/mnras/stz1816} {\bibfield  {journal} {\bibinfo  {journal} {Mon. Not.
  Roy. Astron. Soc.}\ }\textbf {\bibinfo {volume} {488}},\ \bibinfo {pages}
  {1572} (\bibinfo {year} {2019})},\ \Eprint {http://arxiv.org/abs/1812.06981}
  {arXiv:1812.06981 [astro-ph.CO]} \BibitemShut {NoStop}%
%%CITATION = ARXIV:1812.06981;%%
\bibitem [{\citenamefont {Feng}\ \emph {et~al.}(2010)\citenamefont {Feng},
  \citenamefont {Kaplinghat},\ and\ \citenamefont {Yu}}]{Feng:2009hw}%
  \BibitemOpen
  \bibfield  {author} {\bibinfo {author} {\bibfnamefont {J.~L.}\ \bibnamefont
  {Feng}}, \bibinfo {author} {\bibfnamefont {M.}~\bibnamefont {Kaplinghat}}, \
  and\ \bibinfo {author} {\bibfnamefont {H.-B.}\ \bibnamefont {Yu}},\ }\href
  {\doibase 10.1103/PhysRevLett.104.151301} {\bibfield  {journal} {\bibinfo
  {journal} {Phys. Rev. Lett.}\ }\textbf {\bibinfo {volume} {104}},\ \bibinfo
  {pages} {151301} (\bibinfo {year} {2010})},\ \Eprint
  {http://arxiv.org/abs/0911.0422} {arXiv:0911.0422 [hep-ph]} \BibitemShut
  {NoStop}%
%%CITATION = ARXIV:0911.0422;%%
\bibitem [{\citenamefont {Buckley}\ and\ \citenamefont
  {Fox}(2010)}]{Buckley:2009in}%
  \BibitemOpen
  \bibfield  {author} {\bibinfo {author} {\bibfnamefont {M.~R.}\ \bibnamefont
  {Buckley}}\ and\ \bibinfo {author} {\bibfnamefont {P.~J.}\ \bibnamefont
  {Fox}},\ }\href {\doibase 10.1103/PhysRevD.81.083522} {\bibfield  {journal}
  {\bibinfo  {journal} {Phys. Rev.}\ }\textbf {\bibinfo {volume} {D81}},\
  \bibinfo {pages} {083522} (\bibinfo {year} {2010})},\ \Eprint
  {http://arxiv.org/abs/0911.3898} {arXiv:0911.3898 [hep-ph]} \BibitemShut
  {NoStop}%
%%CITATION = ARXIV:0911.3898;%%
\bibitem [{\citenamefont {Tulin}\ \emph
  {et~al.}(2013{\natexlab{a}})\citenamefont {Tulin}, \citenamefont {Yu},\ and\
  \citenamefont {Zurek}}]{Tulin:2012wi}%
  \BibitemOpen
  \bibfield  {author} {\bibinfo {author} {\bibfnamefont {S.}~\bibnamefont
  {Tulin}}, \bibinfo {author} {\bibfnamefont {H.-B.}\ \bibnamefont {Yu}}, \
  and\ \bibinfo {author} {\bibfnamefont {K.~M.}\ \bibnamefont {Zurek}},\ }\href
  {\doibase 10.1103/PhysRevLett.110.111301} {\bibfield  {journal} {\bibinfo
  {journal} {Phys. Rev. Lett.}\ }\textbf {\bibinfo {volume} {110}},\ \bibinfo
  {pages} {111301} (\bibinfo {year} {2013}{\natexlab{a}})},\ \Eprint
  {http://arxiv.org/abs/1210.0900} {arXiv:1210.0900 [hep-ph]} \BibitemShut
  {NoStop}%
%%CITATION = ARXIV:1210.0900;%%
\bibitem [{\citenamefont {Tulin}\ \emph
  {et~al.}(2013{\natexlab{b}})\citenamefont {Tulin}, \citenamefont {Yu},\ and\
  \citenamefont {Zurek}}]{Tulin:2013teo}%
  \BibitemOpen
  \bibfield  {author} {\bibinfo {author} {\bibfnamefont {S.}~\bibnamefont
  {Tulin}}, \bibinfo {author} {\bibfnamefont {H.-B.}\ \bibnamefont {Yu}}, \
  and\ \bibinfo {author} {\bibfnamefont {K.~M.}\ \bibnamefont {Zurek}},\ }\href
  {\doibase 10.1103/PhysRevD.87.115007} {\bibfield  {journal} {\bibinfo
  {journal} {Phys. Rev.}\ }\textbf {\bibinfo {volume} {D87}},\ \bibinfo {pages}
  {115007} (\bibinfo {year} {2013}{\natexlab{b}})},\ \Eprint
  {http://arxiv.org/abs/1302.3898} {arXiv:1302.3898 [hep-ph]} \BibitemShut
  {NoStop}%
%%CITATION = ARXIV:1302.3898;%%
\bibitem [{\citenamefont {Kaplinghat}\ \emph {et~al.}(2014)\citenamefont
  {Kaplinghat}, \citenamefont {Tulin},\ and\ \citenamefont
  {Yu}}]{Kaplinghat:2013yxa}%
  \BibitemOpen
  \bibfield  {author} {\bibinfo {author} {\bibfnamefont {M.}~\bibnamefont
  {Kaplinghat}}, \bibinfo {author} {\bibfnamefont {S.}~\bibnamefont {Tulin}}, \
  and\ \bibinfo {author} {\bibfnamefont {H.-B.}\ \bibnamefont {Yu}},\ }\href
  {\doibase 10.1103/PhysRevD.89.035009} {\bibfield  {journal} {\bibinfo
  {journal} {Phys. Rev.}\ }\textbf {\bibinfo {volume} {D89}},\ \bibinfo {pages}
  {035009} (\bibinfo {year} {2014})},\ \Eprint {http://arxiv.org/abs/1310.7945}
  {arXiv:1310.7945 [hep-ph]} \BibitemShut {NoStop}%
%%CITATION = ARXIV:1310.7945;%%
\bibitem [{\citenamefont {Abdullah}\ \emph {et~al.}(2014)\citenamefont
  {Abdullah}, \citenamefont {DiFranzo}, \citenamefont {Rajaraman},
  \citenamefont {Tait}, \citenamefont {Tanedo},\ and\ \citenamefont
  {Wijangco}}]{Abdullah:2014lla}%
  \BibitemOpen
  \bibfield  {author} {\bibinfo {author} {\bibfnamefont {M.}~\bibnamefont
  {Abdullah}}, \bibinfo {author} {\bibfnamefont {A.}~\bibnamefont {DiFranzo}},
  \bibinfo {author} {\bibfnamefont {A.}~\bibnamefont {Rajaraman}}, \bibinfo
  {author} {\bibfnamefont {T.~M.~P.}\ \bibnamefont {Tait}}, \bibinfo {author}
  {\bibfnamefont {P.}~\bibnamefont {Tanedo}}, \ and\ \bibinfo {author}
  {\bibfnamefont {A.~M.}\ \bibnamefont {Wijangco}},\ }\href {\doibase
  10.1103/PhysRevD.90.035004} {\bibfield  {journal} {\bibinfo  {journal} {Phys.
  Rev.}\ }\textbf {\bibinfo {volume} {D90}},\ \bibinfo {pages} {035004}
  (\bibinfo {year} {2014})},\ \Eprint {http://arxiv.org/abs/1404.6528}
  {arXiv:1404.6528 [hep-ph]} \BibitemShut {NoStop}%
%%CITATION = ARXIV:1404.6528;%%
\bibitem [{\citenamefont {Del~Nobile}\ \emph {et~al.}(2015)\citenamefont
  {Del~Nobile}, \citenamefont {Kaplinghat},\ and\ \citenamefont
  {Yu}}]{DelNobile:2015uua}%
  \BibitemOpen
  \bibfield  {author} {\bibinfo {author} {\bibfnamefont {E.}~\bibnamefont
  {Del~Nobile}}, \bibinfo {author} {\bibfnamefont {M.}~\bibnamefont
  {Kaplinghat}}, \ and\ \bibinfo {author} {\bibfnamefont {H.-B.}\ \bibnamefont
  {Yu}},\ }\href {\doibase 10.1088/1475-7516/2015/10/055} {\bibfield  {journal}
  {\bibinfo  {journal} {JCAP}\ }\textbf {\bibinfo {volume} {1510}},\ \bibinfo
  {pages} {055} (\bibinfo {year} {2015})},\ \Eprint
  {http://arxiv.org/abs/1507.04007} {arXiv:1507.04007 [hep-ph]} \BibitemShut
  {NoStop}%
%%CITATION = ARXIV:1507.04007;%%
\bibitem [{\citenamefont {Bernal}\ \emph {et~al.}(2016)\citenamefont {Bernal},
  \citenamefont {Chu}, \citenamefont {Garcia-Cely}, \citenamefont {Hambye},\
  and\ \citenamefont {Zaldivar}}]{Bernal:2015ova}%
  \BibitemOpen
  \bibfield  {author} {\bibinfo {author} {\bibfnamefont {N.}~\bibnamefont
  {Bernal}}, \bibinfo {author} {\bibfnamefont {X.}~\bibnamefont {Chu}},
  \bibinfo {author} {\bibfnamefont {C.}~\bibnamefont {Garcia-Cely}}, \bibinfo
  {author} {\bibfnamefont {T.}~\bibnamefont {Hambye}}, \ and\ \bibinfo {author}
  {\bibfnamefont {B.}~\bibnamefont {Zaldivar}},\ }\href {\doibase
  10.1088/1475-7516/2016/03/018} {\bibfield  {journal} {\bibinfo  {journal}
  {JCAP}\ }\textbf {\bibinfo {volume} {1603}},\ \bibinfo {pages} {018}
  (\bibinfo {year} {2016})},\ \Eprint {http://arxiv.org/abs/1510.08063}
  {arXiv:1510.08063 [hep-ph]} \BibitemShut {NoStop}%
%%CITATION = ARXIV:1510.08063;%%
\bibitem [{\citenamefont {Bringmann}\ \emph
  {et~al.}(2017{\natexlab{a}})\citenamefont {Bringmann}, \citenamefont
  {Kahlhoefer}, \citenamefont {Schmidt-Hoberg},\ and\ \citenamefont
  {Walia}}]{Bringmann:2016din}%
  \BibitemOpen
  \bibfield  {author} {\bibinfo {author} {\bibfnamefont {T.}~\bibnamefont
  {Bringmann}}, \bibinfo {author} {\bibfnamefont {F.}~\bibnamefont
  {Kahlhoefer}}, \bibinfo {author} {\bibfnamefont {K.}~\bibnamefont
  {Schmidt-Hoberg}}, \ and\ \bibinfo {author} {\bibfnamefont {P.}~\bibnamefont
  {Walia}},\ }\href {\doibase 10.1103/PhysRevLett.118.141802} {\bibfield
  {journal} {\bibinfo  {journal} {Phys. Rev. Lett.}\ }\textbf {\bibinfo
  {volume} {118}},\ \bibinfo {pages} {141802} (\bibinfo {year}
  {2017}{\natexlab{a}})},\ \Eprint {http://arxiv.org/abs/1612.00845}
  {arXiv:1612.00845 [hep-ph]} \BibitemShut {NoStop}%
%%CITATION = ARXIV:1612.00845;%%
\bibitem [{\citenamefont {Duerr}\ \emph {et~al.}(2018)\citenamefont {Duerr},
  \citenamefont {Schmidt-Hoberg},\ and\ \citenamefont {Wild}}]{Duerr:2018mbd}%
  \BibitemOpen
  \bibfield  {author} {\bibinfo {author} {\bibfnamefont {M.}~\bibnamefont
  {Duerr}}, \bibinfo {author} {\bibfnamefont {K.}~\bibnamefont
  {Schmidt-Hoberg}}, \ and\ \bibinfo {author} {\bibfnamefont {S.}~\bibnamefont
  {Wild}},\ }\href {\doibase 10.1088/1475-7516/2018/09/033} {\bibfield
  {journal} {\bibinfo  {journal} {JCAP}\ }\textbf {\bibinfo {volume} {1809}},\
  \bibinfo {pages} {033} (\bibinfo {year} {2018})},\ \Eprint
  {http://arxiv.org/abs/1804.10385} {arXiv:1804.10385 [hep-ph]} \BibitemShut
  {NoStop}%
%%CITATION = ARXIV:1804.10385;%%
\bibitem [{\citenamefont {McDonald}(2002)}]{McDonald:2001vt}%
  \BibitemOpen
  \bibfield  {author} {\bibinfo {author} {\bibfnamefont {J.}~\bibnamefont
  {McDonald}},\ }\href {\doibase 10.1103/PhysRevLett.88.091304} {\bibfield
  {journal} {\bibinfo  {journal} {Phys. Rev. Lett.}\ }\textbf {\bibinfo
  {volume} {88}},\ \bibinfo {pages} {091304} (\bibinfo {year} {2002})},\
  \Eprint {http://arxiv.org/abs/hep-ph/0106249} {arXiv:hep-ph/0106249 [hep-ph]}
  \BibitemShut {NoStop}%
%%CITATION = HEP-PH/0106249;%%
\bibitem [{\citenamefont {Hall}\ \emph {et~al.}(2010)\citenamefont {Hall},
  \citenamefont {Jedamzik}, \citenamefont {March-Russell},\ and\ \citenamefont
  {West}}]{Hall:2009bx}%
  \BibitemOpen
  \bibfield  {author} {\bibinfo {author} {\bibfnamefont {L.~J.}\ \bibnamefont
  {Hall}}, \bibinfo {author} {\bibfnamefont {K.}~\bibnamefont {Jedamzik}},
  \bibinfo {author} {\bibfnamefont {J.}~\bibnamefont {March-Russell}}, \ and\
  \bibinfo {author} {\bibfnamefont {S.~M.}\ \bibnamefont {West}},\ }\href
  {\doibase 10.1007/JHEP03(2010)080} {\bibfield  {journal} {\bibinfo  {journal}
  {JHEP}\ }\textbf {\bibinfo {volume} {03}},\ \bibinfo {pages} {080} (\bibinfo
  {year} {2010})},\ \Eprint {http://arxiv.org/abs/0911.1120} {arXiv:0911.1120
  [hep-ph]} \BibitemShut {NoStop}%
%%CITATION = ARXIV:0911.1120;%%
\bibitem [{\citenamefont {Bernal}\ \emph {et~al.}(2017)\citenamefont {Bernal},
  \citenamefont {Heikinheimo}, \citenamefont {Tenkanen}, \citenamefont
  {Tuominen},\ and\ \citenamefont {Vaskonen}}]{Bernal:2017kxu}%
  \BibitemOpen
  \bibfield  {author} {\bibinfo {author} {\bibfnamefont {N.}~\bibnamefont
  {Bernal}}, \bibinfo {author} {\bibfnamefont {M.}~\bibnamefont {Heikinheimo}},
  \bibinfo {author} {\bibfnamefont {T.}~\bibnamefont {Tenkanen}}, \bibinfo
  {author} {\bibfnamefont {K.}~\bibnamefont {Tuominen}}, \ and\ \bibinfo
  {author} {\bibfnamefont {V.}~\bibnamefont {Vaskonen}},\ }\href {\doibase
  10.1142/S0217751X1730023X} {\bibfield  {journal} {\bibinfo  {journal} {Int.
  J. Mod. Phys.}\ }\textbf {\bibinfo {volume} {A32}},\ \bibinfo {pages}
  {1730023} (\bibinfo {year} {2017})},\ \Eprint
  {http://arxiv.org/abs/1706.07442} {arXiv:1706.07442 [hep-ph]} \BibitemShut
  {NoStop}%
%%CITATION = ARXIV:1706.07442;%%
\bibitem [{\citenamefont {de~Salas}\ \emph {et~al.}(2015)\citenamefont
  {de~Salas}, \citenamefont {Lattanzi}, \citenamefont {Mangano}, \citenamefont
  {Miele}, \citenamefont {Pastor},\ and\ \citenamefont
  {Pisanti}}]{deSalas:2015glj}%
  \BibitemOpen
  \bibfield  {author} {\bibinfo {author} {\bibfnamefont {P.~F.}\ \bibnamefont
  {de~Salas}}, \bibinfo {author} {\bibfnamefont {M.}~\bibnamefont {Lattanzi}},
  \bibinfo {author} {\bibfnamefont {G.}~\bibnamefont {Mangano}}, \bibinfo
  {author} {\bibfnamefont {G.}~\bibnamefont {Miele}}, \bibinfo {author}
  {\bibfnamefont {S.}~\bibnamefont {Pastor}}, \ and\ \bibinfo {author}
  {\bibfnamefont {O.}~\bibnamefont {Pisanti}},\ }\href {\doibase
  10.1103/PhysRevD.92.123534} {\bibfield  {journal} {\bibinfo  {journal} {Phys.
  Rev.}\ }\textbf {\bibinfo {volume} {D92}},\ \bibinfo {pages} {123534}
  (\bibinfo {year} {2015})},\ \Eprint {http://arxiv.org/abs/1511.00672}
  {arXiv:1511.00672 [astro-ph.CO]} \BibitemShut {NoStop}%
%%CITATION = ARXIV:1511.00672;%%
\bibitem [{\citenamefont {Barrow}(1982)}]{Barrow:1982ei}%
  \BibitemOpen
  \bibfield  {author} {\bibinfo {author} {\bibfnamefont {J.~D.}\ \bibnamefont
  {Barrow}},\ }\href {\doibase 10.1016/0550-3213(82)90233-4} {\bibfield
  {journal} {\bibinfo  {journal} {Nucl. Phys.}\ }\textbf {\bibinfo {volume}
  {B208}},\ \bibinfo {pages} {501} (\bibinfo {year} {1982})}\BibitemShut
  {NoStop}%
%%CITATION = NUPHA,B208,501;%%
\bibitem [{\citenamefont {Davoudiasl}\ \emph {et~al.}(2016)\citenamefont
  {Davoudiasl}, \citenamefont {Hooper},\ and\ \citenamefont
  {McDermott}}]{Davoudiasl:2015vba}%
  \BibitemOpen
  \bibfield  {author} {\bibinfo {author} {\bibfnamefont {H.}~\bibnamefont
  {Davoudiasl}}, \bibinfo {author} {\bibfnamefont {D.}~\bibnamefont {Hooper}},
  \ and\ \bibinfo {author} {\bibfnamefont {S.~D.}\ \bibnamefont {McDermott}},\
  }\href {\doibase 10.1103/PhysRevLett.116.031303} {\bibfield  {journal}
  {\bibinfo  {journal} {Phys. Rev. Lett.}\ }\textbf {\bibinfo {volume} {116}},\
  \bibinfo {pages} {031303} (\bibinfo {year} {2016})},\ \Eprint
  {http://arxiv.org/abs/1507.08660} {arXiv:1507.08660 [hep-ph]} \BibitemShut
  {NoStop}%
%%CITATION = ARXIV:1507.08660;%%
\bibitem [{\citenamefont {Randall}\ \emph {et~al.}(2016)\citenamefont
  {Randall}, \citenamefont {Scholtz},\ and\ \citenamefont
  {Unwin}}]{Randall:2015xza}%
  \BibitemOpen
  \bibfield  {author} {\bibinfo {author} {\bibfnamefont {L.}~\bibnamefont
  {Randall}}, \bibinfo {author} {\bibfnamefont {J.}~\bibnamefont {Scholtz}}, \
  and\ \bibinfo {author} {\bibfnamefont {J.}~\bibnamefont {Unwin}},\ }\href
  {\doibase 10.1007/JHEP03(2016)011} {\bibfield  {journal} {\bibinfo  {journal}
  {JHEP}\ }\textbf {\bibinfo {volume} {03}},\ \bibinfo {pages} {011} (\bibinfo
  {year} {2016})},\ \Eprint {http://arxiv.org/abs/1509.08477} {arXiv:1509.08477
  [hep-ph]} \BibitemShut {NoStop}%
%%CITATION = ARXIV:1509.08477;%%
\bibitem [{\citenamefont {Tenkanen}\ and\ \citenamefont
  {Vaskonen}(2016)}]{Tenkanen:2016jic}%
  \BibitemOpen
  \bibfield  {author} {\bibinfo {author} {\bibfnamefont {T.}~\bibnamefont
  {Tenkanen}}\ and\ \bibinfo {author} {\bibfnamefont {V.}~\bibnamefont
  {Vaskonen}},\ }\href {\doibase 10.1103/PhysRevD.94.083516} {\bibfield
  {journal} {\bibinfo  {journal} {Phys. Rev.}\ }\textbf {\bibinfo {volume}
  {D94}},\ \bibinfo {pages} {083516} (\bibinfo {year} {2016})},\ \Eprint
  {http://arxiv.org/abs/1606.00192} {arXiv:1606.00192 [astro-ph.CO]}
  \BibitemShut {NoStop}%
%%CITATION = ARXIV:1606.00192;%%
\bibitem [{\citenamefont {Dror}\ \emph {et~al.}(2016)\citenamefont {Dror},
  \citenamefont {Kuflik},\ and\ \citenamefont {Ng}}]{Dror:2016rxc}%
  \BibitemOpen
  \bibfield  {author} {\bibinfo {author} {\bibfnamefont {J.~A.}\ \bibnamefont
  {Dror}}, \bibinfo {author} {\bibfnamefont {E.}~\bibnamefont {Kuflik}}, \ and\
  \bibinfo {author} {\bibfnamefont {W.~H.}\ \bibnamefont {Ng}},\ }\href
  {\doibase 10.1103/PhysRevLett.117.211801} {\bibfield  {journal} {\bibinfo
  {journal} {Phys. Rev. Lett.}\ }\textbf {\bibinfo {volume} {117}},\ \bibinfo
  {pages} {211801} (\bibinfo {year} {2016})},\ \Eprint
  {http://arxiv.org/abs/1607.03110} {arXiv:1607.03110 [hep-ph]} \BibitemShut
  {NoStop}%
%%CITATION = ARXIV:1607.03110;%%
\bibitem [{\citenamefont {D'Eramo}\ \emph {et~al.}(2017)\citenamefont
  {D'Eramo}, \citenamefont {Fernandez},\ and\ \citenamefont
  {Profumo}}]{DEramo:2017gpl}%
  \BibitemOpen
  \bibfield  {author} {\bibinfo {author} {\bibfnamefont {F.}~\bibnamefont
  {D'Eramo}}, \bibinfo {author} {\bibfnamefont {N.}~\bibnamefont {Fernandez}},
  \ and\ \bibinfo {author} {\bibfnamefont {S.}~\bibnamefont {Profumo}},\ }\href
  {\doibase 10.1088/1475-7516/2017/05/012} {\bibfield  {journal} {\bibinfo
  {journal} {JCAP}\ }\textbf {\bibinfo {volume} {1705}},\ \bibinfo {pages}
  {012} (\bibinfo {year} {2017})},\ \Eprint {http://arxiv.org/abs/1703.04793}
  {arXiv:1703.04793 [hep-ph]} \BibitemShut {NoStop}%
%%CITATION = ARXIV:1703.04793;%%
\bibitem [{\citenamefont {Hamdan}\ and\ \citenamefont
  {Unwin}(2018)}]{Hamdan:2017psw}%
  \BibitemOpen
  \bibfield  {author} {\bibinfo {author} {\bibfnamefont {S.}~\bibnamefont
  {Hamdan}}\ and\ \bibinfo {author} {\bibfnamefont {J.}~\bibnamefont {Unwin}},\
  }\href {\doibase 10.1142/S021773231850181X} {\bibfield  {journal} {\bibinfo
  {journal} {Mod. Phys. Lett.}\ }\textbf {\bibinfo {volume} {A33}},\ \bibinfo
  {pages} {1850181} (\bibinfo {year} {2018})},\ \Eprint
  {http://arxiv.org/abs/1710.03758} {arXiv:1710.03758 [hep-ph]} \BibitemShut
  {NoStop}%
%%CITATION = ARXIV:1710.03758;%%
\bibitem [{\citenamefont {Visinelli}(2018)}]{Visinelli:2017qga}%
  \BibitemOpen
  \bibfield  {author} {\bibinfo {author} {\bibfnamefont {L.}~\bibnamefont
  {Visinelli}},\ }\href {\doibase 10.3390/sym10110546} {\bibfield  {journal}
  {\bibinfo  {journal} {Symmetry}\ }\textbf {\bibinfo {volume} {10}},\ \bibinfo
  {pages} {546} (\bibinfo {year} {2018})},\ \Eprint
  {http://arxiv.org/abs/1710.11006} {arXiv:1710.11006 [astro-ph.CO]}
  \BibitemShut {NoStop}%
%%CITATION = ARXIV:1710.11006;%%
\bibitem [{\citenamefont {Drees}\ and\ \citenamefont
  {Hajkarim}(2018{\natexlab{a}})}]{Drees:2017iod}%
  \BibitemOpen
  \bibfield  {author} {\bibinfo {author} {\bibfnamefont {M.}~\bibnamefont
  {Drees}}\ and\ \bibinfo {author} {\bibfnamefont {F.}~\bibnamefont
  {Hajkarim}},\ }\href {\doibase 10.1088/1475-7516/2018/02/057} {\bibfield
  {journal} {\bibinfo  {journal} {JCAP}\ }\textbf {\bibinfo {volume} {1802}},\
  \bibinfo {pages} {057} (\bibinfo {year} {2018}{\natexlab{a}})},\ \Eprint
  {http://arxiv.org/abs/1711.05007} {arXiv:1711.05007 [hep-ph]} \BibitemShut
  {NoStop}%
%%CITATION = ARXIV:1711.05007;%%
\bibitem [{\citenamefont {Bramante}\ and\ \citenamefont
  {Unwin}(2017)}]{Bramante:2017obj}%
  \BibitemOpen
  \bibfield  {author} {\bibinfo {author} {\bibfnamefont {J.}~\bibnamefont
  {Bramante}}\ and\ \bibinfo {author} {\bibfnamefont {J.}~\bibnamefont
  {Unwin}},\ }\href {\doibase 10.1007/JHEP02(2017)119} {\bibfield  {journal}
  {\bibinfo  {journal} {JHEP}\ }\textbf {\bibinfo {volume} {02}},\ \bibinfo
  {pages} {119} (\bibinfo {year} {2017})},\ \Eprint
  {http://arxiv.org/abs/1701.05859} {arXiv:1701.05859 [hep-ph]} \BibitemShut
  {NoStop}%
%%CITATION = ARXIV:1701.05859;%%
\bibitem [{\citenamefont {Bernal}\ \emph
  {et~al.}(2019{\natexlab{a}})\citenamefont {Bernal}, \citenamefont {Cosme},\
  and\ \citenamefont {Tenkanen}}]{Bernal:2018ins}%
  \BibitemOpen
  \bibfield  {author} {\bibinfo {author} {\bibfnamefont {N.}~\bibnamefont
  {Bernal}}, \bibinfo {author} {\bibfnamefont {C.}~\bibnamefont {Cosme}}, \
  and\ \bibinfo {author} {\bibfnamefont {T.}~\bibnamefont {Tenkanen}},\ }\href
  {\doibase 10.1140/epjc/s10052-019-6608-8} {\bibfield  {journal} {\bibinfo
  {journal} {Eur. Phys. J.}\ }\textbf {\bibinfo {volume} {C79}},\ \bibinfo
  {pages} {99} (\bibinfo {year} {2019}{\natexlab{a}})},\ \Eprint
  {http://arxiv.org/abs/1803.08064} {arXiv:1803.08064 [hep-ph]} \BibitemShut
  {NoStop}%
%%CITATION = ARXIV:1803.08064;%%
\bibitem [{\citenamefont {Di~Marco}\ \emph {et~al.}(2018)\citenamefont
  {Di~Marco}, \citenamefont {Pradisi},\ and\ \citenamefont
  {Cabella}}]{DiMarco:2018bnw}%
  \BibitemOpen
  \bibfield  {author} {\bibinfo {author} {\bibfnamefont {A.}~\bibnamefont
  {Di~Marco}}, \bibinfo {author} {\bibfnamefont {G.}~\bibnamefont {Pradisi}}, \
  and\ \bibinfo {author} {\bibfnamefont {P.}~\bibnamefont {Cabella}},\ }\href
  {\doibase 10.1103/PhysRevD.98.123511} {\bibfield  {journal} {\bibinfo
  {journal} {Phys. Rev.}\ }\textbf {\bibinfo {volume} {D98}},\ \bibinfo {pages}
  {123511} (\bibinfo {year} {2018})},\ \Eprint
  {http://arxiv.org/abs/1807.05916} {arXiv:1807.05916 [astro-ph.CO]}
  \BibitemShut {NoStop}%
%%CITATION = ARXIV:1807.05916;%%
\bibitem [{\citenamefont {D'Eramo}\ \emph {et~al.}(2018)\citenamefont
  {D'Eramo}, \citenamefont {Fernandez},\ and\ \citenamefont
  {Profumo}}]{DEramo:2017ecx}%
  \BibitemOpen
  \bibfield  {author} {\bibinfo {author} {\bibfnamefont {F.}~\bibnamefont
  {D'Eramo}}, \bibinfo {author} {\bibfnamefont {N.}~\bibnamefont {Fernandez}},
  \ and\ \bibinfo {author} {\bibfnamefont {S.}~\bibnamefont {Profumo}},\ }\href
  {\doibase 10.1088/1475-7516/2018/02/046} {\bibfield  {journal} {\bibinfo
  {journal} {JCAP}\ }\textbf {\bibinfo {volume} {1802}},\ \bibinfo {pages}
  {046} (\bibinfo {year} {2018})},\ \Eprint {http://arxiv.org/abs/1712.07453}
  {arXiv:1712.07453 [hep-ph]} \BibitemShut {NoStop}%
%%CITATION = ARXIV:1712.07453;%%
\bibitem [{\citenamefont {Maity}\ and\ \citenamefont
  {Saha}(2018)}]{Maity:2018dgy}%
  \BibitemOpen
  \bibfield  {author} {\bibinfo {author} {\bibfnamefont {D.}~\bibnamefont
  {Maity}}\ and\ \bibinfo {author} {\bibfnamefont {P.}~\bibnamefont {Saha}},\
  }\href {\doibase 10.1103/PhysRevD.98.103525} {\bibfield  {journal} {\bibinfo
  {journal} {Phys. Rev.}\ }\textbf {\bibinfo {volume} {D98}},\ \bibinfo {pages}
  {103525} (\bibinfo {year} {2018})},\ \Eprint
  {http://arxiv.org/abs/1801.03059} {arXiv:1801.03059 [hep-ph]} \BibitemShut
  {NoStop}%
%%CITATION = ARXIV:1801.03059;%%
\bibitem [{\citenamefont {Garcia}\ and\ \citenamefont
  {Amin}(2018)}]{Garcia:2018wtq}%
  \BibitemOpen
  \bibfield  {author} {\bibinfo {author} {\bibfnamefont {M.~A.~G.}\
  \bibnamefont {Garcia}}\ and\ \bibinfo {author} {\bibfnamefont {M.~A.}\
  \bibnamefont {Amin}},\ }\href {\doibase 10.1103/PhysRevD.98.103504}
  {\bibfield  {journal} {\bibinfo  {journal} {Phys. Rev.}\ }\textbf {\bibinfo
  {volume} {D98}},\ \bibinfo {pages} {103504} (\bibinfo {year} {2018})},\
  \Eprint {http://arxiv.org/abs/1806.01865} {arXiv:1806.01865 [hep-ph]}
  \BibitemShut {NoStop}%
%%CITATION = ARXIV:1806.01865;%%
\bibitem [{\citenamefont {Bernal}\ \emph
  {et~al.}(2019{\natexlab{b}})\citenamefont {Bernal}, \citenamefont {Cosme},
  \citenamefont {Tenkanen},\ and\ \citenamefont {Vaskonen}}]{Bernal:2018kcw}%
  \BibitemOpen
  \bibfield  {author} {\bibinfo {author} {\bibfnamefont {N.}~\bibnamefont
  {Bernal}}, \bibinfo {author} {\bibfnamefont {C.}~\bibnamefont {Cosme}},
  \bibinfo {author} {\bibfnamefont {T.}~\bibnamefont {Tenkanen}}, \ and\
  \bibinfo {author} {\bibfnamefont {V.}~\bibnamefont {Vaskonen}},\ }\href
  {\doibase 10.1140/epjc/s10052-019-6550-9} {\bibfield  {journal} {\bibinfo
  {journal} {Eur. Phys. J.}\ }\textbf {\bibinfo {volume} {C79}},\ \bibinfo
  {pages} {30} (\bibinfo {year} {2019}{\natexlab{b}})},\ \Eprint
  {http://arxiv.org/abs/1806.11122} {arXiv:1806.11122 [hep-ph]} \BibitemShut
  {NoStop}%
%%CITATION = ARXIV:1806.11122;%%
\bibitem [{\citenamefont {Arbey}\ \emph {et~al.}(2018)\citenamefont {Arbey},
  \citenamefont {Ellis}, \citenamefont {Mahmoudi},\ and\ \citenamefont
  {Robbins}}]{Arbey:2018uho}%
  \BibitemOpen
  \bibfield  {author} {\bibinfo {author} {\bibfnamefont {A.}~\bibnamefont
  {Arbey}}, \bibinfo {author} {\bibfnamefont {J.}~\bibnamefont {Ellis}},
  \bibinfo {author} {\bibfnamefont {F.}~\bibnamefont {Mahmoudi}}, \ and\
  \bibinfo {author} {\bibfnamefont {G.}~\bibnamefont {Robbins}},\ }\href
  {\doibase 10.1007/JHEP10(2018)132} {\bibfield  {journal} {\bibinfo  {journal}
  {JHEP}\ }\textbf {\bibinfo {volume} {10}},\ \bibinfo {pages} {132} (\bibinfo
  {year} {2018})},\ \Eprint {http://arxiv.org/abs/1807.00554} {arXiv:1807.00554
  [hep-ph]} \BibitemShut {NoStop}%
%%CITATION = ARXIV:1807.00554;%%
\bibitem [{\citenamefont {Drees}\ and\ \citenamefont
  {Hajkarim}(2018{\natexlab{b}})}]{Drees:2018dsj}%
  \BibitemOpen
  \bibfield  {author} {\bibinfo {author} {\bibfnamefont {M.}~\bibnamefont
  {Drees}}\ and\ \bibinfo {author} {\bibfnamefont {F.}~\bibnamefont
  {Hajkarim}},\ }\href {\doibase 10.1007/JHEP12(2018)042} {\bibfield  {journal}
  {\bibinfo  {journal} {JHEP}\ }\textbf {\bibinfo {volume} {12}},\ \bibinfo
  {pages} {042} (\bibinfo {year} {2018}{\natexlab{b}})},\ \Eprint
  {http://arxiv.org/abs/1808.05706} {arXiv:1808.05706 [hep-ph]} \BibitemShut
  {NoStop}%
%%CITATION = ARXIV:1808.05706;%%
\bibitem [{\citenamefont {Betancur}\ and\ \citenamefont
  {Zapata}(2018)}]{Betancur:2018xtj}%
  \BibitemOpen
  \bibfield  {author} {\bibinfo {author} {\bibfnamefont {A.}~\bibnamefont
  {Betancur}}\ and\ \bibinfo {author} {\bibfnamefont {{\'O}.}~\bibnamefont
  {Zapata}},\ }\href {\doibase 10.1103/PhysRevD.98.095003} {\bibfield
  {journal} {\bibinfo  {journal} {Phys. Rev.}\ }\textbf {\bibinfo {volume}
  {D98}},\ \bibinfo {pages} {095003} (\bibinfo {year} {2018})},\ \Eprint
  {http://arxiv.org/abs/1809.04990} {arXiv:1809.04990 [hep-ph]} \BibitemShut
  {NoStop}%
%%CITATION = ARXIV:1809.04990;%%
\bibitem [{\citenamefont {Maldonado}\ and\ \citenamefont
  {Unwin}(2019)}]{Maldonado:2019qmp}%
  \BibitemOpen
  \bibfield  {author} {\bibinfo {author} {\bibfnamefont {C.}~\bibnamefont
  {Maldonado}}\ and\ \bibinfo {author} {\bibfnamefont {J.}~\bibnamefont
  {Unwin}},\ }\href {\doibase 10.1088/1475-7516/2019/06/037} {\bibfield
  {journal} {\bibinfo  {journal} {JCAP}\ }\textbf {\bibinfo {volume} {1906}},\
  \bibinfo {pages} {037} (\bibinfo {year} {2019})},\ \Eprint
  {http://arxiv.org/abs/1902.10746} {arXiv:1902.10746 [hep-ph]} \BibitemShut
  {NoStop}%
%%CITATION = ARXIV:1902.10746;%%
\bibitem [{\citenamefont {Poulin}(2019)}]{Poulin:2019omz}%
  \BibitemOpen
  \bibfield  {author} {\bibinfo {author} {\bibfnamefont {A.}~\bibnamefont
  {Poulin}},\ }\href {\doibase 10.1103/PhysRevD.100.043022} {\bibfield
  {journal} {\bibinfo  {journal} {Phys. Rev.}\ }\textbf {\bibinfo {volume}
  {D100}},\ \bibinfo {pages} {043022} (\bibinfo {year} {2019})},\ \Eprint
  {http://arxiv.org/abs/1905.03126} {arXiv:1905.03126 [hep-ph]} \BibitemShut
  {NoStop}%
%%CITATION = ARXIV:1905.03126;%%
\bibitem [{\citenamefont {Tenkanen}(2019)}]{Tenkanen:2019cik}%
  \BibitemOpen
  \bibfield  {author} {\bibinfo {author} {\bibfnamefont {T.}~\bibnamefont
  {Tenkanen}},\ }\href {\doibase 10.1103/PhysRevD.100.083515} {\bibfield
  {journal} {\bibinfo  {journal} {Phys. Rev.}\ }\textbf {\bibinfo {volume}
  {D100}},\ \bibinfo {pages} {083515} (\bibinfo {year} {2019})},\ \Eprint
  {http://arxiv.org/abs/1905.11737} {arXiv:1905.11737 [astro-ph.CO]}
  \BibitemShut {NoStop}%
%%CITATION = ARXIV:1905.11737;%%
\bibitem [{\citenamefont {Arias}\ \emph {et~al.}(2019)\citenamefont {Arias},
  \citenamefont {Bernal}, \citenamefont {Herrera},\ and\ \citenamefont
  {Maldonado}}]{Arias:2019uol}%
  \BibitemOpen
  \bibfield  {author} {\bibinfo {author} {\bibfnamefont {P.}~\bibnamefont
  {Arias}}, \bibinfo {author} {\bibfnamefont {N.}~\bibnamefont {Bernal}},
  \bibinfo {author} {\bibfnamefont {A.}~\bibnamefont {Herrera}}, \ and\
  \bibinfo {author} {\bibfnamefont {C.}~\bibnamefont {Maldonado}},\ }\href
  {\doibase 10.1088/1475-7516/2019/10/047} {\bibfield  {journal} {\bibinfo
  {journal} {JCAP}\ }\textbf {\bibinfo {volume} {1910}},\ \bibinfo {pages}
  {047} (\bibinfo {year} {2019})},\ \Eprint {http://arxiv.org/abs/1906.04183}
  {arXiv:1906.04183 [hep-ph]} \BibitemShut {NoStop}%
%%CITATION = ARXIV:1906.04183;%%
\bibitem [{\citenamefont {Chanda}\ \emph {et~al.}(2019)\citenamefont {Chanda},
  \citenamefont {Hamdan},\ and\ \citenamefont {Unwin}}]{Chanda:2019xyl}%
  \BibitemOpen
  \bibfield  {author} {\bibinfo {author} {\bibfnamefont {P.}~\bibnamefont
  {Chanda}}, \bibinfo {author} {\bibfnamefont {S.}~\bibnamefont {Hamdan}}, \
  and\ \bibinfo {author} {\bibfnamefont {J.}~\bibnamefont {Unwin}},\
  }\href@noop {} {\  (\bibinfo {year} {2019})},\ \Eprint
  {http://arxiv.org/abs/1911.02616} {arXiv:1911.02616 [hep-ph]} \BibitemShut
  {NoStop}%
%%CITATION = ARXIV:1911.02616;%%
\bibitem [{\citenamefont {Angloher}\ \emph {et~al.}(2016)\citenamefont
  {Angloher} \emph {et~al.}}]{Angloher:2015ewa}%
  \BibitemOpen
  \bibfield  {author} {\bibinfo {author} {\bibfnamefont {G.}~\bibnamefont
  {Angloher}} \emph {et~al.} (\bibinfo {collaboration} {CRESST}),\ }\href
  {\doibase 10.1140/epjc/s10052-016-3877-3} {\bibfield  {journal} {\bibinfo
  {journal} {Eur. Phys. J.}\ }\textbf {\bibinfo {volume} {C76}},\ \bibinfo
  {pages} {25} (\bibinfo {year} {2016})},\ \Eprint
  {http://arxiv.org/abs/1509.01515} {arXiv:1509.01515 [astro-ph.CO]}
  \BibitemShut {NoStop}%
%%CITATION = ARXIV:1509.01515;%%
\bibitem [{\citenamefont {Agnese}\ \emph {et~al.}(2016)\citenamefont {Agnese}
  \emph {et~al.}}]{Agnese:2015nto}%
  \BibitemOpen
  \bibfield  {author} {\bibinfo {author} {\bibfnamefont {R.}~\bibnamefont
  {Agnese}} \emph {et~al.} (\bibinfo {collaboration} {SuperCDMS}),\ }\href
  {\doibase 10.1103/PhysRevLett.116.071301} {\bibfield  {journal} {\bibinfo
  {journal} {Phys. Rev. Lett.}\ }\textbf {\bibinfo {volume} {116}},\ \bibinfo
  {pages} {071301} (\bibinfo {year} {2016})},\ \Eprint
  {http://arxiv.org/abs/1509.02448} {arXiv:1509.02448 [astro-ph.CO]}
  \BibitemShut {NoStop}%
%%CITATION = ARXIV:1509.02448;%%
\bibitem [{\citenamefont {Akerib}\ \emph {et~al.}(2017)\citenamefont {Akerib}
  \emph {et~al.}}]{Akerib:2016vxi}%
  \BibitemOpen
  \bibfield  {author} {\bibinfo {author} {\bibfnamefont {D.~S.}\ \bibnamefont
  {Akerib}} \emph {et~al.} (\bibinfo {collaboration} {LUX}),\ }\href {\doibase
  10.1103/PhysRevLett.118.021303} {\bibfield  {journal} {\bibinfo  {journal}
  {Phys. Rev. Lett.}\ }\textbf {\bibinfo {volume} {118}},\ \bibinfo {pages}
  {021303} (\bibinfo {year} {2017})},\ \Eprint
  {http://arxiv.org/abs/1608.07648} {arXiv:1608.07648 [astro-ph.CO]}
  \BibitemShut {NoStop}%
%%CITATION = ARXIV:1608.07648;%%
\bibitem [{\citenamefont {Aprile}\ \emph {et~al.}(2018)\citenamefont {Aprile}
  \emph {et~al.}}]{Aprile:2018dbl}%
  \BibitemOpen
  \bibfield  {author} {\bibinfo {author} {\bibfnamefont {E.}~\bibnamefont
  {Aprile}} \emph {et~al.} (\bibinfo {collaboration} {XENON}),\ }\href
  {\doibase 10.1103/PhysRevLett.121.111302} {\bibfield  {journal} {\bibinfo
  {journal} {Phys. Rev. Lett.}\ }\textbf {\bibinfo {volume} {121}},\ \bibinfo
  {pages} {111302} (\bibinfo {year} {2018})},\ \Eprint
  {http://arxiv.org/abs/1805.12562} {arXiv:1805.12562 [astro-ph.CO]}
  \BibitemShut {NoStop}%
%%CITATION = ARXIV:1805.12562;%%
\bibitem [{\citenamefont {Essig}\ \emph
  {et~al.}(2012{\natexlab{a}})\citenamefont {Essig}, \citenamefont
  {Manalaysay}, \citenamefont {Mardon}, \citenamefont {Sorensen},\ and\
  \citenamefont {Volansky}}]{Essig:2012yx}%
  \BibitemOpen
  \bibfield  {author} {\bibinfo {author} {\bibfnamefont {R.}~\bibnamefont
  {Essig}}, \bibinfo {author} {\bibfnamefont {A.}~\bibnamefont {Manalaysay}},
  \bibinfo {author} {\bibfnamefont {J.}~\bibnamefont {Mardon}}, \bibinfo
  {author} {\bibfnamefont {P.}~\bibnamefont {Sorensen}}, \ and\ \bibinfo
  {author} {\bibfnamefont {T.}~\bibnamefont {Volansky}},\ }\href {\doibase
  10.1103/PhysRevLett.109.021301} {\bibfield  {journal} {\bibinfo  {journal}
  {Phys. Rev. Lett.}\ }\textbf {\bibinfo {volume} {109}},\ \bibinfo {pages}
  {021301} (\bibinfo {year} {2012}{\natexlab{a}})},\ \Eprint
  {http://arxiv.org/abs/1206.2644} {arXiv:1206.2644 [astro-ph.CO]} \BibitemShut
  {NoStop}%
%%CITATION = ARXIV:1206.2644;%%
\bibitem [{\citenamefont {Essig}\ \emph {et~al.}(2017)\citenamefont {Essig},
  \citenamefont {Volansky},\ and\ \citenamefont {Yu}}]{Essig:2017kqs}%
  \BibitemOpen
  \bibfield  {author} {\bibinfo {author} {\bibfnamefont {R.}~\bibnamefont
  {Essig}}, \bibinfo {author} {\bibfnamefont {T.}~\bibnamefont {Volansky}}, \
  and\ \bibinfo {author} {\bibfnamefont {T.-T.}\ \bibnamefont {Yu}},\ }\href
  {\doibase 10.1103/PhysRevD.96.043017} {\bibfield  {journal} {\bibinfo
  {journal} {Phys. Rev.}\ }\textbf {\bibinfo {volume} {D96}},\ \bibinfo {pages}
  {043017} (\bibinfo {year} {2017})},\ \Eprint
  {http://arxiv.org/abs/1703.00910} {arXiv:1703.00910 [hep-ph]} \BibitemShut
  {NoStop}%
%%CITATION = ARXIV:1703.00910;%%
\bibitem [{\citenamefont {Abramoff}\ \emph {et~al.}(2019)\citenamefont
  {Abramoff} \emph {et~al.}}]{Abramoff:2019dfb}%
  \BibitemOpen
  \bibfield  {author} {\bibinfo {author} {\bibfnamefont {O.}~\bibnamefont
  {Abramoff}} \emph {et~al.} (\bibinfo {collaboration} {SENSEI}),\ }\href
  {\doibase 10.1103/PhysRevLett.122.161801} {\bibfield  {journal} {\bibinfo
  {journal} {Phys. Rev. Lett.}\ }\textbf {\bibinfo {volume} {122}},\ \bibinfo
  {pages} {161801} (\bibinfo {year} {2019})},\ \Eprint
  {http://arxiv.org/abs/1901.10478} {arXiv:1901.10478 [hep-ex]} \BibitemShut
  {NoStop}%
%%CITATION = ARXIV:1901.10478;%%
\bibitem [{\citenamefont {Aprile}\ \emph {et~al.}(2019)\citenamefont {Aprile}
  \emph {et~al.}}]{Aprile:2019xxb}%
  \BibitemOpen
  \bibfield  {author} {\bibinfo {author} {\bibfnamefont {E.}~\bibnamefont
  {Aprile}} \emph {et~al.} (\bibinfo {collaboration} {XENON}),\ }\href@noop {}
  {\  (\bibinfo {year} {2019})},\ \Eprint {http://arxiv.org/abs/1907.11485}
  {arXiv:1907.11485 [hep-ex]} \BibitemShut {NoStop}%
%%CITATION = ARXIV:1907.11485;%%
\bibitem [{\citenamefont {Bringmann}\ \emph
  {et~al.}(2017{\natexlab{b}})\citenamefont {Bringmann} \emph
  {et~al.}}]{Workgroup:2017lvb}%
  \BibitemOpen
  \bibfield  {author} {\bibinfo {author} {\bibfnamefont {T.}~\bibnamefont
  {Bringmann}} \emph {et~al.} (\bibinfo {collaboration} {The GAMBIT Dark Matter
  Workgroup}),\ }\href {\doibase 10.1140/epjc/s10052-017-5155-4} {\bibfield
  {journal} {\bibinfo  {journal} {Eur. Phys. J.}\ }\textbf {\bibinfo {volume}
  {C77}},\ \bibinfo {pages} {831} (\bibinfo {year} {2017}{\natexlab{b}})},\
  \Eprint {http://arxiv.org/abs/1705.07920} {arXiv:1705.07920 [hep-ph]}
  \BibitemShut {NoStop}%
%%CITATION = ARXIV:1705.07920;%%
\bibitem [{\citenamefont {Athron}\ \emph {et~al.}(2019)\citenamefont {Athron}
  \emph {et~al.}}]{Athron:2018hpc}%
  \BibitemOpen
  \bibfield  {author} {\bibinfo {author} {\bibfnamefont {P.}~\bibnamefont
  {Athron}} \emph {et~al.} (\bibinfo {collaboration} {GAMBIT}),\ }\href
  {\doibase 10.1140/epjc/s10052-018-6513-6} {\bibfield  {journal} {\bibinfo
  {journal} {Eur. Phys. J.}\ }\textbf {\bibinfo {volume} {C79}},\ \bibinfo
  {pages} {38} (\bibinfo {year} {2019})},\ \Eprint
  {http://arxiv.org/abs/1808.10465} {arXiv:1808.10465 [hep-ph]} \BibitemShut
  {NoStop}%
%%CITATION = ARXIV:1808.10465;%%
\bibitem [{\citenamefont {Emken}\ \emph {et~al.}(2019)\citenamefont {Emken},
  \citenamefont {Essig}, \citenamefont {Kouvaris},\ and\ \citenamefont
  {Sholapurkar}}]{Emken:2019tni}%
  \BibitemOpen
  \bibfield  {author} {\bibinfo {author} {\bibfnamefont {T.}~\bibnamefont
  {Emken}}, \bibinfo {author} {\bibfnamefont {R.}~\bibnamefont {Essig}},
  \bibinfo {author} {\bibfnamefont {C.}~\bibnamefont {Kouvaris}}, \ and\
  \bibinfo {author} {\bibfnamefont {M.}~\bibnamefont {Sholapurkar}},\ }\href
  {\doibase 10.1088/1475-7516/2019/09/070} {\bibfield  {journal} {\bibinfo
  {journal} {JCAP}\ }\textbf {\bibinfo {volume} {1909}},\ \bibinfo {pages}
  {070} (\bibinfo {year} {2019})},\ \Eprint {http://arxiv.org/abs/1905.06348}
  {arXiv:1905.06348 [hep-ph]} \BibitemShut {NoStop}%
%%CITATION = ARXIV:1905.06348;%%
\bibitem [{\citenamefont {Essig}\ \emph
  {et~al.}(2012{\natexlab{b}})\citenamefont {Essig}, \citenamefont {Mardon},\
  and\ \citenamefont {Volansky}}]{Essig:2011nj}%
  \BibitemOpen
  \bibfield  {author} {\bibinfo {author} {\bibfnamefont {R.}~\bibnamefont
  {Essig}}, \bibinfo {author} {\bibfnamefont {J.}~\bibnamefont {Mardon}}, \
  and\ \bibinfo {author} {\bibfnamefont {T.}~\bibnamefont {Volansky}},\ }\href
  {\doibase 10.1103/PhysRevD.85.076007} {\bibfield  {journal} {\bibinfo
  {journal} {Phys. Rev.}\ }\textbf {\bibinfo {volume} {D85}},\ \bibinfo {pages}
  {076007} (\bibinfo {year} {2012}{\natexlab{b}})},\ \Eprint
  {http://arxiv.org/abs/1108.5383} {arXiv:1108.5383 [hep-ph]} \BibitemShut
  {NoStop}%
%%CITATION = ARXIV:1108.5383;%%
\bibitem [{\citenamefont {Cappiello}\ \emph {et~al.}(2019)\citenamefont
  {Cappiello}, \citenamefont {Ng},\ and\ \citenamefont
  {Beacom}}]{Cappiello:2018hsu}%
  \BibitemOpen
  \bibfield  {author} {\bibinfo {author} {\bibfnamefont {C.~V.}\ \bibnamefont
  {Cappiello}}, \bibinfo {author} {\bibfnamefont {K.~C.~Y.}\ \bibnamefont
  {Ng}}, \ and\ \bibinfo {author} {\bibfnamefont {J.~F.}\ \bibnamefont
  {Beacom}},\ }\href {\doibase 10.1103/PhysRevD.99.063004} {\bibfield
  {journal} {\bibinfo  {journal} {Phys. Rev.}\ }\textbf {\bibinfo {volume}
  {D99}},\ \bibinfo {pages} {063004} (\bibinfo {year} {2019})},\ \Eprint
  {http://arxiv.org/abs/1810.07705} {arXiv:1810.07705 [hep-ph]} \BibitemShut
  {NoStop}%
%%CITATION = ARXIV:1810.07705;%%
\bibitem [{\citenamefont {B{\oe}hm}\ \emph {et~al.}(2001)\citenamefont
  {B{\oe}hm}, \citenamefont {Fayet},\ and\ \citenamefont
  {Schaeffer}}]{Boehm:2000gq}%
  \BibitemOpen
  \bibfield  {author} {\bibinfo {author} {\bibfnamefont {C.}~\bibnamefont
  {B{\oe}hm}}, \bibinfo {author} {\bibfnamefont {P.}~\bibnamefont {Fayet}}, \
  and\ \bibinfo {author} {\bibfnamefont {R.}~\bibnamefont {Schaeffer}},\ }\href
  {\doibase 10.1016/S0370-2693(01)01060-7} {\bibfield  {journal} {\bibinfo
  {journal} {Phys. Lett.}\ }\textbf {\bibinfo {volume} {B518}},\ \bibinfo
  {pages} {8} (\bibinfo {year} {2001})},\ \Eprint
  {http://arxiv.org/abs/astro-ph/0012504} {arXiv:astro-ph/0012504 [astro-ph]}
  \BibitemShut {NoStop}%
%%CITATION = ASTRO-PH/0012504;%%
\bibitem [{\citenamefont {B{\oe}hm}\ and\ \citenamefont
  {Schaeffer}(2005)}]{Boehm:2004th}%
  \BibitemOpen
  \bibfield  {author} {\bibinfo {author} {\bibfnamefont {C.}~\bibnamefont
  {B{\oe}hm}}\ and\ \bibinfo {author} {\bibfnamefont {R.}~\bibnamefont
  {Schaeffer}},\ }\href {\doibase 10.1051/0004-6361:20042238} {\bibfield
  {journal} {\bibinfo  {journal} {Astron. Astrophys.}\ }\textbf {\bibinfo
  {volume} {438}},\ \bibinfo {pages} {419} (\bibinfo {year} {2005})},\ \Eprint
  {http://arxiv.org/abs/astro-ph/0410591} {arXiv:astro-ph/0410591 [astro-ph]}
  \BibitemShut {NoStop}%
%%CITATION = ASTRO-PH/0410591;%%
\bibitem [{\citenamefont {Dvorkin}\ \emph {et~al.}(2014)\citenamefont
  {Dvorkin}, \citenamefont {Blum},\ and\ \citenamefont
  {Kamionkowski}}]{Dvorkin:2013cea}%
  \BibitemOpen
  \bibfield  {author} {\bibinfo {author} {\bibfnamefont {C.}~\bibnamefont
  {Dvorkin}}, \bibinfo {author} {\bibfnamefont {K.}~\bibnamefont {Blum}}, \
  and\ \bibinfo {author} {\bibfnamefont {M.}~\bibnamefont {Kamionkowski}},\
  }\href {\doibase 10.1103/PhysRevD.89.023519} {\bibfield  {journal} {\bibinfo
  {journal} {Phys. Rev.}\ }\textbf {\bibinfo {volume} {D89}},\ \bibinfo {pages}
  {023519} (\bibinfo {year} {2014})},\ \Eprint {http://arxiv.org/abs/1311.2937}
  {arXiv:1311.2937 [astro-ph.CO]} \BibitemShut {NoStop}%
%%CITATION = ARXIV:1311.2937;%%
\bibitem [{\citenamefont {Xu}\ \emph {et~al.}(2018)\citenamefont {Xu},
  \citenamefont {Dvorkin},\ and\ \citenamefont {Chael}}]{Xu:2018efh}%
  \BibitemOpen
  \bibfield  {author} {\bibinfo {author} {\bibfnamefont {W.~L.}\ \bibnamefont
  {Xu}}, \bibinfo {author} {\bibfnamefont {C.}~\bibnamefont {Dvorkin}}, \ and\
  \bibinfo {author} {\bibfnamefont {A.}~\bibnamefont {Chael}},\ }\href
  {\doibase 10.1103/PhysRevD.97.103530} {\bibfield  {journal} {\bibinfo
  {journal} {Phys. Rev.}\ }\textbf {\bibinfo {volume} {D97}},\ \bibinfo {pages}
  {103530} (\bibinfo {year} {2018})},\ \Eprint
  {http://arxiv.org/abs/1802.06788} {arXiv:1802.06788 [astro-ph.CO]}
  \BibitemShut {NoStop}%
%%CITATION = ARXIV:1802.06788;%%
\bibitem [{\citenamefont {Boddy}\ \emph {et~al.}(2018)\citenamefont {Boddy},
  \citenamefont {Gluscevic}, \citenamefont {Poulin}, \citenamefont {Kovetz},
  \citenamefont {Kamionkowski},\ and\ \citenamefont {Barkana}}]{Boddy:2018wzy}%
  \BibitemOpen
  \bibfield  {author} {\bibinfo {author} {\bibfnamefont {K.~K.}\ \bibnamefont
  {Boddy}}, \bibinfo {author} {\bibfnamefont {V.}~\bibnamefont {Gluscevic}},
  \bibinfo {author} {\bibfnamefont {V.}~\bibnamefont {Poulin}}, \bibinfo
  {author} {\bibfnamefont {E.~D.}\ \bibnamefont {Kovetz}}, \bibinfo {author}
  {\bibfnamefont {M.}~\bibnamefont {Kamionkowski}}, \ and\ \bibinfo {author}
  {\bibfnamefont {R.}~\bibnamefont {Barkana}},\ }\href {\doibase
  10.1103/PhysRevD.98.123506} {\bibfield  {journal} {\bibinfo  {journal} {Phys.
  Rev.}\ }\textbf {\bibinfo {volume} {D98}},\ \bibinfo {pages} {123506}
  (\bibinfo {year} {2018})},\ \Eprint {http://arxiv.org/abs/1808.00001}
  {arXiv:1808.00001 [astro-ph.CO]} \BibitemShut {NoStop}%
%%CITATION = ARXIV:1808.00001;%%
\bibitem [{\citenamefont {Nadler}\ \emph {et~al.}(2019)\citenamefont {Nadler},
  \citenamefont {Gluscevic}, \citenamefont {Boddy},\ and\ \citenamefont
  {Wechsler}}]{Nadler:2019zrb}%
  \BibitemOpen
  \bibfield  {author} {\bibinfo {author} {\bibfnamefont {E.~O.}\ \bibnamefont
  {Nadler}}, \bibinfo {author} {\bibfnamefont {V.}~\bibnamefont {Gluscevic}},
  \bibinfo {author} {\bibfnamefont {K.~K.}\ \bibnamefont {Boddy}}, \ and\
  \bibinfo {author} {\bibfnamefont {R.~H.}\ \bibnamefont {Wechsler}},\ }\href
  {\doibase 10.3847/2041-8213/ab1eb2} {\bibfield  {journal} {\bibinfo
  {journal} {Astrophys. J.}\ }\textbf {\bibinfo {volume} {878}},\ \bibinfo
  {pages} {L32} (\bibinfo {year} {2019})},\ \bibinfo {note} {[Astrophys. J.
  Lett.878,32(2019)]},\ \Eprint {http://arxiv.org/abs/1904.10000}
  {arXiv:1904.10000 [astro-ph.CO]} \BibitemShut {NoStop}%
%%CITATION = ARXIV:1904.10000;%%
\bibitem [{\citenamefont {McDermott}\ \emph {et~al.}(2011)\citenamefont
  {McDermott}, \citenamefont {Yu},\ and\ \citenamefont
  {Zurek}}]{McDermott:2010pa}%
  \BibitemOpen
  \bibfield  {author} {\bibinfo {author} {\bibfnamefont {S.~D.}\ \bibnamefont
  {McDermott}}, \bibinfo {author} {\bibfnamefont {H.-B.}\ \bibnamefont {Yu}}, \
  and\ \bibinfo {author} {\bibfnamefont {K.~M.}\ \bibnamefont {Zurek}},\ }\href
  {\doibase 10.1103/PhysRevD.83.063509} {\bibfield  {journal} {\bibinfo
  {journal} {Phys. Rev.}\ }\textbf {\bibinfo {volume} {D83}},\ \bibinfo {pages}
  {063509} (\bibinfo {year} {2011})},\ \Eprint {http://arxiv.org/abs/1011.2907}
  {arXiv:1011.2907 [hep-ph]} \BibitemShut {NoStop}%
%%CITATION = ARXIV:1011.2907;%%
\bibitem [{\citenamefont {Kadota}\ \emph {et~al.}(2016)\citenamefont {Kadota},
  \citenamefont {Sekiguchi},\ and\ \citenamefont {Tashiro}}]{Kadota:2016tqq}%
  \BibitemOpen
  \bibfield  {author} {\bibinfo {author} {\bibfnamefont {K.}~\bibnamefont
  {Kadota}}, \bibinfo {author} {\bibfnamefont {T.}~\bibnamefont {Sekiguchi}}, \
  and\ \bibinfo {author} {\bibfnamefont {H.}~\bibnamefont {Tashiro}},\
  }\href@noop {} {\  (\bibinfo {year} {2016})},\ \Eprint
  {http://arxiv.org/abs/1602.04009} {arXiv:1602.04009 [astro-ph.CO]}
  \BibitemShut {NoStop}%
%%CITATION = ARXIV:1602.04009;%%
\bibitem [{\citenamefont {Stebbins}\ and\ \citenamefont
  {Krnjaic}(2019)}]{Stebbins:2019xjr}%
  \BibitemOpen
  \bibfield  {author} {\bibinfo {author} {\bibfnamefont {A.}~\bibnamefont
  {Stebbins}}\ and\ \bibinfo {author} {\bibfnamefont {G.}~\bibnamefont
  {Krnjaic}},\ }\href {\doibase 10.1088/1475-7516/2019/12/003} {\bibfield
  {journal} {\bibinfo  {journal} {JCAP}\ }\textbf {\bibinfo {volume} {1912}},\
  \bibinfo {pages} {003} (\bibinfo {year} {2019})},\ \Eprint
  {http://arxiv.org/abs/1908.05275} {arXiv:1908.05275 [astro-ph.CO]}
  \BibitemShut {NoStop}%
%%CITATION = ARXIV:1908.05275;%%
\bibitem [{\citenamefont {An}\ \emph {et~al.}(2018)\citenamefont {An},
  \citenamefont {Pospelov}, \citenamefont {Pradler},\ and\ \citenamefont
  {Ritz}}]{An:2017ojc}%
  \BibitemOpen
  \bibfield  {author} {\bibinfo {author} {\bibfnamefont {H.}~\bibnamefont
  {An}}, \bibinfo {author} {\bibfnamefont {M.}~\bibnamefont {Pospelov}},
  \bibinfo {author} {\bibfnamefont {J.}~\bibnamefont {Pradler}}, \ and\
  \bibinfo {author} {\bibfnamefont {A.}~\bibnamefont {Ritz}},\ }\href {\doibase
  10.1103/PhysRevLett.120.141801, 10.1103/PhysRevLett.121.259903} {\bibfield
  {journal} {\bibinfo  {journal} {Phys. Rev. Lett.}\ }\textbf {\bibinfo
  {volume} {120}},\ \bibinfo {pages} {141801} (\bibinfo {year} {2018})},\
  \bibinfo {note} {[Erratum: Phys. Rev. Lett.121,no.25,259903(2018)]},\ \Eprint
  {http://arxiv.org/abs/1708.03642} {arXiv:1708.03642 [hep-ph]} \BibitemShut
  {NoStop}%
%%CITATION = ARXIV:1708.03642;%%
\bibitem [{\citenamefont {Emken}\ \emph {et~al.}(2018)\citenamefont {Emken},
  \citenamefont {Kouvaris},\ and\ \citenamefont {Nielsen}}]{Emken:2017hnp}%
  \BibitemOpen
  \bibfield  {author} {\bibinfo {author} {\bibfnamefont {T.}~\bibnamefont
  {Emken}}, \bibinfo {author} {\bibfnamefont {C.}~\bibnamefont {Kouvaris}}, \
  and\ \bibinfo {author} {\bibfnamefont {N.~G.}\ \bibnamefont {Nielsen}},\
  }\href {\doibase 10.1103/PhysRevD.97.063007} {\bibfield  {journal} {\bibinfo
  {journal} {Phys. Rev.}\ }\textbf {\bibinfo {volume} {D97}},\ \bibinfo {pages}
  {063007} (\bibinfo {year} {2018})},\ \Eprint
  {http://arxiv.org/abs/1709.06573} {arXiv:1709.06573 [hep-ph]} \BibitemShut
  {NoStop}%
%%CITATION = ARXIV:1709.06573;%%
\bibitem [{\citenamefont {Bringmann}\ and\ \citenamefont
  {Pospelov}(2019)}]{Bringmann:2018cvk}%
  \BibitemOpen
  \bibfield  {author} {\bibinfo {author} {\bibfnamefont {T.}~\bibnamefont
  {Bringmann}}\ and\ \bibinfo {author} {\bibfnamefont {M.}~\bibnamefont
  {Pospelov}},\ }\href {\doibase 10.1103/PhysRevLett.122.171801} {\bibfield
  {journal} {\bibinfo  {journal} {Phys. Rev. Lett.}\ }\textbf {\bibinfo
  {volume} {122}},\ \bibinfo {pages} {171801} (\bibinfo {year} {2019})},\
  \Eprint {http://arxiv.org/abs/1810.10543} {arXiv:1810.10543 [hep-ph]}
  \BibitemShut {NoStop}%
%%CITATION = ARXIV:1810.10543;%%
\bibitem [{\citenamefont {Ema}\ \emph {et~al.}(2019)\citenamefont {Ema},
  \citenamefont {Sala},\ and\ \citenamefont {Sato}}]{Ema:2018bih}%
  \BibitemOpen
  \bibfield  {author} {\bibinfo {author} {\bibfnamefont {Y.}~\bibnamefont
  {Ema}}, \bibinfo {author} {\bibfnamefont {F.}~\bibnamefont {Sala}}, \ and\
  \bibinfo {author} {\bibfnamefont {R.}~\bibnamefont {Sato}},\ }\href {\doibase
  10.1103/PhysRevLett.122.181802} {\bibfield  {journal} {\bibinfo  {journal}
  {Phys. Rev. Lett.}\ }\textbf {\bibinfo {volume} {122}},\ \bibinfo {pages}
  {181802} (\bibinfo {year} {2019})},\ \Eprint
  {http://arxiv.org/abs/1811.00520} {arXiv:1811.00520 [hep-ph]} \BibitemShut
  {NoStop}%
%%CITATION = ARXIV:1811.00520;%%
\bibitem [{\citenamefont {Cappiello}\ and\ \citenamefont
  {Beacom}(2019)}]{Cappiello:2019qsw}%
  \BibitemOpen
  \bibfield  {author} {\bibinfo {author} {\bibfnamefont {C.}~\bibnamefont
  {Cappiello}}\ and\ \bibinfo {author} {\bibfnamefont {J.~F.}\ \bibnamefont
  {Beacom}},\ }\href {\doibase 10.1103/PhysRevD.100.103011} {\bibfield
  {journal} {\bibinfo  {journal} {Phys. Rev.}\ }\textbf {\bibinfo {volume}
  {D100}},\ \bibinfo {pages} {103011} (\bibinfo {year} {2019})},\ \Eprint
  {http://arxiv.org/abs/1906.11283} {arXiv:1906.11283 [hep-ph]} \BibitemShut
  {NoStop}%
%%CITATION = ARXIV:1906.11283;%%
\bibitem [{\citenamefont {Bondarenko}\ \emph {et~al.}(2019)\citenamefont
  {Bondarenko}, \citenamefont {Boyarsky}, \citenamefont {Bringmann},
  \citenamefont {Hufnagel}, \citenamefont {Schmidt-Hoberg},\ and\ \citenamefont
  {Sokolenko}}]{Bondarenko:2019vrb}%
  \BibitemOpen
  \bibfield  {author} {\bibinfo {author} {\bibfnamefont {K.}~\bibnamefont
  {Bondarenko}}, \bibinfo {author} {\bibfnamefont {A.}~\bibnamefont
  {Boyarsky}}, \bibinfo {author} {\bibfnamefont {T.}~\bibnamefont {Bringmann}},
  \bibinfo {author} {\bibfnamefont {M.}~\bibnamefont {Hufnagel}}, \bibinfo
  {author} {\bibfnamefont {K.}~\bibnamefont {Schmidt-Hoberg}}, \ and\ \bibinfo
  {author} {\bibfnamefont {A.}~\bibnamefont {Sokolenko}},\ }\href@noop {} {\
  (\bibinfo {year} {2019})},\ \Eprint {http://arxiv.org/abs/1909.08632}
  {arXiv:1909.08632 [hep-ph]} \BibitemShut {NoStop}%
%%CITATION = ARXIV:1909.08632;%%
\bibitem [{\citenamefont {An}\ \emph {et~al.}(2013)\citenamefont {An},
  \citenamefont {Pospelov},\ and\ \citenamefont {Pradler}}]{An:2013yfc}%
  \BibitemOpen
  \bibfield  {author} {\bibinfo {author} {\bibfnamefont {H.}~\bibnamefont
  {An}}, \bibinfo {author} {\bibfnamefont {M.}~\bibnamefont {Pospelov}}, \ and\
  \bibinfo {author} {\bibfnamefont {J.}~\bibnamefont {Pradler}},\ }\href
  {\doibase 10.1016/j.physletb.2013.07.008} {\bibfield  {journal} {\bibinfo
  {journal} {Phys. Lett.}\ }\textbf {\bibinfo {volume} {B725}},\ \bibinfo
  {pages} {190} (\bibinfo {year} {2013})},\ \Eprint
  {http://arxiv.org/abs/1302.3884} {arXiv:1302.3884 [hep-ph]} \BibitemShut
  {NoStop}%
%%CITATION = ARXIV:1302.3884;%%
\bibitem [{\citenamefont {Redondo}\ and\ \citenamefont
  {Postma}(2009)}]{Redondo:2008ec}%
  \BibitemOpen
  \bibfield  {author} {\bibinfo {author} {\bibfnamefont {J.}~\bibnamefont
  {Redondo}}\ and\ \bibinfo {author} {\bibfnamefont {M.}~\bibnamefont
  {Postma}},\ }\href {\doibase 10.1088/1475-7516/2009/02/005} {\bibfield
  {journal} {\bibinfo  {journal} {JCAP}\ }\textbf {\bibinfo {volume} {0902}},\
  \bibinfo {pages} {005} (\bibinfo {year} {2009})},\ \Eprint
  {http://arxiv.org/abs/0811.0326} {arXiv:0811.0326 [hep-ph]} \BibitemShut
  {NoStop}%
%%CITATION = ARXIV:0811.0326;%%
\bibitem [{\citenamefont {Chang}\ \emph {et~al.}(2017)\citenamefont {Chang},
  \citenamefont {Essig},\ and\ \citenamefont {McDermott}}]{Chang:2016ntp}%
  \BibitemOpen
  \bibfield  {author} {\bibinfo {author} {\bibfnamefont {J.~H.}\ \bibnamefont
  {Chang}}, \bibinfo {author} {\bibfnamefont {R.}~\bibnamefont {Essig}}, \ and\
  \bibinfo {author} {\bibfnamefont {S.~D.}\ \bibnamefont {McDermott}},\ }\href
  {\doibase 10.1007/JHEP01(2017)107} {\bibfield  {journal} {\bibinfo  {journal}
  {JHEP}\ }\textbf {\bibinfo {volume} {01}},\ \bibinfo {pages} {107} (\bibinfo
  {year} {2017})},\ \Eprint {http://arxiv.org/abs/1611.03864} {arXiv:1611.03864
  [hep-ph]} \BibitemShut {NoStop}%
%%CITATION = ARXIV:1611.03864;%%
\bibitem [{\citenamefont {Ilten}\ \emph {et~al.}(2018)\citenamefont {Ilten},
  \citenamefont {Soreq}, \citenamefont {Williams},\ and\ \citenamefont
  {Xue}}]{Ilten:2018crw}%
  \BibitemOpen
  \bibfield  {author} {\bibinfo {author} {\bibfnamefont {P.}~\bibnamefont
  {Ilten}}, \bibinfo {author} {\bibfnamefont {Y.}~\bibnamefont {Soreq}},
  \bibinfo {author} {\bibfnamefont {M.}~\bibnamefont {Williams}}, \ and\
  \bibinfo {author} {\bibfnamefont {W.}~\bibnamefont {Xue}},\ }\href {\doibase
  10.1007/JHEP06(2018)004} {\bibfield  {journal} {\bibinfo  {journal} {JHEP}\
  }\textbf {\bibinfo {volume} {06}},\ \bibinfo {pages} {004} (\bibinfo {year}
  {2018})},\ \Eprint {http://arxiv.org/abs/1801.04847} {arXiv:1801.04847
  [hep-ph]} \BibitemShut {NoStop}%
%%CITATION = ARXIV:1801.04847;%%
\bibitem [{\citenamefont {Aaij}\ \emph {et~al.}(2019)\citenamefont {Aaij} \emph
  {et~al.}}]{Aaij:2019bvg}%
  \BibitemOpen
  \bibfield  {author} {\bibinfo {author} {\bibfnamefont {R.}~\bibnamefont
  {Aaij}} \emph {et~al.} (\bibinfo {collaboration} {LHCb}),\ }\href@noop {} {\
  (\bibinfo {year} {2019})},\ \Eprint {http://arxiv.org/abs/1910.06926}
  {arXiv:1910.06926 [hep-ex]} \BibitemShut {NoStop}%
%%CITATION = ARXIV:1910.06926;%%
\bibitem [{\citenamefont {DeRocco}\ \emph
  {et~al.}(2019{\natexlab{a}})\citenamefont {DeRocco}, \citenamefont {Graham},
  \citenamefont {Kasen}, \citenamefont {Marques-Tavares},\ and\ \citenamefont
  {Rajendran}}]{DeRocco:2019njg}%
  \BibitemOpen
  \bibfield  {author} {\bibinfo {author} {\bibfnamefont {W.}~\bibnamefont
  {DeRocco}}, \bibinfo {author} {\bibfnamefont {P.~W.}\ \bibnamefont {Graham}},
  \bibinfo {author} {\bibfnamefont {D.}~\bibnamefont {Kasen}}, \bibinfo
  {author} {\bibfnamefont {G.}~\bibnamefont {Marques-Tavares}}, \ and\ \bibinfo
  {author} {\bibfnamefont {S.}~\bibnamefont {Rajendran}},\ }\href {\doibase
  10.1007/JHEP02(2019)171} {\bibfield  {journal} {\bibinfo  {journal} {JHEP}\
  }\textbf {\bibinfo {volume} {02}},\ \bibinfo {pages} {171} (\bibinfo {year}
  {2019}{\natexlab{a}})},\ \Eprint {http://arxiv.org/abs/1901.08596}
  {arXiv:1901.08596 [hep-ph]} \BibitemShut {NoStop}%
%%CITATION = ARXIV:1901.08596;%%
\bibitem [{\citenamefont {Sung}\ \emph {et~al.}(2019)\citenamefont {Sung},
  \citenamefont {Tu},\ and\ \citenamefont {Wu}}]{Sung:2019xie}%
  \BibitemOpen
  \bibfield  {author} {\bibinfo {author} {\bibfnamefont {A.}~\bibnamefont
  {Sung}}, \bibinfo {author} {\bibfnamefont {H.}~\bibnamefont {Tu}}, \ and\
  \bibinfo {author} {\bibfnamefont {M.-R.}\ \bibnamefont {Wu}},\ }\href
  {\doibase 10.1103/PhysRevD.99.121305} {\bibfield  {journal} {\bibinfo
  {journal} {Phys. Rev.}\ }\textbf {\bibinfo {volume} {D99}},\ \bibinfo {pages}
  {121305} (\bibinfo {year} {2019})},\ \Eprint
  {http://arxiv.org/abs/1903.07923} {arXiv:1903.07923 [hep-ph]} \BibitemShut
  {NoStop}%
%%CITATION = ARXIV:1903.07923;%%
\bibitem [{\citenamefont {Fradette}\ \emph {et~al.}(2014)\citenamefont
  {Fradette}, \citenamefont {Pospelov}, \citenamefont {Pradler},\ and\
  \citenamefont {Ritz}}]{Fradette:2014sza}%
  \BibitemOpen
  \bibfield  {author} {\bibinfo {author} {\bibfnamefont {A.}~\bibnamefont
  {Fradette}}, \bibinfo {author} {\bibfnamefont {M.}~\bibnamefont {Pospelov}},
  \bibinfo {author} {\bibfnamefont {J.}~\bibnamefont {Pradler}}, \ and\
  \bibinfo {author} {\bibfnamefont {A.}~\bibnamefont {Ritz}},\ }\href {\doibase
  10.1103/PhysRevD.90.035022} {\bibfield  {journal} {\bibinfo  {journal} {Phys.
  Rev.}\ }\textbf {\bibinfo {volume} {D90}},\ \bibinfo {pages} {035022}
  (\bibinfo {year} {2014})},\ \Eprint {http://arxiv.org/abs/1407.0993}
  {arXiv:1407.0993 [hep-ph]} \BibitemShut {NoStop}%
%%CITATION = ARXIV:1407.0993;%%
\bibitem [{\citenamefont {Pospelov}\ \emph {et~al.}(2008)\citenamefont
  {Pospelov}, \citenamefont {Ritz},\ and\ \citenamefont
  {Voloshin}}]{Pospelov:2007mp}%
  \BibitemOpen
  \bibfield  {author} {\bibinfo {author} {\bibfnamefont {M.}~\bibnamefont
  {Pospelov}}, \bibinfo {author} {\bibfnamefont {A.}~\bibnamefont {Ritz}}, \
  and\ \bibinfo {author} {\bibfnamefont {M.~B.}\ \bibnamefont {Voloshin}},\
  }\href {\doibase 10.1016/j.physletb.2008.02.052} {\bibfield  {journal}
  {\bibinfo  {journal} {Phys. Lett.}\ }\textbf {\bibinfo {volume} {B662}},\
  \bibinfo {pages} {53} (\bibinfo {year} {2008})},\ \Eprint
  {http://arxiv.org/abs/0711.4866} {arXiv:0711.4866 [hep-ph]} \BibitemShut
  {NoStop}%
%%CITATION = ARXIV:0711.4866;%%
\bibitem [{\citenamefont {McDermott}\ \emph {et~al.}(2018)\citenamefont
  {McDermott}, \citenamefont {Patel},\ and\ \citenamefont
  {Ramani}}]{McDermott:2017qcg}%
  \BibitemOpen
  \bibfield  {author} {\bibinfo {author} {\bibfnamefont {S.~D.}\ \bibnamefont
  {McDermott}}, \bibinfo {author} {\bibfnamefont {H.~H.}\ \bibnamefont
  {Patel}}, \ and\ \bibinfo {author} {\bibfnamefont {H.}~\bibnamefont
  {Ramani}},\ }\href {\doibase 10.1103/PhysRevD.97.073005} {\bibfield
  {journal} {\bibinfo  {journal} {Phys. Rev.}\ }\textbf {\bibinfo {volume}
  {D97}},\ \bibinfo {pages} {073005} (\bibinfo {year} {2018})},\ \Eprint
  {http://arxiv.org/abs/1705.00619} {arXiv:1705.00619 [hep-ph]} \BibitemShut
  {NoStop}%
%%CITATION = ARXIV:1705.00619;%%
\bibitem [{\citenamefont {Cassel}(2010)}]{Cassel:2009wt}%
  \BibitemOpen
  \bibfield  {author} {\bibinfo {author} {\bibfnamefont {S.}~\bibnamefont
  {Cassel}},\ }\href {\doibase 10.1088/0954-3899/37/10/105009} {\bibfield
  {journal} {\bibinfo  {journal} {J. Phys.}\ }\textbf {\bibinfo {volume}
  {G37}},\ \bibinfo {pages} {105009} (\bibinfo {year} {2010})},\ \Eprint
  {http://arxiv.org/abs/0903.5307} {arXiv:0903.5307 [hep-ph]} \BibitemShut
  {NoStop}%
%%CITATION = ARXIV:0903.5307;%%
\bibitem [{\citenamefont {Blum}\ \emph {et~al.}(2016)\citenamefont {Blum},
  \citenamefont {Sato},\ and\ \citenamefont {Slatyer}}]{Blum:2016nrz}%
  \BibitemOpen
  \bibfield  {author} {\bibinfo {author} {\bibfnamefont {K.}~\bibnamefont
  {Blum}}, \bibinfo {author} {\bibfnamefont {R.}~\bibnamefont {Sato}}, \ and\
  \bibinfo {author} {\bibfnamefont {T.~R.}\ \bibnamefont {Slatyer}},\ }\href
  {\doibase 10.1088/1475-7516/2016/06/021} {\bibfield  {journal} {\bibinfo
  {journal} {JCAP}\ }\textbf {\bibinfo {volume} {1606}},\ \bibinfo {pages}
  {021} (\bibinfo {year} {2016})},\ \Eprint {http://arxiv.org/abs/1603.01383}
  {arXiv:1603.01383 [hep-ph]} \BibitemShut {NoStop}%
%%CITATION = ARXIV:1603.01383;%%
\bibitem [{\citenamefont {Pospelov}\ and\ \citenamefont
  {Ritz}(2009)}]{Pospelov:2008jd}%
  \BibitemOpen
  \bibfield  {author} {\bibinfo {author} {\bibfnamefont {M.}~\bibnamefont
  {Pospelov}}\ and\ \bibinfo {author} {\bibfnamefont {A.}~\bibnamefont
  {Ritz}},\ }\href {\doibase 10.1016/j.physletb.2008.12.012} {\bibfield
  {journal} {\bibinfo  {journal} {Phys. Lett.}\ }\textbf {\bibinfo {volume}
  {B671}},\ \bibinfo {pages} {391} (\bibinfo {year} {2009})},\ \Eprint
  {http://arxiv.org/abs/0810.1502} {arXiv:0810.1502 [hep-ph]} \BibitemShut
  {NoStop}%
%%CITATION = ARXIV:0810.1502;%%
\bibitem [{\citenamefont {Petraki}\ \emph {et~al.}(2017)\citenamefont
  {Petraki}, \citenamefont {Postma},\ and\ \citenamefont
  {de~Vries}}]{Petraki:2016cnz}%
  \BibitemOpen
  \bibfield  {author} {\bibinfo {author} {\bibfnamefont {K.}~\bibnamefont
  {Petraki}}, \bibinfo {author} {\bibfnamefont {M.}~\bibnamefont {Postma}}, \
  and\ \bibinfo {author} {\bibfnamefont {J.}~\bibnamefont {de~Vries}},\ }\href
  {\doibase 10.1007/JHEP04(2017)077} {\bibfield  {journal} {\bibinfo  {journal}
  {JHEP}\ }\textbf {\bibinfo {volume} {04}},\ \bibinfo {pages} {077} (\bibinfo
  {year} {2017})},\ \Eprint {http://arxiv.org/abs/1611.01394} {arXiv:1611.01394
  [hep-ph]} \BibitemShut {NoStop}%
%%CITATION = ARXIV:1611.01394;%%
\bibitem [{\citenamefont {Cirelli}\ \emph {et~al.}(2017)\citenamefont
  {Cirelli}, \citenamefont {Panci}, \citenamefont {Petraki}, \citenamefont
  {Sala},\ and\ \citenamefont {Taoso}}]{Cirelli:2016rnw}%
  \BibitemOpen
  \bibfield  {author} {\bibinfo {author} {\bibfnamefont {M.}~\bibnamefont
  {Cirelli}}, \bibinfo {author} {\bibfnamefont {P.}~\bibnamefont {Panci}},
  \bibinfo {author} {\bibfnamefont {K.}~\bibnamefont {Petraki}}, \bibinfo
  {author} {\bibfnamefont {F.}~\bibnamefont {Sala}}, \ and\ \bibinfo {author}
  {\bibfnamefont {M.}~\bibnamefont {Taoso}},\ }\href {\doibase
  10.1088/1475-7516/2017/05/036} {\bibfield  {journal} {\bibinfo  {journal}
  {JCAP}\ }\textbf {\bibinfo {volume} {1705}},\ \bibinfo {pages} {036}
  (\bibinfo {year} {2017})},\ \Eprint {http://arxiv.org/abs/1612.07295}
  {arXiv:1612.07295 [hep-ph]} \BibitemShut {NoStop}%
%%CITATION = ARXIV:1612.07295;%%
\bibitem [{\citenamefont {Pospelov}\ and\ \citenamefont
  {Pradler}(2010)}]{Pospelov:2010hj}%
  \BibitemOpen
  \bibfield  {author} {\bibinfo {author} {\bibfnamefont {M.}~\bibnamefont
  {Pospelov}}\ and\ \bibinfo {author} {\bibfnamefont {J.}~\bibnamefont
  {Pradler}},\ }\href {\doibase 10.1146/annurev.nucl.012809.104521} {\bibfield
  {journal} {\bibinfo  {journal} {Ann. Rev. Nucl. Part. Sci.}\ }\textbf
  {\bibinfo {volume} {60}},\ \bibinfo {pages} {539} (\bibinfo {year} {2010})},\
  \Eprint {http://arxiv.org/abs/1011.1054} {arXiv:1011.1054 [hep-ph]}
  \BibitemShut {NoStop}%
%%CITATION = ARXIV:1011.1054;%%
\bibitem [{\citenamefont {Kawasaki}\ \emph {et~al.}(2015)\citenamefont
  {Kawasaki}, \citenamefont {Kohri}, \citenamefont {Moroi},\ and\ \citenamefont
  {Takaesu}}]{Kawasaki:2015yya}%
  \BibitemOpen
  \bibfield  {author} {\bibinfo {author} {\bibfnamefont {M.}~\bibnamefont
  {Kawasaki}}, \bibinfo {author} {\bibfnamefont {K.}~\bibnamefont {Kohri}},
  \bibinfo {author} {\bibfnamefont {T.}~\bibnamefont {Moroi}}, \ and\ \bibinfo
  {author} {\bibfnamefont {Y.}~\bibnamefont {Takaesu}},\ }\href {\doibase
  10.1016/j.physletb.2015.10.048} {\bibfield  {journal} {\bibinfo  {journal}
  {Phys. Lett.}\ }\textbf {\bibinfo {volume} {B751}},\ \bibinfo {pages} {246}
  (\bibinfo {year} {2015})},\ \Eprint {http://arxiv.org/abs/1509.03665}
  {arXiv:1509.03665 [hep-ph]} \BibitemShut {NoStop}%
%%CITATION = ARXIV:1509.03665;%%
\bibitem [{\citenamefont {Depta}\ \emph {et~al.}(2019)\citenamefont {Depta},
  \citenamefont {Hufnagel}, \citenamefont {Schmidt-Hoberg},\ and\ \citenamefont
  {Wild}}]{Depta:2019lbe}%
  \BibitemOpen
  \bibfield  {author} {\bibinfo {author} {\bibfnamefont {P.~F.}\ \bibnamefont
  {Depta}}, \bibinfo {author} {\bibfnamefont {M.}~\bibnamefont {Hufnagel}},
  \bibinfo {author} {\bibfnamefont {K.}~\bibnamefont {Schmidt-Hoberg}}, \ and\
  \bibinfo {author} {\bibfnamefont {S.}~\bibnamefont {Wild}},\ }\href {\doibase
  10.1088/1475-7516/2019/04/029} {\bibfield  {journal} {\bibinfo  {journal}
  {JCAP}\ }\textbf {\bibinfo {volume} {1904}},\ \bibinfo {pages} {029}
  (\bibinfo {year} {2019})},\ \Eprint {http://arxiv.org/abs/1901.06944}
  {arXiv:1901.06944 [hep-ph]} \BibitemShut {NoStop}%
%%CITATION = ARXIV:1901.06944;%%
\bibitem [{\citenamefont {Slatyer}(2016)}]{Slatyer:2015jla}%
  \BibitemOpen
  \bibfield  {author} {\bibinfo {author} {\bibfnamefont {T.~R.}\ \bibnamefont
  {Slatyer}},\ }\href {\doibase 10.1103/PhysRevD.93.023527} {\bibfield
  {journal} {\bibinfo  {journal} {Phys. Rev.}\ }\textbf {\bibinfo {volume}
  {D93}},\ \bibinfo {pages} {023527} (\bibinfo {year} {2016})},\ \Eprint
  {http://arxiv.org/abs/1506.03811} {arXiv:1506.03811 [hep-ph]} \BibitemShut
  {NoStop}%
%%CITATION = ARXIV:1506.03811;%%
\bibitem [{\citenamefont {Ade}\ \emph {et~al.}(2016)\citenamefont {Ade} \emph
  {et~al.}}]{Ade:2015xua}%
  \BibitemOpen
  \bibfield  {author} {\bibinfo {author} {\bibfnamefont {P.~A.~R.}\
  \bibnamefont {Ade}} \emph {et~al.} (\bibinfo {collaboration} {Planck}),\
  }\href {\doibase 10.1051/0004-6361/201525830} {\bibfield  {journal} {\bibinfo
   {journal} {Astron. Astrophys.}\ }\textbf {\bibinfo {volume} {594}},\
  \bibinfo {pages} {A13} (\bibinfo {year} {2016})},\ \Eprint
  {http://arxiv.org/abs/1502.01589} {arXiv:1502.01589 [astro-ph.CO]}
  \BibitemShut {NoStop}%
%%CITATION = ARXIV:1502.01589;%%
\bibitem [{\citenamefont {Elor}\ \emph {et~al.}(2016)\citenamefont {Elor},
  \citenamefont {Rodd}, \citenamefont {Slatyer},\ and\ \citenamefont
  {Xue}}]{Elor:2015bho}%
  \BibitemOpen
  \bibfield  {author} {\bibinfo {author} {\bibfnamefont {G.}~\bibnamefont
  {Elor}}, \bibinfo {author} {\bibfnamefont {N.~L.}\ \bibnamefont {Rodd}},
  \bibinfo {author} {\bibfnamefont {T.~R.}\ \bibnamefont {Slatyer}}, \ and\
  \bibinfo {author} {\bibfnamefont {W.}~\bibnamefont {Xue}},\ }\href {\doibase
  10.1088/1475-7516/2016/06/024} {\bibfield  {journal} {\bibinfo  {journal}
  {JCAP}\ }\textbf {\bibinfo {volume} {1606}},\ \bibinfo {pages} {024}
  (\bibinfo {year} {2016})},\ \Eprint {http://arxiv.org/abs/1511.08787}
  {arXiv:1511.08787 [hep-ph]} \BibitemShut {NoStop}%
%%CITATION = ARXIV:1511.08787;%%
\bibitem [{\citenamefont {Ackermann}\ \emph {et~al.}(2015)\citenamefont
  {Ackermann} \emph {et~al.}}]{Ackermann:2015tah}%
  \BibitemOpen
  \bibfield  {author} {\bibinfo {author} {\bibfnamefont {M.}~\bibnamefont
  {Ackermann}} \emph {et~al.} (\bibinfo {collaboration} {Fermi-LAT}),\ }\href
  {\doibase 10.1088/1475-7516/2015/09/008} {\bibfield  {journal} {\bibinfo
  {journal} {JCAP}\ }\textbf {\bibinfo {volume} {1509}},\ \bibinfo {pages}
  {008} (\bibinfo {year} {2015})},\ \Eprint {http://arxiv.org/abs/1501.05464}
  {arXiv:1501.05464 [astro-ph.CO]} \BibitemShut {NoStop}%
%%CITATION = ARXIV:1501.05464;%%
\bibitem [{\citenamefont {Liu}\ \emph {et~al.}(2017)\citenamefont {Liu},
  \citenamefont {Bi}, \citenamefont {Lin},\ and\ \citenamefont
  {Yin}}]{Liu:2016ngs}%
  \BibitemOpen
  \bibfield  {author} {\bibinfo {author} {\bibfnamefont {W.}~\bibnamefont
  {Liu}}, \bibinfo {author} {\bibfnamefont {X.-J.}\ \bibnamefont {Bi}},
  \bibinfo {author} {\bibfnamefont {S.-J.}\ \bibnamefont {Lin}}, \ and\
  \bibinfo {author} {\bibfnamefont {P.-F.}\ \bibnamefont {Yin}},\ }\href
  {\doibase 10.1088/1674-1137/41/4/045104} {\bibfield  {journal} {\bibinfo
  {journal} {Chin. Phys.}\ }\textbf {\bibinfo {volume} {C41}},\ \bibinfo
  {pages} {045104} (\bibinfo {year} {2017})},\ \Eprint
  {http://arxiv.org/abs/1602.01012} {arXiv:1602.01012 [astro-ph.CO]}
  \BibitemShut {NoStop}%
%%CITATION = ARXIV:1602.01012;%%
\bibitem [{\citenamefont {Slatyer}\ and\ \citenamefont
  {Wu}(2017)}]{Slatyer:2016qyl}%
  \BibitemOpen
  \bibfield  {author} {\bibinfo {author} {\bibfnamefont {T.~R.}\ \bibnamefont
  {Slatyer}}\ and\ \bibinfo {author} {\bibfnamefont {C.-L.}\ \bibnamefont
  {Wu}},\ }\href {\doibase 10.1103/PhysRevD.95.023010} {\bibfield  {journal}
  {\bibinfo  {journal} {Phys. Rev.}\ }\textbf {\bibinfo {volume} {D95}},\
  \bibinfo {pages} {023010} (\bibinfo {year} {2017})},\ \Eprint
  {http://arxiv.org/abs/1610.06933} {arXiv:1610.06933 [astro-ph.CO]}
  \BibitemShut {NoStop}%
%%CITATION = ARXIV:1610.06933;%%
\bibitem [{\citenamefont {Slatyer}(2013)}]{Slatyer:2012yq}%
  \BibitemOpen
  \bibfield  {author} {\bibinfo {author} {\bibfnamefont {T.~R.}\ \bibnamefont
  {Slatyer}},\ }\href {\doibase 10.1103/PhysRevD.87.123513} {\bibfield
  {journal} {\bibinfo  {journal} {Phys. Rev.}\ }\textbf {\bibinfo {volume}
  {D87}},\ \bibinfo {pages} {123513} (\bibinfo {year} {2013})},\ \Eprint
  {http://arxiv.org/abs/1211.0283} {arXiv:1211.0283 [astro-ph.CO]} \BibitemShut
  {NoStop}%
%%CITATION = ARXIV:1211.0283;%%
\bibitem [{\citenamefont {Essig}\ \emph {et~al.}(2013)\citenamefont {Essig},
  \citenamefont {Kuflik}, \citenamefont {McDermott}, \citenamefont {Volansky},\
  and\ \citenamefont {Zurek}}]{Essig:2013goa}%
  \BibitemOpen
  \bibfield  {author} {\bibinfo {author} {\bibfnamefont {R.}~\bibnamefont
  {Essig}}, \bibinfo {author} {\bibfnamefont {E.}~\bibnamefont {Kuflik}},
  \bibinfo {author} {\bibfnamefont {S.~D.}\ \bibnamefont {McDermott}}, \bibinfo
  {author} {\bibfnamefont {T.}~\bibnamefont {Volansky}}, \ and\ \bibinfo
  {author} {\bibfnamefont {K.~M.}\ \bibnamefont {Zurek}},\ }\href {\doibase
  10.1007/JHEP11(2013)193} {\bibfield  {journal} {\bibinfo  {journal} {JHEP}\
  }\textbf {\bibinfo {volume} {11}},\ \bibinfo {pages} {193} (\bibinfo {year}
  {2013})},\ \Eprint {http://arxiv.org/abs/1309.4091} {arXiv:1309.4091
  [hep-ph]} \BibitemShut {NoStop}%
%%CITATION = ARXIV:1309.4091;%%
\bibitem [{\citenamefont {Abdallah}\ \emph {et~al.}(2016)\citenamefont
  {Abdallah} \emph {et~al.}}]{Abdallah:2016ygi}%
  \BibitemOpen
  \bibfield  {author} {\bibinfo {author} {\bibfnamefont {H.}~\bibnamefont
  {Abdallah}} \emph {et~al.} (\bibinfo {collaboration} {H.E.S.S.}),\ }\href
  {\doibase 10.1103/PhysRevLett.117.111301} {\bibfield  {journal} {\bibinfo
  {journal} {Phys. Rev. Lett.}\ }\textbf {\bibinfo {volume} {117}},\ \bibinfo
  {pages} {111301} (\bibinfo {year} {2016})},\ \Eprint
  {http://arxiv.org/abs/1607.08142} {arXiv:1607.08142 [astro-ph.HE]}
  \BibitemShut {NoStop}%
%%CITATION = ARXIV:1607.08142;%%
\bibitem [{\citenamefont {Chu}\ \emph {et~al.}(2017)\citenamefont {Chu},
  \citenamefont {Kulkarni},\ and\ \citenamefont {Salati}}]{Chu:2017vao}%
  \BibitemOpen
  \bibfield  {author} {\bibinfo {author} {\bibfnamefont {X.}~\bibnamefont
  {Chu}}, \bibinfo {author} {\bibfnamefont {S.}~\bibnamefont {Kulkarni}}, \
  and\ \bibinfo {author} {\bibfnamefont {P.}~\bibnamefont {Salati}},\ }\href
  {\doibase 10.1088/1475-7516/2017/11/023} {\bibfield  {journal} {\bibinfo
  {journal} {JCAP}\ }\textbf {\bibinfo {volume} {1711}},\ \bibinfo {pages}
  {023} (\bibinfo {year} {2017})},\ \Eprint {http://arxiv.org/abs/1706.08543}
  {arXiv:1706.08543 [hep-ph]} \BibitemShut {NoStop}%
%%CITATION = ARXIV:1706.08543;%%
\bibitem [{\citenamefont {Ajello}\ \emph {et~al.}(2011)\citenamefont {Ajello}
  \emph {et~al.}}]{Ajello:2011dq}%
  \BibitemOpen
  \bibfield  {author} {\bibinfo {author} {\bibfnamefont {M.}~\bibnamefont
  {Ajello}} \emph {et~al.} (\bibinfo {collaboration} {The Fermi LAT}),\ }\href
  {\doibase 10.1103/PhysRevD.84.032007} {\bibfield  {journal} {\bibinfo
  {journal} {Phys. Rev.}\ }\textbf {\bibinfo {volume} {D84}},\ \bibinfo {pages}
  {032007} (\bibinfo {year} {2011})},\ \Eprint {http://arxiv.org/abs/1107.4272}
  {arXiv:1107.4272 [astro-ph.HE]} \BibitemShut {NoStop}%
%%CITATION = ARXIV:1107.4272;%%
\bibitem [{\citenamefont {Feng}\ \emph {et~al.}(2016)\citenamefont {Feng},
  \citenamefont {Smolinsky},\ and\ \citenamefont {Tanedo}}]{Feng:2016ijc}%
  \BibitemOpen
  \bibfield  {author} {\bibinfo {author} {\bibfnamefont {J.~L.}\ \bibnamefont
  {Feng}}, \bibinfo {author} {\bibfnamefont {J.}~\bibnamefont {Smolinsky}}, \
  and\ \bibinfo {author} {\bibfnamefont {P.}~\bibnamefont {Tanedo}},\ }\href
  {\doibase 10.1103/PhysRevD.93.115036, 10.1103/PhysRevD.96.099903} {\bibfield
  {journal} {\bibinfo  {journal} {Phys. Rev.}\ }\textbf {\bibinfo {volume}
  {D93}},\ \bibinfo {pages} {115036} (\bibinfo {year} {2016})},\ \bibinfo
  {note} {[Erratum: Phys. Rev.D96,no.9,099903(2017)]},\ \Eprint
  {http://arxiv.org/abs/1602.01465} {arXiv:1602.01465 [hep-ph]} \BibitemShut
  {NoStop}%
%%CITATION = ARXIV:1602.01465;%%
\bibitem [{\citenamefont {Leane}\ \emph {et~al.}(2017)\citenamefont {Leane},
  \citenamefont {Ng},\ and\ \citenamefont {Beacom}}]{Leane:2017vag}%
  \BibitemOpen
  \bibfield  {author} {\bibinfo {author} {\bibfnamefont {R.~K.}\ \bibnamefont
  {Leane}}, \bibinfo {author} {\bibfnamefont {K.~C.~Y.}\ \bibnamefont {Ng}}, \
  and\ \bibinfo {author} {\bibfnamefont {J.~F.}\ \bibnamefont {Beacom}},\
  }\href {\doibase 10.1103/PhysRevD.95.123016} {\bibfield  {journal} {\bibinfo
  {journal} {Phys. Rev.}\ }\textbf {\bibinfo {volume} {D95}},\ \bibinfo {pages}
  {123016} (\bibinfo {year} {2017})},\ \Eprint
  {http://arxiv.org/abs/1703.04629} {arXiv:1703.04629 [astro-ph.HE]}
  \BibitemShut {NoStop}%
%%CITATION = ARXIV:1703.04629;%%
\bibitem [{\citenamefont {Robertson}\ and\ \citenamefont
  {Albuquerque}(2018)}]{Robertson:2017hdw}%
  \BibitemOpen
  \bibfield  {author} {\bibinfo {author} {\bibfnamefont {D.~S.}\ \bibnamefont
  {Robertson}}\ and\ \bibinfo {author} {\bibfnamefont {I.~F.~M.}\ \bibnamefont
  {Albuquerque}},\ }\href {\doibase 10.1088/1475-7516/2018/02/056} {\bibfield
  {journal} {\bibinfo  {journal} {JCAP}\ }\textbf {\bibinfo {volume} {1802}},\
  \bibinfo {pages} {056} (\bibinfo {year} {2018})},\ \Eprint
  {http://arxiv.org/abs/1711.02052} {arXiv:1711.02052 [astro-ph.CO]}
  \BibitemShut {NoStop}%
%%CITATION = ARXIV:1711.02052;%%
\bibitem [{\citenamefont {Albert}\ \emph {et~al.}(2018)\citenamefont {Albert}
  \emph {et~al.}}]{Albert:2018jwh}%
  \BibitemOpen
  \bibfield  {author} {\bibinfo {author} {\bibfnamefont {A.}~\bibnamefont
  {Albert}} \emph {et~al.} (\bibinfo {collaboration} {HAWC}),\ }\href {\doibase
  10.1103/PhysRevD.98.123012} {\bibfield  {journal} {\bibinfo  {journal} {Phys.
  Rev.}\ }\textbf {\bibinfo {volume} {D98}},\ \bibinfo {pages} {123012}
  (\bibinfo {year} {2018})},\ \Eprint {http://arxiv.org/abs/1808.05624}
  {arXiv:1808.05624 [hep-ph]} \BibitemShut {NoStop}%
%%CITATION = ARXIV:1808.05624;%%
\bibitem [{\citenamefont {Zentner}(2009)}]{Zentner:2009is}%
  \BibitemOpen
  \bibfield  {author} {\bibinfo {author} {\bibfnamefont {A.~R.}\ \bibnamefont
  {Zentner}},\ }\href {\doibase 10.1103/PhysRevD.80.063501} {\bibfield
  {journal} {\bibinfo  {journal} {Phys. Rev.}\ }\textbf {\bibinfo {volume}
  {D80}},\ \bibinfo {pages} {063501} (\bibinfo {year} {2009})},\ \Eprint
  {http://arxiv.org/abs/0907.3448} {arXiv:0907.3448 [astro-ph.HE]} \BibitemShut
  {NoStop}%
%%CITATION = ARXIV:0907.3448;%%
\bibitem [{\citenamefont {Kouvaris}\ \emph {et~al.}(2016)\citenamefont
  {Kouvaris}, \citenamefont {Langæble},\ and\ \citenamefont
  {Nielsen}}]{Kouvaris:2016ltf}%
  \BibitemOpen
  \bibfield  {author} {\bibinfo {author} {\bibfnamefont {C.}~\bibnamefont
  {Kouvaris}}, \bibinfo {author} {\bibfnamefont {K.}~\bibnamefont {Langæble}},
  \ and\ \bibinfo {author} {\bibfnamefont {N.~G.}\ \bibnamefont {Nielsen}},\
  }\href {\doibase 10.1088/1475-7516/2016/10/012} {\bibfield  {journal}
  {\bibinfo  {journal} {JCAP}\ }\textbf {\bibinfo {volume} {1610}},\ \bibinfo
  {pages} {012} (\bibinfo {year} {2016})},\ \Eprint
  {http://arxiv.org/abs/1607.00374} {arXiv:1607.00374 [hep-ph]} \BibitemShut
  {NoStop}%
%%CITATION = ARXIV:1607.00374;%%
\bibitem [{\citenamefont {Gaidau}\ and\ \citenamefont
  {Shelton}(2019)}]{Gaidau:2018yws}%
  \BibitemOpen
  \bibfield  {author} {\bibinfo {author} {\bibfnamefont {C.}~\bibnamefont
  {Gaidau}}\ and\ \bibinfo {author} {\bibfnamefont {J.}~\bibnamefont
  {Shelton}},\ }\href {\doibase 10.1088/1475-7516/2019/06/022} {\bibfield
  {journal} {\bibinfo  {journal} {JCAP}\ }\textbf {\bibinfo {volume} {1906}},\
  \bibinfo {pages} {022} (\bibinfo {year} {2019})},\ \Eprint
  {http://arxiv.org/abs/1811.00557} {arXiv:1811.00557 [hep-ph]} \BibitemShut
  {NoStop}%
%%CITATION = ARXIV:1811.00557;%%
\bibitem [{\citenamefont {Agashe}\ \emph {et~al.}(2014)\citenamefont {Agashe},
  \citenamefont {Cui}, \citenamefont {Necib},\ and\ \citenamefont
  {Thaler}}]{Agashe:2014yua}%
  \BibitemOpen
  \bibfield  {author} {\bibinfo {author} {\bibfnamefont {K.}~\bibnamefont
  {Agashe}}, \bibinfo {author} {\bibfnamefont {Y.}~\bibnamefont {Cui}},
  \bibinfo {author} {\bibfnamefont {L.}~\bibnamefont {Necib}}, \ and\ \bibinfo
  {author} {\bibfnamefont {J.}~\bibnamefont {Thaler}},\ }\href {\doibase
  10.1088/1475-7516/2014/10/062} {\bibfield  {journal} {\bibinfo  {journal}
  {JCAP}\ }\textbf {\bibinfo {volume} {1410}},\ \bibinfo {pages} {062}
  (\bibinfo {year} {2014})},\ \Eprint {http://arxiv.org/abs/1405.7370}
  {arXiv:1405.7370 [hep-ph]} \BibitemShut {NoStop}%
%%CITATION = ARXIV:1405.7370;%%
\bibitem [{\citenamefont {Cui}\ \emph {et~al.}(2018)\citenamefont {Cui},
  \citenamefont {Pospelov},\ and\ \citenamefont {Pradler}}]{Cui:2017ytb}%
  \BibitemOpen
  \bibfield  {author} {\bibinfo {author} {\bibfnamefont {Y.}~\bibnamefont
  {Cui}}, \bibinfo {author} {\bibfnamefont {M.}~\bibnamefont {Pospelov}}, \
  and\ \bibinfo {author} {\bibfnamefont {J.}~\bibnamefont {Pradler}},\ }\href
  {\doibase 10.1103/PhysRevD.97.103004} {\bibfield  {journal} {\bibinfo
  {journal} {Phys. Rev.}\ }\textbf {\bibinfo {volume} {D97}},\ \bibinfo {pages}
  {103004} (\bibinfo {year} {2018})},\ \Eprint
  {http://arxiv.org/abs/1711.04531} {arXiv:1711.04531 [hep-ph]} \BibitemShut
  {NoStop}%
%%CITATION = ARXIV:1711.04531;%%
\bibitem [{\citenamefont {DeRocco}\ \emph
  {et~al.}(2019{\natexlab{b}})\citenamefont {DeRocco}, \citenamefont {Graham},
  \citenamefont {Kasen}, \citenamefont {Marques-Tavares},\ and\ \citenamefont
  {Rajendran}}]{DeRocco:2019jti}%
  \BibitemOpen
  \bibfield  {author} {\bibinfo {author} {\bibfnamefont {W.}~\bibnamefont
  {DeRocco}}, \bibinfo {author} {\bibfnamefont {P.~W.}\ \bibnamefont {Graham}},
  \bibinfo {author} {\bibfnamefont {D.}~\bibnamefont {Kasen}}, \bibinfo
  {author} {\bibfnamefont {G.}~\bibnamefont {Marques-Tavares}}, \ and\ \bibinfo
  {author} {\bibfnamefont {S.}~\bibnamefont {Rajendran}},\ }\href {\doibase
  10.1103/PhysRevD.100.075018} {\bibfield  {journal} {\bibinfo  {journal}
  {Phys. Rev.}\ }\textbf {\bibinfo {volume} {D100}},\ \bibinfo {pages} {075018}
  (\bibinfo {year} {2019}{\natexlab{b}})},\ \Eprint
  {http://arxiv.org/abs/1905.09284} {arXiv:1905.09284 [hep-ph]} \BibitemShut
  {NoStop}%
%%CITATION = ARXIV:1905.09284;%%
\bibitem [{\citenamefont {Essig}\ \emph {et~al.}(2019)\citenamefont {Essig},
  \citenamefont {Mcdermott}, \citenamefont {Yu},\ and\ \citenamefont
  {Zhong}}]{Essig:2018pzq}%
  \BibitemOpen
  \bibfield  {author} {\bibinfo {author} {\bibfnamefont {R.}~\bibnamefont
  {Essig}}, \bibinfo {author} {\bibfnamefont {S.~D.}\ \bibnamefont
  {Mcdermott}}, \bibinfo {author} {\bibfnamefont {H.-B.}\ \bibnamefont {Yu}}, \
  and\ \bibinfo {author} {\bibfnamefont {Y.-M.}\ \bibnamefont {Zhong}},\ }\href
  {\doibase 10.1103/PhysRevLett.123.121102} {\bibfield  {journal} {\bibinfo
  {journal} {Phys. Rev. Lett.}\ }\textbf {\bibinfo {volume} {123}},\ \bibinfo
  {pages} {121102} (\bibinfo {year} {2019})},\ \Eprint
  {http://arxiv.org/abs/1809.01144} {arXiv:1809.01144 [hep-ph]} \BibitemShut
  {NoStop}%
%%CITATION = ARXIV:1809.01144;%%
\bibitem [{\citenamefont {Chang}\ \emph {et~al.}(2019)\citenamefont {Chang},
  \citenamefont {Egana-Ugrinovic}, \citenamefont {Essig},\ and\ \citenamefont
  {Kouvaris}}]{Chang:2018bgx}%
  \BibitemOpen
  \bibfield  {author} {\bibinfo {author} {\bibfnamefont {J.~H.}\ \bibnamefont
  {Chang}}, \bibinfo {author} {\bibfnamefont {D.}~\bibnamefont
  {Egana-Ugrinovic}}, \bibinfo {author} {\bibfnamefont {R.}~\bibnamefont
  {Essig}}, \ and\ \bibinfo {author} {\bibfnamefont {C.}~\bibnamefont
  {Kouvaris}},\ }\href {\doibase 10.1088/1475-7516/2019/03/036} {\bibfield
  {journal} {\bibinfo  {journal} {JCAP}\ }\textbf {\bibinfo {volume} {1903}},\
  \bibinfo {pages} {036} (\bibinfo {year} {2019})},\ \Eprint
  {http://arxiv.org/abs/1812.07000} {arXiv:1812.07000 [hep-ph]} \BibitemShut
  {NoStop}%
%%CITATION = ARXIV:1812.07000;%%
\bibitem [{\citenamefont {Rothstein}\ \emph {et~al.}(2009)\citenamefont
  {Rothstein}, \citenamefont {Schwetz},\ and\ \citenamefont
  {Zupan}}]{Rothstein:2009pm}%
  \BibitemOpen
  \bibfield  {author} {\bibinfo {author} {\bibfnamefont {I.~Z.}\ \bibnamefont
  {Rothstein}}, \bibinfo {author} {\bibfnamefont {T.}~\bibnamefont {Schwetz}},
  \ and\ \bibinfo {author} {\bibfnamefont {J.}~\bibnamefont {Zupan}},\ }\href
  {\doibase 10.1088/1475-7516/2009/07/018} {\bibfield  {journal} {\bibinfo
  {journal} {JCAP}\ }\textbf {\bibinfo {volume} {0907}},\ \bibinfo {pages}
  {018} (\bibinfo {year} {2009})},\ \Eprint {http://arxiv.org/abs/0903.3116}
  {arXiv:0903.3116 [astro-ph.HE]} \BibitemShut {NoStop}%
%%CITATION = ARXIV:0903.3116;%%
\bibitem [{\citenamefont {P\'erez}\ and\ \citenamefont
  {Granger}(2007)}]{Perez:2007emg}%
  \BibitemOpen
  \bibfield  {author} {\bibinfo {author} {\bibfnamefont {F.}~\bibnamefont
  {P\'erez}}\ and\ \bibinfo {author} {\bibfnamefont {B.~E.}\ \bibnamefont
  {Granger}},\ }\href {\doibase 10.1109/MCSE.2007.53} {\bibfield  {journal}
  {\bibinfo  {journal} {Comput. Sci. Eng.}\ }\textbf {\bibinfo {volume} {9}},\
  \bibinfo {pages} {21} (\bibinfo {year} {2007})}\BibitemShut {NoStop}%
%%CITATION = CSENF,9,21;%%
\bibitem [{\citenamefont {Hunter}(2007)}]{Hunter:2007ouj}%
  \BibitemOpen
  \bibfield  {author} {\bibinfo {author} {\bibfnamefont {J.~D.}\ \bibnamefont
  {Hunter}},\ }\href {\doibase 10.1109/MCSE.2007.55} {\bibfield  {journal}
  {\bibinfo  {journal} {Comput. Sci. Eng.}\ }\textbf {\bibinfo {volume} {9}},\
  \bibinfo {pages} {90} (\bibinfo {year} {2007})}\BibitemShut {NoStop}%
%%CITATION = CSENF,9,90;%%
\bibitem [{\citenamefont {Jones}\ \emph {et~al.}(01  )\citenamefont {Jones},
  \citenamefont {Oliphant}, \citenamefont {Peterson} \emph {et~al.}}]{SciPy}%
  \BibitemOpen
  \bibfield  {author} {\bibinfo {author} {\bibfnamefont {E.}~\bibnamefont
  {Jones}}, \bibinfo {author} {\bibfnamefont {T.}~\bibnamefont {Oliphant}},
  \bibinfo {author} {\bibfnamefont {P.}~\bibnamefont {Peterson}},  \emph
  {et~al.},\ }\href {http://www.scipy.org/} {\enquote {\bibinfo {title}
  {{SciPy}: Open source scientific tools for {Python}},}\ } (\bibinfo {year}
  {2001--})\BibitemShut {NoStop}%
\end{thebibliography}%

\end{document}